%% file: main.tex
\documentclass[a4paper]{article}

\usepackage[english]{babel}
\usepackage[hidelinks]{hyperref}

\input{packages}

\newcommand{\Oof}{\mathcal{O}}
\newcommand{\Oh}{\Oof}
\renewcommand{\preceq}{\preccurlyeq}

\input{macros}
\usepackage{palatino}
\usepackage{mathpazo}
\usepackage{comment}
\usepackage{geometry}
\usepackage{todonotes}

\usepackage{tikz}
\usetikzlibrary{arrows.meta}

\usepackage[breakable,skins]{tcolorbox}
\tcbset{skin=enhanced, fonttitle=\bfseries,breakable, coltitle=black,colbacktitle=orange!10!white, colback=yellow!3!white}

\def\xInd#1#2{#1\setbox0=\hbox{$#1x$}\kern\wd0\hbox to 0pt{\hss$#1\mid$\hss}
	\lower.9\ht0\hbox to 0pt{\hss$#1\smile$\hss}\kern\wd0}

\def\xind{\mathop{\mathpalette\xInd{}}} 

\def\xnotind#1#2{#1\setbox0=\hbox{$#1x$}\kern\wd0
	\hbox to 0pt{\mathchardef\nn=12854\hss$#1\nn$\kern1.4\wd0\hss}
	\hbox to 0pt{\hss$#1\mid$\hss}\lower.9\ht0 \hbox to 0pt{\hss$#1\smile$\hss}\kern\wd0}

\def\xnind{\mathop{\mathpalette\xnotind{}}} 
\newcommand{\ind}[2][]{\xind_{#1}^{#2}}
\newcommand{\nind}[2][]{\xnind_{#1}^{#2}}
  \newcommand{\Types}[1][]{\mathrm{Types}^{#1}}
  
  \newcommand{\flip}[1]{\mathsf{#1}}

\newcommand{\tup}{\bar}

\newcommand{\from}{\colon}
\newcommand{\str}[1]{\mathbf{#1}}
\renewcommand{\cal}[1]{\mathcal {#1}}
\newcommand{\F}{\cal F}
\renewcommand{\le}{\leqslant}
\renewcommand{\ge}{\geqslant}
\renewcommand{\leq}{\leqslant}
\renewcommand{\geq}{\geqslant}
\renewcommand{\phi}{\varphi}

\newcommand{\setof}[2]{\set{#1\mid#2}}

\DeclareMathOperator{\dist}{dist}

\newcommand{\ERCagreement}{
\ifx\anonymous\undefined
This paper is part of projects that have received funding from the European Research Council (ERC) (grant agreement No 948057 -- {\sc BOBR}) and from the German
Research Foundation (DFG) with grant agreement
No 444419611.
\else
Anonymous funding.
\fi
}

\begin{document}

\title{Flipper games for monadically stable graph classes\thanks{\ERCagreement}}

\ifx\anonymous\undefined
\author{Jakub Gajarsk{\'y}\thanks{University of Warsaw, Poland} \and Nikolas M\"ahlmann\thanks{University of Bremen, Germany} \and Rose McCarty\thanks{Princeton University, USA. Supported by European Research Council (ERC) grant No. 714704 -- {\sc CUTACOMBS} and National Science Foundation (NSF) grant No. DMS-2202961.}  \and Pierre Ohlmann$^\dagger$\and
Micha{\l} Pilipczuk$^\dagger$\and Wojciech Przybyszewski$^\dagger$ \and Sebastian Siebertz$^\ddagger$ \and Marek Sokołowski$^\dagger$ \and Szymon Toru{\'n}czyk$^\dagger$ }
\else
\author{Anonymous flippers}
\fi

\date{}

\maketitle
\begin{abstract}
 A class of graphs $\Cc$ is {\em{monadically stable}} if for any unary expansion $\widehat{\Cc}$ of $\Cc$, one cannot interpret, in first-order logic, arbitrarily long linear orders in graphs from $\widehat{\Cc}$. It is known that nowhere dense graph classes are monadically stable; these encompass most of the studied concepts of sparsity in graphs, including classes of graphs that exclude a fixed topological minor. On the other hand, monadic stability is a property expressed in purely model-theoretic terms and hence it is also suited for capturing structure in dense graphs.
 
 For several years, it has been suspected that one can construct a structure theory for monadically stable graph classes that mirrors the theory of nowhere dense graph classes in the dense setting. In this work we provide a next step in this direction by giving a characterization of monadic stability through the {\em{Flipper game}}: a game on a graph played by {\em{Flipper}}, who in each round can complement the edge relation between any pair of vertex subsets, and {\em{Localizer}}, who in each round is forced to localize the game to a ball of bounded radius. This is an analog of the {\em{Splitter game}}, which characterizes nowhere dense classes of graphs (Grohe, Kreutzer, and Siebertz, J.~ACM~'17).
 
 We give two different proofs of our main result. The first proof is based on tools borrowed from model theory, and it exposes an additional property of monadically stable graph classes that is close in spirit to definability of types. Also, as a byproduct, we give an alternative proof of the recent result of Braunfeld and Laskowski (arXiv 2209.05120) that monadic stability for graph classes coincides with existential monadic stability. The second proof relies on the recently introduced notion of {\em{flip-flatness}} (Dreier, M\"ahlmann, Siebertz, and Toruńczyk, ICALP 2023) and provides an efficient algorithm to compute Flipper's moves in a winning strategy.
\end{abstract}

\paragraph{Acknowledgements.} 
\ifx\anonymous\undefined
We thank Patrice Ossona de Mendez for his valuable contributions to this paper. 
\else
Anonymous acknowledgements.
\fi

\begin{picture}(0,0)
\put(422,-140)
{\hbox{\includegraphics[width=40px]{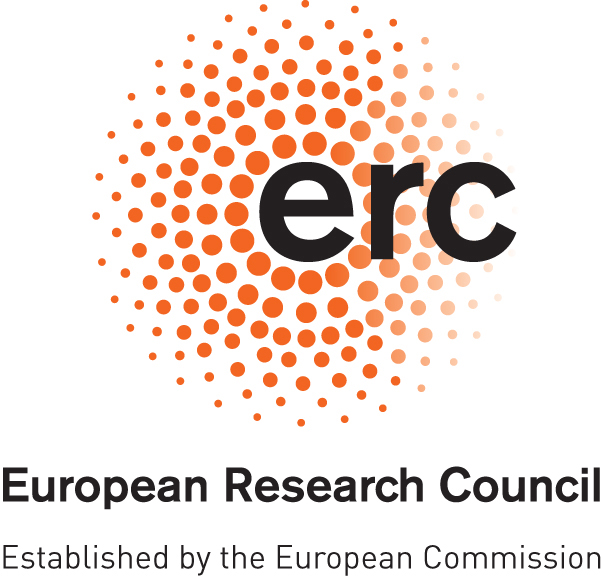}}}
\put(412,-200)
{\hbox{\includegraphics[width=60px]{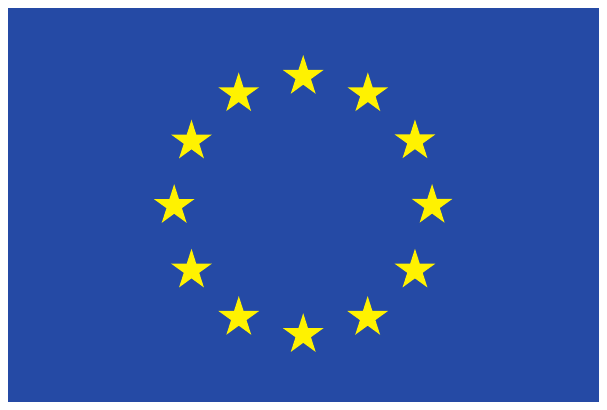}}}
\end{picture}

\thispagestyle{empty}

\newpage
\tableofcontents
\newpage
\newpage


\input{intro}

\input{prelims}

\input{flipper_game}
\input{equivalences}

\input{flip_to_mstab}

\part{Model-theoretic proof}\label{part:equivalences}
In this part we prove the 
implications 
 \eqref{it:ems}$\rightarrow$\eqref{it:pat}$\rightarrow$\eqref{it:sep}$\rightarrow$\eqref{it:sg} of \cref{thm:main} using elementary techniques from model theory.

In Section~\ref{sec:mt-prelims}
we give additional model-theoretic preliminaries: we discuss models, theories, compactness, the Tarski-Vaught test, 
definability of types, a variant of Morley sequences, 
and some basic lemmas about ladders. 

In Section~\ref{sec:pattern-free} we introduce pattern-free classes, and prove 
several simple facts about them. In particular,
we prove that every existentially monadically dependent class is pattern-free, proving the implication \eqref{it:ems}$\rightarrow$\eqref{it:pat} in Theorem~\ref{thm:main}.
Section~\ref{sec:separators} presents the main technical step, 
the implication \eqref{it:pat}$\rightarrow$\eqref{it:sep}. In a nutshell, from pattern-freeness and edge-stability we derive a model-theoretic property, a variant of definability of types, which allows us to control elements in elementary extensions through finite sets in the ground model.

In Section~\ref{sec:sep-game}, 
we use this definability property  
together with compactness of first-order logic to give a strategy for Flipper that ensures victory in the confining game with qf-definable separation in a bounded number of rounds. This is the implication \eqref{it:sep}$\rightarrow$\eqref{it:sg}.
 
 We start with introducing the necessary tools.

\input{mt-prelims}
\input{patterns}
\input{separators}

\input{separation_game}

\part{Algorithmic Flipper game}\label{part:afg}
\input{afg}

\bibliographystyle{alpha}
\bibliography{ref}

\end{document}

%% file: packages.tex
\usepackage{amsmath, amsfonts,xfrac, amsthm,amssymb,fancyhdr}
\usepackage[capitalize]{cleveref}

\usepackage{mathtools}
\usepackage{setspace}

\usepackage{geometry}
\usepackage{xypic,colortbl}

\usepackage{fixltx2e, ifthen, sidecap, wrapfig, xspace}

\usepackage{mathrsfs, stmaryrd}

\usepackage{sectsty}

\usepackage{
 calc, enumerate, makeidx,bbold
}

\usepackage[font=small,format=plain,labelfont=sc,textfont=it]{caption}
\usepackage[T1]{fontenc}
\usepackage[utf8]{inputenc}
\usepackage[all]{xy}
\usepackage[geometry]{ifsym}
\usepackage[cmyk]{xcolor}
\usepackage{thm-restate}

%% file: macros.tex
\newcommand{\lift}[1]{\widehat{\str{#1}}}

\newcommand{\mathsym}[1]{{}}

{\begin{figure}[#1]
\centering 
\includegraphics[scale=#3]{#2}
}
{\end{figure}}

\newenvironment{romanlist'}[0]
{\begin{list}{\makebox[0.5cm][l]{\textit{\roman{enumi}')}}}{\usecounter{enumi}}}
{\end{list}}

\newcommand{\savelabel}[2]{\expandafter\newtoks\csname#1\endcsname
  \global\csname#1\endcsname={#2} \label{#1} #2}
\newcommand{\loadlabel}[1]{\noindent {\bf Lemma~\ref{#1}. } \textit{\the\csname#1\endcsname}
\medskip

}
\renewcommand{\setminus}{-}

\newcommand{\loadlabelthm}[1]{\medskip\noindent {\bf Theorem~\ref{#1}. }
  \noindent  \textit{\the\csname#1\endcsname}
\medskip
}

\newcommand{\loadlabelprop}[1]{\noindent {\bf Proposition~\ref{#1}. }
  \noindent  \textit{\the\csname#1\endcsname}
\medskip

}

\newcommand{\N}{\mathbb{N}}

\newcommand{\eps}{\varepsilon}

\renewcommand{\subset}{\subseteq}

\newcommand{\atleast}[1]{{\ge n}}
\newcommand{\less}[1]{{<n}}

	\newcommand{\notacol}[2]{}

\newsavebox{\quoteitbox}

{\hspace*{\fill}{\upshape(\usebox{\quoteitbox})}\end{quote}%
\end{sloppypar}\noindent\ignorespacesafterend}

\newenvironment{quoteit*} 
{\begin{sloppypar}\noindent\slshape\begin{quote}\itshape} 
	{\end{quote}\ignorespaces\end{sloppypar}\noindent\ignorespacesafterend}

\newenvironment{quotetag*}
{~\par
	\begingroup                  
	\begin{equation*}
		 \begin{minipage}[c]{115mm}
			\it\noindent{\par}
}
{
		\end{minipage}
	\end{equation*}
	\endgroup                        
\par
\textnormal
\medskip
}

\newcommand{\Aa}{\ensuremath{\mathcal A}\xspace}
\newcommand{\Bb}{{\mathcal B}}
\newcommand{\Cc}{{\mathscr C}}
\newcommand{\DD}[0]{\mathrm{\mathscr{D}}}

\newcommand{\CC}{\Cc}

\newcommand{\Qq}{{\mathcal Q}}

\newcommand\set[1]{\ensuremath{\{#1\}}}

\newtheoremstyle{theoremstyle}
  {3pt}
  {3pt}
  {\itshape}
  {0pt}
  {\bfseries}
  {}
  {4pt}
  {}

\theoremstyle{theoremstyle}
\newtheorem{theorem}{Theorem}[section]
\newtheorem{proposition}[theorem]{Proposition}
\newtheorem{conjecture}{Conjecture}[section]
\newtheorem*{theorem*}{Theorem}

\newtheorem{lemma}[theorem]{Lemma}
\newtheorem{corollary}[theorem]{Corollary}
\newtheorem{observation}[theorem]{Observation}

\newtheorem{claim}[theorem]{$\rhd$ Claim}
\newtheoremstyle{remarkstyle}
  {3pt}
  {10pt}
  {}
  {0pt}
  {\itshape}
  {}
  {4pt}
  {\thmname{#1}\thmnumber{ #2}\thmnote{ (#3)}.}

\theoremstyle{remarkstyle}

\newtheorem{remark}{Remark}[section]

\newtheoremstyle{definitionstyle}
  {3pt}
  {3pt}
  {}
  {0pt}
  {\itshape}
  {}
  {4pt}
  {\thmname{#1}\thmnumber{ #2}\thmnote{ (#3)}.}

\theoremstyle{definitionstyle}
\newtheorem{definition}{Definition}

\numberwithin{equation}{section}

\crefname{claim}{claim}{claims}
\newcommand{\cqed}{\ensuremath{\lhd}}
\makeatletter
\newenvironment{claimproof}{\par
	\pushQED{\cqed}%
	\normalfont \topsep6\p@\@plus6\p@\relax
	\trivlist
	\item\relax
	{\itshape
		Proof of the claim\@addpunct{.}}\hspace\labelsep\ignorespaces
}{%
	\hfill\popQED\endtrivlist\@endpefalse
}
\makeatother

\newcommand{\wh}{\widehat}

\newlength{\wideaslength}

\renewcommand{\subset}{\subseteq}

\renewcommand{\P}{\mathbb{P}}

\newcommand{\seta}[1]{}

\def\lsim{\mathrel{\rlap{\lower4pt\hbox{\hskip1pt$\sim$}}
    \raise1pt\hbox{$<$}}}                

\definecolor{gray1}{rgb}{0.99,0.99,0.99}
\definecolor{gray2}{rgb}{0.97,0.97,0.97}
\definecolor{gray3}{rgb}{0.95,0.95,0.95}
\definecolor{gray4}{rgb}{0.93,0.93,0.93}
\definecolor{gray5}{rgb}{0.91,0.91,0.91}
\definecolor{gray6}{rgb}{0.89,0.89,0.89}
\definecolor{gray7}{rgb}{0.87,0.87,0.87}
\definecolor{gray8}{rgb}{0.85,0.85,0.85}
\definecolor{gray9}{rgb}{0.83,0.83,0.83}
\definecolor{gray10}{rgb}{0.81,0.81,0.81}
\definecolor{gray20}{rgb}{0.71,0.71,0.71}
\definecolor{gray40}{rgb}{0.51,0.51,0.51}
\DeclareMathOperator{\Flip}{\text{Flip}}

\newcommand{\sdiff}{\mathbin{\triangle}}

\newcommand{\strat}[1]{\mathsf{#1}}

\newcommand{\stratcon}{\strat{con}}
\newcommand{\stratflip}{\strat{flip}}
\newcommand{\flipstrat}{\operatorname{flip}^\star}
\newcommand{\guessflips}{\mathsf{Predict}}
\newcommand{\wsize}{\alpha}
\newcommand{\wflips}{\lambda}

%% file: intro.tex
\section{Introduction}

Monadic stability is a notion of logical tameness for classes of structures. Introduced by Baldwin and Shelah~\cite{BS1985monadic} in the context of model theory\footnote{Formally, Baldwin and Shelah~\cite{BS1985monadic}, as well as~Braunfeld and Laskowski~\cite{braunfeld2022existential}, study monadically dependent and monadically stable {\em{theories}}, rather than classes of structures. Some of their results transfer to the more general setting of monadically dependent/stable classes of structures.}, it has recently attracted  attention in the field of structural graph theory. We recall the logical definition below. One of the main contributions of this paper is to provide a purely combinatorial 
characterization of monadically stable classes of graphs. Our characterization is effective, and can be employed in algorithmic applications, as we explain later.

In this paper we focus on (undirected, simple) graphs, rather than arbitrary structures. A graph is modelled as a relational structure with one symmetric binary relation signifying adjacency. By a {\em{class}} of graphs we mean any set of graphs. 
For a class of graphs $\Cc$, a {\em{unary expansion}} of $\Cc$ is any class $\widehat{\Cc}$ of structures such that each $\widehat G\in \widehat{\Cc}$ is obtained from some graph in $G\in \Cc$ by adding some unary predicates. Thus, the elements of $\widehat{\Cc}$ can be regarded as vertex-colored graphs from $\Cc$.
A class of graphs $\Cc$ is called {\em{monadically stable}} if one cannot interpret, using a fixed formula
$\varphi(\tup x,\tup y)$ of first-order logic, arbitrarily long linear orders in any unary expansion~$\widehat\CC$ of $\Cc$. 
More precisely, for every unary expansion $\widehat{\CC}$ and formula $\phi(\tup x,\tup y)$ with $|\tup x|=|\tup y|$ (over the signature of $\widehat \Cc$) there is a bound $\ell$ such that there is no structure $\widehat G\in \widehat{\CC}$
and tuples $\tup a_1,\ldots,\tup a_\ell\in V(\widehat G)^{\tup x}$  such that $\widehat G\models \phi(\tup a_i,\tup a_j)$ if and only if $i\le j$.
More generally, $\Cc$ is {\em{monadically dependent}} (or \emph{monadically NIP}) if one cannot interpret, using a fixed formula $\phi(\tup x,\tup y)$ of first-order logic, all finite graphs in any unary expansion of $\Cc$. Thus, from the model-theoretic perspective, the intuition is that being monadically dependent is being non-trivially constrained: for any fixed interpretation, one cannot interpret arbitrarily complicated structures in vertex-colored graphs from the considered class. On the other hand, graphs from monadically stable classes are ``orderless'', in the sense that one cannot totally order any large part of them using a fixed first-order~formula.

Baldwin and Shelah proved that in the definitions, one can alternatively rely on formulas $\phi(x,y)$ with just a pair of free variables, instead of a pair of tuples of variables~\cite[Lemma~8.1.3, Theorem~8.1.8]{BS1985monadic}. 
Moreover, they proved that monadically stable theories are \emph{tree decomposable}~\cite[Theorem 4.2.17]{BS1985monadic}, providing a structure theorem for such theories,
although one of a very infinitary nature.
A more explicit, combinatorial structure theorem for monadically stable and monadically dependent is desirable for obtaining algorithmic results for the considered classes, as we discuss later.

On the other hand, Braunfeld and Laskowski~\cite{braunfeld2022existential} very recently proved that for \emph{hereditary} classes of structures $\CC$
that are not monadically stable or monadically dependent,
the required obstructions (total orders or arbitrary graphs)
can be exhibited by an existential formula $\phi(\tup x,\tup y)$ in the signature of $\CC$, without any additional unary predicates.
Among other things, this shows that for hereditary classes of structures, the notions of monadic stability coincides with the more well-known notion of stability, and similarly, monadic dependence coincides with dependence (NIP).
Furthermore, since the formulas are existential, this result can be seen as a rather explicit, combinatorial non-structure theorem for hereditary classes that are not monadically stable (resp. monadically~dependent). Still, they do not provide explicit structural results for classes that are monadically stable or monadically dependent.



Explicit, combinatorial and algorithmic structural results for monadically dependent and monadically stable classes are not only
desired, but also expected to exist, based on the known examples of such classes that have been studied in graph theory and computer science.
 As observed by Adler and Adler~\cite{AdlerA14} based on the work of Podewski and Ziegler~\cite{podewski1978stable}, all {\em{nowhere dense}} graph classes are monadically stable. A class $\Cc$ is nowhere dense if for every fixed $r\in \N$, one cannot 
 find $r$-subdivisions\footnote{The {\em{$r$-subdivision}} of a graph $H$ is the graph obtained from $H$ by replacing every edge with a path of length $r+1$.} of arbitrarily large cliques
 as subgraphs of graphs in $\Cc$. In particular, every class excluding a fixed topological minor (so also the class of planar graphs, or the class of subcubic graphs) is monadically stable. In fact, it follows from the results of Adler and Adler~\cite{AdlerA14} and of Dvo\v{r}\'ak~\cite{Dvorak18} that monadic stability and monadic dependence are both equivalent to nowhere denseness if one assumes that we work with a class of sparse graphs (formally, with a class of graphs that excludes a fixed biclique as a subgraph). However, since they are defined in only model-theoretic terms, monadic stability and monadic dependence are not bound to sparsity; they can be used to understand and quantify structure in dense graphs as well.

The pinnacle of the theory of nowhere dense graph classes is the result of Grohe, Kreutzer, and Siebertz~\cite{GroheKS17} that the model-checking problem for first-order logic is fixed-parameter tractable on any nowhere dense class of graphs.

\begin{theorem}[\cite{GroheKS17}]\label{thm:mc-nd}
 For every nowhere dense graph class $\Cc$, first-order sentence $\varphi$, and $\eps>0$, there exists an algorithm that given an $n$-vertex graph $G\in \Cc$ decides whether $G\models \varphi$ in time $\Oh_{\Cc,\varphi,\eps}(n^{1+\eps})$. 
\end{theorem}

Monadically dependent classes include all monadically stable classes,
in particular all nowhere dense classes, but also for instance all classes of bounded twin-width~\cite{tww1}. An analogous result, with $1+\eps$ replaced by $3$, holds for all classes $\CC$ of \emph{ordered} graphs\footnote{\emph{Ordered graphs} are graphs equipped with a total order.} of bounded twin-width~\cite{tww4}.

\medskip
In light of the discussion above, monadic stability and monadic dependence seem to be well-behaved generalizations of nowhere denseness that are defined in purely model-theoretic terms; hence these concepts may be even better suited for treating the model-checking problem for first-order logic. This motivated the following conjecture~\cite{warwick-problems}, which has been a subject of intensive study over the last few years\footnote{To the best of our knowledge the conjecture was first explicitly discussed during the open problem session of the  Algorithms, Logic and Structure Workshop in Warwick, in 2016, see \cite{warwick-problems}.}.
\begin{conjecture}\label{conj:mc-NIP}
Let $\Cc$ be a monadically dependent graph class. There exists a constant $c\in \N$ depending only on $\Cc$ and, for every first-order sentence $\varphi$, an algorithm that, given a $n$-vertex graph $G\in \Cc$, decides whether $G\models \varphi$ in time $\Oh_{\Cc,\varphi}(n^c)$.
\end{conjecture}

\cref{conj:mc-NIP} is not even resolved for monadically stable classes. To approach this conjecture, it is imperative to obtain explicit, combinatorial structure theorems for monadically stable and in monadically dependent graph classes, with a particular focus on finding analogs of the tools used in the proof of \cref{thm:mc-nd}. Our work contributes in this direction. We provide 
 certain tree-like decompositions for graphs in monadically stable graph classes, which can be most intuitively explained in terms of games. On the one hand, our decompositions generalize a similar  
 result for nowhere dense classes, recalled below.
  On the other hand, they are remininiscent of the tree decomposability property proved by Baldwin and Shelah, but are more explicit and finitary in nature.

\paragraph{Splitter game.} The cornerstone of the proof of \cref{thm:mc-nd} is a game-theoretic characterization of nowhere denseness, through the {\em{Splitter game}}. This game has a fixed radius parameter $r\in \N$ and is played on a graph $G$ between two players, {\em{Splitter}} and {\em{Localizer}}, who make moves in rounds alternately. In each round, Splitter first chooses any vertex $u$ and removes it from the graph. Next, Localizer has to select any other vertex $v$, and the game gets restricted to the subgraph induced by the ball of radius $r$ with center at $v$. The game ends with Splitter's victory when there are no vertices left in the graph.

\begin{theorem}[\cite{GroheKS17}]\label{thm:splitter}
 A class $\Cc$ of graphs is nowhere dense if and only if for every $r\in \N$ there exists $k\in \N$ such that for every $G\in \Cc$, Splitter can win the radius-$r$ Splitter game on $G$ within $k$~rounds.
\end{theorem}

Very roughly speaking, \cref{thm:splitter} shows that any graph from a nowhere dense class can be hierarchically decomposed into smaller and smaller parts so that the decomposition has height bounded by a constant $k$ depending only on the class and the locality parameter~$r$. This decomposition is used in the algorithm of \cref{thm:mc-nd} to guide model-checking.

\paragraph{Flipper game.}
In this work we introduce an analog of the Splitter game for monadically stable graph classes: the {\em{Flipper game}}. Similarly to before, the game is played on a graph $G$ and there is a fixed radius parameter $r\in \N$. There are two players, {\em{Flipper}} and {\em{Localizer}}, which make moves in rounds alternately. In her move, Flipper selects any pair of vertex subsets $A,B$ (possibly non-disjoint) and applies the {\em{flip}} between $A$ and~$B$: inverts the adjacency between any pair $(a,b)$ of vertices with $a\in A$ and $b\in B$. Localizer's moves are exactly as in the Splitter game; in every round, he needs to select a ball of radius $r$, and the game is restricted to the subgraph induced by this ball. The game is won by Flipper once there is only one vertex left.

We remark that the Flipper game is a radius-constrained variant of the natural game for graph parameter {\em{SC-depth}}, which is functionally equivalent to {\em{shrubdepth}}, in the same way that the Splitter game is a radius-constrained variant of the natural game for treedepth. SC-depth and shrubdepth were introduced and studied by Ganian et al. in~\cite{GanianHNOMR12,GanianHNOM19}.

Our main result is the following analog of \cref{thm:splitter} for monadically stable classes.

\begin{theorem}\label{thm:main-main}
 A class $\Cc$ of graphs is monadically stable if and only if for every $r\in \N$ there exists $k\in \N$ such that for every graph $G\in \Cc$, Flipper can win the radius-$r$ Flipper game on $G$ within $k$~rounds. 
\end{theorem}


Let us compare \cref{thm:main-main} with another recent characterization of monadic stability, proposed by Gajarsk{\'y} and Kreutzer, and proved by Dreier, M\"ahlmann, Siebertz, and Toru{\'n}czyk \cite{dreier2022indiscernibles}, through the notion of {\em{flip-flatness}}. This notion is an analog of {\em{uniform quasi-wideness}}, introduced by Dawar~\cite{Dawar10}. Without going into technical details, a class of graphs $\Cc$ is {\em{uniformly quasi-wide}} if for any graph $G\in \Cc$ and any  large enough set of vertices $A$ in $G$, one can find many vertices in $A$ that are pairwise far from each other after the removal of a constant number of vertices from $G$. As proved by Ne\v{s}et\v{r}il and Ossona de Mendez~\cite{NesetrilM11a}, a class of graphs is uniformly quasi-wide if and only if it is nowhere dense. Flip-flatness is an analog of uniform quasi-wideness obtained similarly to the Flipper game: by replacing the concept of deleting a vertex with applying a flip; see \cref{def:wideness} for a formal definition. The fact that monadic stability is equivalent to flip-flatness (as proved in~\cite{dreier2022indiscernibles}) and to the existence of a short winning strategy in the Flipper game (as proved in this paper) suggests the following: the structural theory of monadically stable graph classes mirrors that of nowhere dense graph classes, where the flip operation is the analog of the operation of removing a~vertex.

We give two very different proofs of \cref{thm:main-main}. The first proof is based on elementary model-theoretic techniques, and it provides new insight into the properties of monadically stable graph classes. As a side effect, it gives a new (though non-algorithmic) proof of the main result of \cite{dreier2022indiscernibles}: equivalence of monadic stability and flip-flatness. 
On the other hand, 
the second proof relies on the combinatorial techniques developed in \cite{dreier2022indiscernibles}. 
It has the advantage of being effective, and provides an efficient algorithm for computing Flipper's moves in a winning~strategy.

\paragraph{Model-theoretic proof.} The following statement lists properties equivalent to monadic stability uncovered in our model-theoretic proof. Notions not defined so far will be explained later.

\begin{theorem}\label{thm:main}
  Let $\Cc$ be a class of graphs. Then the following conditions are equivalent:
  \begin{enumerate}
    \item\label{it:ms} $\CC$ is monadically stable.
    \item\label{it:ems} 
    $\CC$ has a stable edge relation and is  monadically dependent with respect 
    to existential formulas $\phi(x,y)$ with two free variables.
    \item\label{it:pat} 
    $\CC$ has a stable edge relation and is  pattern-free.
    \item \label{it:sep} 
For every $r\in\N$ every  model $G$ of the theory of $\CC$, every elementary extension $H$ of $G$, and every vertex $v\in V(H)\setminus V(G)$, there is a finite set $S\subset V(G)$ that $r$-separates $v$ from $G$.
    \item\label{it:sg} For every $r\in\N$ there is $k\in\N$ such that Flipper wins the confining Flipper game with qf-definable separation of radius $r$ on every $G\in \CC$ in at most $k$ rounds.
    \item\label{it:fg} For every $r\in\N$ there is $k\in\N$ such that Flipper wins the Flipper game of radius $r$ on every $G\in \CC$ in at most $k$ rounds.
   \item\label{it:fw} $\CC$ is flip-flat.
  \end{enumerate}
\end{theorem}

\input{implications-diagram.tex}

The implications that constitute \cref{thm:main} are illustrated in \cref{fig:implications}.
Note that \cref{thm:main-main} is the equivalence \eqref{it:ms}$\leftrightarrow$\eqref{it:fg}.
Let us give a brief overview of the presented conditions. 

%

Conditions~\eqref{it:ms} and~\eqref{it:ems}, respectively, are monadic stability and a weak form of {\em{existential monadic stability}}. 
Recall that Baldwin and Shelah proved that it is sufficient to consider formulas $\phi(x,y)$ with two free variables in the definition of monadic stability (instead of formulas $\phi(\tup x,\tup y)$),
whereas Braunfeld and Laskowski proved that it is sufficient to consider existential formulas $\phi(\tup x,\tup y)$ that do not involve additional unary predicates. The condition \eqref{it:ems} lies somewhere in between: it implies that
it is sufficient to consider existential formulas $\phi(x,y)$ with two variables, possibly involving additional unary predicates.
In particular, it implies the result of Baldwin and Shelah
(in the case of graph classes) and is incomparable with the result of Braunfeld and Laskowski. Our proof uses different techniques.

Condition~\eqref{it:pat}
is a rather explicit combinatorial condition. Roughly, a class $\CC$  is \emph{pattern-free}
if it does not encode, using a quantifier-free formula $\phi(x,y)$, the class of $r$-subdivided cliques, for any fixed $r\ge 1$. See Definition \ref{def:pattern-free} for details.

Condition~\eqref{it:sep} is phrased in the language of model theory and serves a key role in our proof. It resembles a fundamental property called ``definability of types'', and in essence it says the following: whenever working with a model $G$ of the theory of $\Cc$, every element of any elementary extension of $G$ can be robustly ``controlled'' by a finite subset of $G$. We believe that the new notion of $r$-separation used here is of independent interest. It refers to non-existence of short paths after applying some flips governed by $S$. The implication \eqref{it:pat}$\rightarrow$\eqref{it:sep} is the core part of our~proof.

Conditions~\eqref{it:sg} and~\eqref{it:fg} assert the existence of a short winning strategy in the two variants of the Flipper game. The implication~\eqref{it:sep}$\rightarrow$\eqref{it:sg} is proved by proposing a strategy for Flipper in the confining game with qf-definable separation and using compactness combined with~\eqref{it:sep} to argue that it leads to a victory within a bounded number of rounds.

Finally, condition~\eqref{it:fw} is the notion of flip-flatness, whose equivalence with monadic stability was proved by Dreier et al.~\cite{dreier2022indiscernibles}. We prove the implication~\eqref{it:sg}$\rightarrow$\eqref{it:fw} by (essentially) providing a strategy for Localizer in the Flipper game when the class is not flip-flat. Then we rely on the implication~\eqref{it:fw}$\rightarrow$\eqref{it:ms} from~\cite{dreier2022indiscernibles} to close the circle of implications; this proves the equivalence of \eqref{it:ms}-\eqref{it:fw} with the exception of~\eqref{it:fg}. We remark that \eqref{it:fw}$\rightarrow$\eqref{it:ms} is the easy implication of~\cite{dreier2022indiscernibles}, hence our reasoning can also serve as an alternative proof of the flip-flatness characterization given in~\cite{dreier2022indiscernibles}.

To put the Flipper game into the picture, we separately prove the implications \eqref{it:sg}$\rightarrow$\eqref{it:fg}$\rightarrow$\eqref{it:ems}. The implication \eqref{it:sg}$\rightarrow$\eqref{it:fg} relies on a conceptually easy, but technically not-so-trivial translation of the strategies. In the implication \eqref{it:fg}$\rightarrow$\eqref{it:ems} we use obstructions to existential monadic stability to give a strategy for Localizer in the Flipper game that enables her to endure for arbitrarily~long.
%

\paragraph{Algorithmic proof.} We also give a purely combinatorial proof of (the forward implication of) \cref{thm:main-main}, which in particular provides a way to efficiently compute Flipper's moves in a winning strategy. Formally, we show the following.

\begin{restatable}{theorem}{afgmain}
  \label{thm:afg_main}
    Let $\Cc$ be a monadically stable class of graphs. Then for every radius $r\in \N$ there exist $k\in\N$ and a Flipper strategy $\flipstrat$ such that the following holds:
    \begin{itemize}
     \item When playing according to $\flipstrat$ in the Flipper game of radius $r$ on any graph $G\in \Cc$, Flipper wins within at most $k$ rounds.
     \item The moves of $\flipstrat$ on an $n$-vertex graph $G\in \CC$ can be computed in time $\Oof_{\CC,r}(n^2)$.
    \end{itemize}
\end{restatable}

The main idea behind the proof of \cref{thm:afg_main} is to rely on the result of Dreier et al.\ that monadically stable graph classes are flip-flat~\cite{dreier2022indiscernibles}. Using the combinatorial tools developed in \cite{dreier2022indiscernibles}, we  strengthen this property: we prove that the set of flips $F$ whose application uncovers a large {\em{scattered set}} $Y$ (a set of vertices that are pairwise far from each other) can be selected in a somewhat canonical way, so that knowing any $5$-tuple of vertices in $Y$ is enough to uniquely determine $F$. We can then use such strengthened flip-flatness to provide a winning strategy for Flipper; this roughly resembles the Splitter's strategy used by Grohe et al. in their proof of \cref{thm:splitter}, which in turn relies on uniform quasi-wideness. 

\Cref{thm:afg_main}, the algorithmic version of \Cref{thm:main-main}, is the key to any algorithmic applications of the Flipper game. In particular, it was very recently already used by Dreier, M\"{a}hlmann, and Siebertz~\cite{mcss} to approach the first-order model checking problem on monadically stable graph classes and even solve it on structurally nowhere dense classes, an important subclass of monadically stable classes. 
On a high level, the proof of~\cite{mcss} follows the approach of Grohe et al.~\cite{GroheKS17} on nowhere dense classes. Essentially, by Gaifman's Locality Theorem the model checking problem reduces to computing which formulas up to a certain quantifier rank $q$ are true in the local $r$-neighborhoods of the input graph. Here, the numbers $q$ and $r$ depend only on the input formula. The set of formulas of quantifier rank $q$ that are true in a local neighborhood is called the \emph{local $q$-type} of the neighborhood. The computation of local $q$-types is done recursively in a recursion guided by the Splitter game, which hence terminates after a bounded number of steps. A naive branching into all local neighborhoods, however, leads to a too high running time. This issue is solved by grouping close by elements into clusters and computing the local types of all neighborhoods that are grouped in one cluster in one recursive call. Whenever the clusters can be collected into a \emph{sparse neighborhood cover}, which is the case in nowhere dense graph classes, this leads to an efficient model checking algorithm. Dreier et al.~\cite{mcss} showed that the Splitter game in the above approach can be replaced by the Flipper game and present an efficient model checking algorithm on all monadically stable classes that admit sparse (weak) neighborhood covers. Their result does not fully solve the model checking problem on monadically stable classes as the question whether sparse weak neighborhood covers for monadically stable classes exist remains an open problem. Dreier et al.\ only showed the existence of such covers for structurally nowhere dense classes. We refer to \cite{mcss} for the details.


\paragraph{Organization.}
The paper is split into three parts.

\Cref{part:prelude} is devoted to introducing the basic notions  and relations between them. In particular, we prove the implications \eqref{it:sg}$\rightarrow$\eqref{it:fg}$\rightarrow$\eqref{it:ems} of
and  \eqref{it:sg}$\rightarrow$\eqref{it:fw} of Theorem~\ref{thm:main}.

In \Cref{part:equivalences} we present the model-theoretic proof
of Theorem~\ref{thm:main-main}.
More precisely, we prove the implications \eqref{it:ems}$\rightarrow$\eqref{it:pat}$\rightarrow$\eqref{it:sep}$\rightarrow$\eqref{it:sg}
of Theorem~\ref{thm:main}.
Together with the implication \eqref{it:sg}$\rightarrow$\eqref{it:fw} proved in \Cref{part:prelude},
the 
(easy) implication \eqref{it:fw}$\rightarrow$\eqref{it:ms} proved in \cite{dreier2022indiscernibles}, and the 
trivial implication \eqref{it:ms}$\rightarrow$\eqref{it:ems}, 
this closes the cycle of implications between the conditions \eqref{it:ms}, \eqref{it:ems}, \eqref{it:pat}, \eqref{it:sep}, \eqref{it:sg}, \eqref{it:fw}.
And with the implications \eqref{it:sg}$\rightarrow$\eqref{it:fg}$\rightarrow$\eqref{it:ems} proved in \cref{part:prelude}, this completes the proof of Theorem~\ref{thm:main}.

In \Cref{part:afg} we prove \Cref{thm:afg_main}.

%% file: implications-diagram.tex
\begin{figure}[ht]
\centering
\scalebox{0.57}{
\begin{tikzpicture}[%
  node distance=27mm,>=Latex,
  every edge/.style={draw=black,thin},
  block/.style={
    draw,
    fill=white,
    rectangle, 
    minimum width={2.5cm},
    minimum height={1.5cm},
    align = center
  }
  ]

\node[block] (ms) 
{\eqref{it:ms}\\ mon. stable};

\node[block, right = 1.3cm of ms] (ems) 
{\eqref{it:ems}\\ \small stable edge relation \\ \small + ex. mon. dependent};

\node[block, right = 1cm of ems] (pat) 
{\eqref{it:pat}\\ \small stable edge relation \\ \small + pattern-free};

\node[block, right = 1cm of pat] (sep) 
{\eqref{it:sep}\\ separable};

\node[block, right = 1.3cm of sep] (sg) 
{\eqref{it:sg}\\ \small Flipper wins with \\\small qf-definable separation \\ \small and confining localization};

\node[block, above right = 0.9cm of sg] (fg) 
{\eqref{it:fg}\\ Flipper wins};

\node[block, below right = 0.9cm of sg] (fw) 
{\eqref{it:fw}\\ flip-flat};

\path[->] 
(ms) edge node[above] {trivial} (ems);

\path[->] 
(ems) edge node[above] {Sec. \ref{sec:pattern-free}} (pat);

\path[->] 
(pat) edge node[above] {Sec. \ref{sec:separators}} (sep);

\path[->] 
(sep) edge node[above] {Sec. \ref{sec:sep-game}} (sg);

\draw[->]  (ms.south)+(6mm, 0) -- node {} ++(
  6mm,-0.75cm) -| (sep.south) node[pos=0.25, above] {Sec. \ref{sec:separators}} node[pos=0.75] {};


\draw[->] 
(sg.east)+(0, 3mm) -| node[near start,above] {Sec.\ \ref{sec:variants}} (fg.south);

\draw[->] 
(sg.east)+(0, -3mm) -| node[near start,below] {Sec.\ \ref{sec:pfg_to_fw}} (fw.north);

\draw[->]    
(fg.west) -| node[near start,below] {Sec.\ \ref{sec:flip-to-mstab-new}} (ems.north); 

\draw[->] 
(fw.west) -| node[near start,above] {\cite{dreier2022indiscernibles}} (ms.south);

\end{tikzpicture}   
   
}
\caption{The implications that constitute \cref{thm:main}.
\vspace{-2mm}
}
\label{fig:implications}
\end{figure}

%% file: prelims.tex
\part{Prelude}\label{part:prelude}
In this part, we introduce the basic notions of interest:
monadic stability, variants of the Flipper game, flip-flatness,
and prove some relations between them.

\medskip

In \cref{sec:defs} we establish notation and common definitions, and present basic model-theoretic background. Also, we discuss the notions of flips, of flip-connectivity, and of flip-flatness.

In \cref{sec:flippers} we define variants of the Flipper game, and prove results relating some of them to the others.
In particular, the implication \eqref{it:sg}$\rightarrow$\eqref{it:fg} 
of Theorem~\ref{thm:main}, which is relatively easy, is proved in Section~\ref{sec:variants}.

In \cref{sec:flip-to-mstab} we relate variants of the Flipper game to other notions around monadic stability: flip-flatness and existential monadic stability. In particular, we prove the implication \eqref{it:sg}$\rightarrow$\eqref{it:fw}
of Theorem~\ref{thm:main}. Also we derive existential monadic stability from a strategy of Flipper in the Flipper game (this is the implication
\eqref{it:fg}$\rightarrow$\eqref{it:ems}).

\section{Preliminaries}\label{sec:defs}
All graphs in this paper are simple and loopless but not necessarily finite.
For a vertex $v$ of a graph $G$, we write $N(v)$ for the (open) neighborhood of $v$ in $G$; so $N(v) \coloneqq \setof{u \in V(G)} {uv \in E(G)}$. For a set of vertices $X \subseteq V(G)$, we write $G[X]$ for the subgraph of $G$ induced by $X$, and $G-X$ for the subgraph of $G$ induced by $V(G)-X$. 
For two sets $X,Y\subset V(G)$
the bipartite graph \emph{semi-induced} by $X$ and $Y$
in $G$, denoted $G[X,Y]$, is the bipartite graph with 
parts $X$ and $Y$, and edges $uv$ for $u\in X$, $v\in Y$ with $uv\in E(G)$.
By $|G|$ we denote the cardinality of the vertex set of $G$.

For vertices $a,b \in V(G)$, an \emph{$(a,b)$-path} is a path with ends $a$ and $b$. Similarly, for sets $A,B \subseteq V(G)$, an \emph{$(A,B)$-path} is a path where one end is in $A$ and the other end is in $B$. We write $\triangle$ for the symmetric difference of two sets. Similarly, for two graphs $G$ and $G'$ on the same vertex-set, we write $G \triangle G'$ for the graph with vertex-set $V(G)=V(G')$ and edge-set $E(G)\triangle E(G')$.

\subsection{Model theory}
We work with first-order logic over a fixed signature $\Sigma$ that consists of (possibly infinitely many) constant symbols and of relation symbols. 
A \emph{model} is a $\Sigma$-structure, and is typically denoted $\str M,\str N$, etc. We usually do not distinguish between a model and its domain, when writing, for instance, $m\in\str M$ or $f\from \str M\to X$, or $X\subset \str M$. A graph $G$ is viewed as a model over the signature consisting of one binary relation denoted $E$, indicating adjacency between vertices.

If $\tup x$ is a finite set of variables, then we write $\phi(\tup x)$ to denote a first-order formula $\phi$ with free variables contained in $\tup x$. We may also write $\phi(\tup x_1,\ldots,\tup x_k)$ to denote a formula 
whose free variables are contained in $\tup x_1\cup\ldots\cup \tup x_k$.
We will write $x$ instead of $\set x$
in case of a singleton set of variables, e.g. 
$\phi(x,y)$ will always refer to a formula with two free variables $x$ and $y$.
We sometimes write $\phi(\tup x;\tup y)$ to distinguish a partition of the set of free variables of $\phi$ into two parts, $\tup x$ and $\tup y$; this partition plays an implicit role in some definitions. 

If $U$ is a set and $\tup x$ is a set of variables, then $U^{\tup x}$ denotes the set of all \emph{valuations} $\tup a\from \tup x\to U$ of $\tup x$ in $U$. Such a valuation is also called an \emph{$\tup x$-tuple}. For a formula $\phi(\tup x)$ and an $\tup x$-tuple $\tup m\in \str M^{\tup x}$, we write $\str M\models \phi(\tup m)$, or $\str M,\tup m\models\phi(\tup x)$, if the valuation $\tup m$ satisfies the formula $\phi(\tup x)$ in $\str M$.

A \emph{quantifier-free} formula is a formula that does not use quantifiers.
An \emph{existential formula}
is a formula of the form $\exists y_1\ldots \exists y_l.\alpha$, where $\alpha$ is quantifier-free.

We will use the notion of an \emph{atomic type} only
in the context of a finite relational signature $\Sigma$.
In this case, an \emph{atomic type} with variables $\tup x$ is a formula $\alpha(\tup x)$ 
that is a conjunction of clauses of the form $R(x_1,\ldots,x_k)$ or $\neg R(x_1,\ldots,x_k)$,
where $R\in \Sigma\cup\set{=}$ is a relation symbol of arity $k$ and $x_1,\ldots,x_k$ are variables from $\tup x$, such that each possible such clause or its negation occurs as a conjunct in $\alpha$.
The \emph{atomic type} of a tuple $\tup a\in \str M^{\tup x}$ is the unique atomic type $\alpha(\tup x)$ that is satisfied by $\tup a$ in $\str M$.

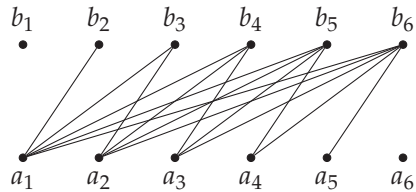
\begin{figure}[h]
  \centering
  
  \begin{tikzpicture}
   
\tikzstyle{vertex} = [draw, circle, fill, inner sep=1pt]

  \foreach \i in {1,2,3,4,5,6} {
    \node[vertex] (a\i) at (\i,0) [label=below:$a_\i$] {};
    \node[vertex] (b\i) at (\i,1.5) [label=above:$b_\i$] {};
  }
  \foreach \i/\j in {1/2,1/3,1/4,1/5,1/6,2/3,2/4,2/5,2/6,3/4,3/5,3/6,4/5,4/6,5/6} {
    \draw (a\i) -- (b\j);
  }
  \end{tikzpicture}

  \caption{A ladder (half-graph) of order $6$.}
  \label{fig:ladder}
\end{figure}

\paragraph{Stability and dependence.}
A formula $\phi(\tup x;\tup y)$ is 
\emph{stable} in a class $\CC$ of structures if there exists $k \in \N$ such that for every $\str M\in \CC$, there are no sequences $\tup a_1,\ldots\tup a_{k}\in \str M^{\tup x}$ and $\tup b_1,\ldots,\tup b_k\in\str M^{\tup y}$ such that \[\str M\models\phi(\tup a_i;\tup b_j)\quad\iff\quad i < j,\quad\text{for $1\le i,j\le k$}.\]
We say that a class $\CC$ of graphs has a \emph{stable edge relation} if the formula $E(x;y)$ is stable  in $\CC$. Equivalently, $\CC$ excludes some ladder as a semi-induced subgraph, where a {\em{ladder}} (often called also {\em{half-graph}}) of order $k$ is the graph with vertices $a_1,\ldots,a_k,b_1,\ldots,b_k$ and edges $a_ib_j$ for all $1\le i< j\le k$; see \cref{fig:ladder}.
Note that replacing $<$ by $\leq$ in the above definitions does not change them.

 A formula $\phi(\tup x;\tup y)$ is \emph{dependent}, or \emph{NIP} (standing for ``not the independence property'') in a class $\CC$ if there exists $k \in \N$ such that for every $\str M\in\CC$, there are no tuples $\tup a_1, \ldots, \tup a_k\in \str M^{\tup x}$ and $\tup b_J\in\str M^{\tup y}$  for $J\subset \set{1,\ldots,k}$ such that 
\[\str M\models\phi(\tup a_i;\tup b_J)\quad\iff\quad i\in J,\quad\text{for $1\le i\le k$ and $J\subset \set{1,\ldots,k}$}.\]
Observe that a formula which is stable is also dependent.
A class $\CC$ is \emph{stable} (resp. \emph{dependent}) if every formula $\phi(\tup x;\tup y)$ is stable (resp. dependent) in $\CC$.

Let ${\Sigma}$ be a signature and let $\widehat{\Sigma}$ be a signature extending $\Sigma$ by (possibly infinitely many) unary relation symbols and constant symbols. A $\widehat{\Sigma}$-structure $\widehat{\str M}$ is a \emph{lift} of a $\Sigma$-structure $\str M$ if $\str M$ is obtained from ${\widehat{\str M}}$ by forgetting the symbols from $\widehat{\Sigma}\setminus\Sigma$. A class of $\widehat{\Sigma}$-structures $\widehat{\CC}$ is a \emph{unary expansion} of a class of $\Sigma$-structures $\CC$ if every structure $\widehat{\str M}\in \widehat{\CC}$ is a lift of some structure $\str M\in\CC$. A class $\CC$ of structures is \emph{monadically stable} if every unary expansion $\widehat\CC$ of $\CC$ is stable. Similarly, $\CC$ is \emph{monadically dependent} (or \emph{monadically NIP}) if every unary expansion $\widehat\CC$ of $\CC$ is dependent. 
A class $\CC$ is \emph{simply existentially monadically dependent} (resp.\ \emph{simply existentially monadically stable}) 
if every existential formula $\phi(x,y)$ (with two free variables) is dependent (resp.\ stable), in every unary expansion $\widehat{\CC}$ of~$\CC$.
A single structure $\str M$ is monadically stable (resp.\ monadically dependent) if the class $\set{\str M}$ is.
Note that a class which is monadically stable (resp.\ monadically dependent) is stable (resp.\ dependent).
Also, a class which is simply existentially monadically stable 
is simply existentially monadically dependent, and has a stable edge relation.

We remark that even though monadic dependence and monadic stability are defined in terms of formulas $\phi(\tup x,\tup y)$, where $\tup x$ and $\tup y$ are tuples of free variables, rather than single variables, by the results of Baldwin and Shelah~\cite[Lemma 8.1.3, Theorem 8.1.8]{BS1985monadic}, it is enough to consider formulas $\phi(x,y)$ with two free variables instead.
Our definitions of existential monadic stability and existential monadic dependence involves existential formulas $\phi(x,y)$ with two free variables only. As follows from Theorem~\ref{thm:main},
for graph classes monadic stability is equivalent to existential monadic stability, thus strengthening the result of Baldwin and Shelah, when restricted to graph classes, by additionally showing that it suffices to consider existential formulas.


\subsection{Flips}

For the purpose of the following definition, it is convenient to assume that 
all considered graphs have a domain contained in some fixed universe $\Omega$.

\paragraph{Atomic flips.}
An \emph{atomic flip}  is an operation  $\flip{F}$ specified by a pair $(A,B)$  of (possibly intersecting) subsets of $\Omega$, which complements the adjacency relation between the sets $A$ and $B$ in a given graph $G$. Formally, for a graph $G$, the graph obtained from $G$ by applying the atomic flip~$\flip F$ is the graph
denoted $G\oplus\flip{F}$ with vertex set $V(G)$, where, for distinct vertices $u,v$ in $V(G)$,
\[
	uv\in E(G\oplus\flip{F})\qquad\iff\qquad \begin{cases}
		uv\notin E(G),&\text{ if }(u,v)\in (A\times B)\cup (B\times A);\\
		uv\in E(G),&\text{ otherwise.}
	\end{cases}
\]
In the remainder of this section, it will be convenient to identify an atomic flip $\flip F$ with the pair $(A,B)$ of sets defining it and to write $G\oplus(A,B)$ instead of $G\oplus\flip F$.
Note that, in our definition, we do not require  $A$ and $B$ to be subsets of $V(G)$.
Instead, we will allow $A$ and $B$ to be any subset of $\Omega$.
 In particular,  $G\oplus (A,B) = G\oplus (A \cap V(G), B\cap V(G))$. 
This is useful as we often work with induced subgraphs.
As an example, for every graph $G$ and vertex $v\in V(G)$, the atomic flip $\flip F_v=(\{v\}, N(v))$ isolates the vertex $v$ in $G$, meaning that $v$ is an isolated vertex of $G\oplus \flip F_v$.
We denote by $\Flip(G)=\{(A,B)\mid A,B\subseteq V(G)\}$ the set of all the atomic flips defined by subsets of vertices of $G$.

\begin{observation}\label{obs:flips_commute}
  For every graph $G$ and atomic flips $\flip F_1, \flip F_2$ we have $G \oplus \flip F_1 \oplus \flip F_2 = G \oplus \flip F_2 \oplus \flip F_1$.
\end{observation}

Furthermore, when considering a sequence of atomic flips, we may restrict to the case where no atomic flip appears more than once in the sequence
since performing twice the same atomic flip twice leaves the graph unchanged,
thanks to the following.

\begin{observation}\label{obs:flips_cancel}
  For every graph $G$ and atomic flip $\flip F$ we have $G \oplus \flip F \oplus \flip F = G$.
\end{observation}

This justifies the following definition.

\paragraph{Sets of flips.}
A set of flips $\set{\flip F_1,\ldots,\flip F_k}$ defines an operation $F$
that, given a graph $G$, results in the graph 
\[G\oplus F:=G \oplus \flip F_1 \oplus \dots \oplus \flip F_k.\]
Note that the order in which we carry out the atomic flips does not matter, according to \cref{obs:flips_commute} and that it would be useless to consider mutlisets, according \cref{obs:flips_cancel}. Abusing terminology, we will often just say that the operation $F$ is a set of flips, and write $F=\set{\flip F_1,\ldots,\flip F_k}$.
\begin{observation}
If $F$  and $F'$ are sets of flips, and $F\sdiff F'$ is the symmetric difference of $F$ and $F'$, then
\[
G\oplus F\oplus F'=G\oplus(F\sdiff F').
\]	
\end{observation}

The next observation is elementary but crucial.

\begin{observation}
	\label{obs_non_separated_twins}
	Let $G$ be a graph, let $X$ be a subset of the vertices of $G$, and let $\mathsf F=(A,B)$ be an atomic  flip.
	Suppose two vertices $u,v$ have the same neighborhood on $X$ in $G$
	and have the same membership in $A$ and $B$ (i.e. $u\in A\iff v\in A$ and $u\in B\iff v\in B$). Then $u$ and $v$ have the same neighborhood on $X$ in $G\oplus\mathsf F$.
\end{observation}

Finally, flips commute with vertex deletions. As a consequence, we have
\begin{observation}\label{obs:subgraph_flip}
	For every graph $G$, every subset $A$ of $V(G)$, and every set of flips $F$, we have $G[A] \oplus F =(G\oplus F)[A]$.
\end{observation}

\begin{remark}
Specifying an atomic flip by a pair of subsets of $\Omega$ will be convenient for carrying out our proofs.
We note that we could also define an atomic flip by a single set $A\subseteq \Omega$ producing the graph $G\oplus (A,A)$ that is obtained by complementing all adjacencies within $A$. 
Up to a constant factor, both definitions yield the same expressive power: 
every atomic flip $(A,B)$ is equivalent to the
set of flips defined by $\set{(A\cup B,A\cup B),(A,A),(B,B)}$.
\end{remark}

\paragraph{Flip terminology.}
For conciseness, we will use the term \emph{flip} to refer to an \emph{atomic flip}.
Therefore, a flip may be identified with a pair $(A,B)$ of vertex sets.

\paragraph{$\cal F$-flips.}
Let $\cal F$ be a family of subsets of $\Omega$.
Then an $\cal F$-flip is a set of flips of the form $\set{\mathsf F_1,\ldots,\mathsf F_k}$, where each flip $\flip F_i$ is a pair $(A,B)$ with $A,B\in\cal F$.
Note that there are at most $2^{|\cal F|^2}$ different $\cal F$-flips.
In our context, the family $\cal F$ will usually be a partition of the set $V(G)\subset \Omega$ of vertices of some graph $G$. 
An \emph{$\cal F$-flip} of a graph $G$,
where $\cal F$ is a family of subsets of $V(G)$,
is a graph $G'$ obtained from $G$ after applying an $\cal F$-flip. Whenever we speak about an $\cal F$-flip, it will be always clear from the context whether we mean a graph or the family of flips used to obtain it.

\paragraph{$S$-classes.}
Let $G$ be a graph, and let $S\subset V(G)$ be a finite set of vertices. 
Consider the equivalence relation $\sim_S$ on $V(G)$, in which two vertices $a,b$ are equivalent if either $a,b\in S$ and $a=b$,
or $a,b\notin S$ and $N(a)\cap S=N(b)\cap S$.
An \emph{$S$-class} is an equivalence class of $\sim_S$.
 In other words, it is a set of vertices either of the form $\{s\}$ for some $s \in S$, or of the form
\[\setof{v\in V(G)\setminus S}{N(v)\cap S=T}\] for some $T\subset S$. The \emph{$S$-class} of a vertex $v \in V(G)$ is the unique $S$-class which contains $v$. Hence, $V(G)$ is partitioned into $S$-classes, and the number of $S$-classes is at most $|S|+2^{|S|}$.

\paragraph{$S$-flips.}
An \emph{$S$-flip} of a graph is an $\cal F$-flip $G'$ of $G$, where $\cal F$ is the partition of $V(G)$ into $S$-classes.
Note that there are $2^{2^{\Oof(|S|)}}$ many $S$-flips of a given graph $G$. 

Note that a direct consequence of \cref{obs_non_separated_twins} is that $S$-flips preserve $S$-classes.
Moreover, $S$-flips satisfy the following transitivity property. 

\begin{lemma}\label{lem:basic-flip-properties2}
	If $G$ is a graph, $S$ and $T$ are finite subsets of $V(G)$, $G'$ is an $S$-flip of $G$, and $G''$ is a $T$-flip of $G'$, then $G''$ is an $(S \cup T)$-flip of $G$. 
\end{lemma}
\begin{proof}
	Let $\mathcal{S}$ be the partition of $V(G)$ into $S$-classes in $G$ and let $\mathcal{T}$ be the partition of $V(G)$ into $T$-classes in $G'$.
	Let $\cal F$ be the common refinement of $\mathcal{S}$ and $\mathcal{T}$.
	It is easy to see that $G''$ is an $\mathcal F$-flip of $G$.
	It therefore remains to argue that the partition $\cal F'$ of $V(G)$ into $(S\cup T)$-classes in~$G$ is a refinement of $\cal F$.

	Assume two vertices $u$ and $v$ are in the same part of $\cal F'$.
	Then $u$ and $v$ are in the same $T$-class in $G$.
	As $u$ and $v$ are in the same part of $\cal S$, \cref{obs_non_separated_twins} yields that $u$ and $v$ are also in the same $T$-class in the $S$-flip $G'$, so in the same part of $\cal T$.
	This means that $u$ and $v$ are in the same part of $\cal F'$.
	As we made no assumptions on $u$ and $v$, this proves that $\cal F'$ is a refinement of $\cal F$ as desired.
\end{proof}

Finally, the $S$-flips behave well with the extraction of an induced subgraph.

\begin{lemma}\label{lem:basic-flip-properties1}
    Let $G$ be a graph and $S\subset X\subset V(G)$ be sets with $S$ finite. If $G'$ is an $S$-flip of $G$, then $G'[X]$ is an $S$-flip of $G[X]$.
\end{lemma}
\begin{proof}
Let  $F$ be the set of flips (between $S$-classes of $G$) used to obtain $G'$ from $G$, so that $G'=G\oplus F$.
By Observation~\ref{obs:subgraph_flip} we have $G'[X] = G[X] \oplus F'$, where $F'$ is obtained from $F$ by replacing each $S$-class $C$ used in any flip by $C \cap X$.
Now just note that the $S$-classes of $G[X]$ are 
exactly the sets of the form $C\cap X$, where $C$ is an $S$-class of $G$. Therefore, $F'$ applies flips between $S$-classes in $G[X]$, so $G'[X]$ is an $S$-flip of $G[X]$.
\end{proof}

\subsection{Flip-flatness}
The following notion of flip-flatness was introduced in~\cite{dreier2022indiscernibles}.
 Given a graph $G$ and a set of vertices $A\subseteq V(G)$ we call $A$ \emph{distance-$r$ independent} if all vertices in~$A$ are pairwise at distance greater than $r$ in $G$.

\begin{definition}[Flip-flatness]\label{def:wideness}
	A class of graphs $\CC$ is \emph{flip-flat}
	if for every $r \in \N$ there exists a function $N_r \colon \N \rightarrow \N$ and a constant $s_r\in \N$ such that for all $m\in\N$, $G\in \CC$, and $A\subseteq V(G)$ with $|A|\geq N_r(m)$, there exists a set $F$ of flips with $|F| \le s_r$ and $B\subseteq A$ with $|B|\geq m$ such that $B$ is distance-$r$ independent in~$G \oplus F$. 
\end{definition}

Flip-flatness is known to be equivalent to monadic stability, with a polynomial time algorithmic version.
\begin{theorem}[\cite{dreier2022indiscernibles}]\label{thm:fuqw}
	Let $\CC$ be a class of graphs. Then, the following are equivalent:
	\begin{enumerate}
		\item $\CC$ is monadically stable;
		\item $\CC$ is flip-flat;
		\item  for every $r \in \N$ there exists a function $N_r \from \N \rightarrow \N$ and a constant $s_r\in \N$ such that for all $m\in\N$, $G\in \CC$ with $n$ vertices, and $A\subseteq V(G)$ with $|A|\geq N_r(m)$, we can compute in time $f_\CC(r) \cdot n^3$ (for some function~$f_\CC$) a set $F$ of flips 
		 with $|F|\le s_R$ and a set $B\subseteq A$ with $|B|\geq m$ such that $B$ is distance-$r$ independent in~$G\oplus F$. 
	\end{enumerate}
\end{theorem}

\subsection{Flip-connectivity}

Let $G$ be a graph and $\mathcal{P}$ a partition of $V(G)$.
We will mostly be interested in the case when $\cal P$ is the partition of $V(G)$ into $S$-classes, for some finite $S\subset V(G)$. In this case, in the notation below we will write $S$ instead of $\cal P$.

The following notion is central in our proof of \cref{thm:main}.
We say that vertices $a$ and $b$ of $G$ are \emph{$r$-separated over $\cal P$}, denoted by\footnote{The symbol $\ind{}$ denotes forking independence in stable theories. Its use here is justified by the relationship of $r$-separation and forking independence in monadically stable theories, see next footnote.} 
\[ a \ind[\cal P]r b,\]
if there exists a $\cal P$-flip $H$ of $G$ such that $\dist_H(a,b) > r$.
We set \[ B^r_{\cal P}(a)\coloneqq \{b~|~b \text{ is not $r$-separated from $a$ over $\cal P$}\}.\] 
We define a more general notion below, where instead of single vertices $a$ and $b$ we may have sets of vertices. We also justify the use of the notation $B^r_\cal P(a)$,
by observing that this is indeed a ball in some metric.

\medskip

For any $r \in \N$ and any graph $G$, finite partition $\cal P$ of $V(G)$, and sets $A,B\in V(G)$, we write \[A\ind[\cal P]r B\] if there exists a $\cal P$-flip $H$ of $G$ such that $H$ has no $(A,B)$-path of length at most $r$. If $A\ind[\cal P]r B$ we say that $A$ and $B$ are \emph{$r$-separated over $\cal P$}. Note that when $A \cap B \neq \emptyset$, $A$ and $B$ are not $r$-separated over any partition $\cal P$. 
Note also that in the case when $\cal P$ is the partition into $S$-classes, we can assume that every vertex in $S$ is isolated in $H$, because this can be achieved through an $S$-flip. 

If $A$ consists of a single vertex $a$ and $B$ of a single vertex $b$, then we write $a\ind[\cal P]r b$ for $A\ind[\cal P]r B$. We use similar notation for $a\ind[\cal P]r B$ and $A\ind[\cal P]r b$. Note that $A\ind[\cal P]r B$ is a stronger condition than $a\ind[\cal P]r b$ for all $a\in A$ and $b\in B$, since we require that the same set $\cal P$ and the same $\cal P$-flip $H$ are used for all $a\in A$ and $b\in B$.
We write $\nind[\cal P]r$ to denote the negation of the relation $\ind[\cal P]r$. If $A\nind[\cal P]r B$ we say that $A$ and $B$ are \emph{$r$-connected over $\cal P$}.

For two vertices $u,v$, denote by $\dist_\cal P(u,v)$ the smallest number $r\in\N$
such that $u$ and $v$ are $r$-connected over $\cal P$, or $+\infty$ if no such number exists.
This can be equivalently described as follows.
For $u,v \in V(G)$, a \emph{$\cal P$-flip-path} from $u$ to $v$ in $G$ is a collection consisting of one $(u,v)$-path in $H$ for each $\cal P$-flip $H$ of $G$.
Note that a $\cal P$-flip-path might not exist. The \emph{length} of a $\cal P$-flip path is the supremum of the lengths of its paths. Then, equivalently, $\dist_{\cal P}(u,v)$ is the infimum of the lengths of all $\cal P$-flip-paths (where if there is no $\cal P$-flip-path then $\dist_\cal P(u,v)=+\infty$).

Now we prove that $\dist_\cal P(\cdot,\cdot)$ is a metric, where we allow metrics to take on the value of $+\infty$. Thus, in particular, the relation $\dist_\cal P(\cdot,\cdot)<{+\infty}$ is an equivalence relation on $V(G)$:
the set of all pairs $(u,v)\in V(G)^2$ such that $\dist_\cal P(u,v)<{+\infty}$
is transitive, reflexive, and symmetric.

\begin{lemma}
\label{lem:metric}
For any graph $G$ and finite partition  $\cal P$ of $V(G)$, $\dist_\cal P(\cdot,\cdot)$ is a metric on $V(G)$.
\end{lemma}
\begin{proof}
    First of all, notice that for any different $u,v \in V(G)$ it holds that $u\ind[\cal P]1 v$, because we have $\dist_H(u,v)>1$ either for $H=G$ or for $H$ being the  complement of $G$, which is a $\cal P$-flip.
    So in general $\dist_\cal P(a,b)=0$ if and only if $a=b$. From the definition we also have that $\dist_\cal P(\cdot,\cdot)$ is symmetric. Finally, a $\cal P$-flip-path from $a$ to $b$ can be naturally composed with a $\cal P$-flip-path from $b$ to $c$ in a path-by-path manner, yielding a $\cal P$-flip-path from $a$ to $c$. This proves the triangle inequality.
\end{proof}

Now that we know $\dist_\cal P(\cdot,\cdot)$ is a metric, it is sensible to define an analog of the ``ball of radius $r$ around a vertex''. This is exactly 
the notion $B^r_\cal P(v)$ defined earlier, as 
\[B^r_\cal P(v)=\setof{w\in V(G)}{\dist_\cal P(v,w)\le r}.\]

In case $\cal P$ is the partition into $S$-classes, for some finite $S\subset V(G)$, we write $A\ind[S] r B$ and $B^r_S(v)$ to denote 
$A\ind[\cal P]r B$ and $B^r_\cal P(v)$, respectively
\footnote{We comment on the relationship between $r$-separation and forking independence in monadically stable graphs,
thus justifying the use of the symbol $\ind{}$.
As a subset of authors will show in subsequent work, for a monadically stable graph $G$, its elementary superstructure $H$, and two vertices $a,b\in V(H)$,
$a$ and $b$ are forking independent over $H$ if and only 
if for every $r\in \N$ there is a finite $S\subset G$ 
such that $a\ind[S]r b$. This is related to a result of Ivanov, characterising forking independence in nowhere dense (or superflat) graphs~\cite{ivanov}.}.
This  notion behaves well under taking subsets of $S$ as follows.

\begin{lemma}
\label{lem:refininig-partitions}  
    For any $r\in\N\cup\set{+\infty}$, graph $G$, and finite sets $T\subset S\subset V(G)$, if $a$ is $r$-separated from $b$ over $T$, then $a$ is also $r$-separated from $b$ over $S$.
\end{lemma}

\begin{proof}
    Since $a\ind[T]rb$, there exists a $T$-flip $H$ of $G$ such that $H$ contains no $(a,b)$-path of length at most $r$. As $T\subseteq S$, $H$ is also an $S$-flip. Hence $a\ind[S]rb$.
\end{proof}



    

%% file: flipper_game.tex
\section{Flipper game}\label{sec:flippers}

We are now ready to introduce the Flipper game and the different variants we will use.

\subsection{The game and its variants}

Fix a radius $r\geq 1$ and a graph $G$.
The Flipper game over $G$ is played by two players, Flipper and Localizer.
At the level of intuition, one may think of it as a pursuit-evasion game: Flipper aims at capturing the Localizer, while Localizer aimes at surviving as long as possible.
At round $i$, Localizer is bound to the current arena $A_i \subseteq V(G)$, and Flipper wins when the arena is reduced to a singleton.

It is convenient to work with different variants of the game (which we will eventually show to be all equivalent); which correspond to different ways for Flipper to alter the metric in which the localisation is performed.
Initially, we set $A_0=V(G)$.
Then in the $i$th round, for $i>0$, the game proceeds as follows.
\begin{itemize}
\item If $|A_{i-1}|=1$ then Flipper wins.
\item Otherwise, Localizer chooses a vertex $c_i \in A_{i-1}$ and we let $A_i \subseteq A_{i-1}$ be the set of vertices in $A_{i-1}$ at distance at most $r$ from $c_i$ in the current metric.
\item Then Flipper updates the current metric following one of the possible variants described below.
\end{itemize}
In the definition above, it is crucial that the arena $A_i$ is updated \emph{before} the metric.
We now provide different ways of updating the metric.

\paragraph{Shrinking \emph{vs} confining.}
The most significant variation in the definition concerns whether we compute distance within the subgraph induced over the arena, or within the whole vertex set.
Stated differently, whether or not we allow the Localizer to travel through vertices which are not part of the current arena.
A formal definition follows.
We refer to this variation as the chosen \emph{localization mode}, where the first one (where distances are induced) is the \emph{shrinking mode}, and the second one (where distances are computed in the whole vertex set) is the \emph{confining mode}.

\paragraph{Atomic flips \emph{vs} separation.}
In the atomic-flips variant, we initially set $G_0 = G$, and at each round $i>0$, Flipper chooses an atomic flip $\flip F_i$ and sets $G_i = G_{i-1} \oplus \flip F_i$.
The metric is then set to be the graph metric over $G_i[A_i]$ in the shrinking variant, and over $G_i$ in the confining variant.
In the separation variant, we initially set $\cal P_0$ to be the trivial partition of $V(G)$, and at each round $i>0$, Flipper produces partition $\cal P_i$ by intersecting\footnote{The intersection of two partitions $\cal P$ and $\cal Q$ is the set of non-empty $P\cap Q$, where $P \in \cal P$ and $Q \in \cal Q$.} $\cal P_{i-1}$ with a bi-partition of $V(G)$.
The metric is then set to be the separating metric over $\cal P_i$, within $G[A_i]$ in the shrinking variant, or within $G$ in the confining variant.

Roughly speaking, playing in the separating variant is similar to applying all possible flips between the parts featured in the current partition.
This intuition is made formal in~\Cref{lem:atomic_flip_to_separation} below.

\paragraph{Definable separation.} In the quantifier-free definable (abbreviated qf-definable) separation variant, which is a special case of the separation variant, we initially set $S_0=\emptyset$ and at each round $i>0$, Flipper picks a vertex $s_i \in V(G)$ and lets $S_i=S_{i-1} \cup \{s_i\}$.
The metric is then set to be the separating metric over $S_i$ (or equivalently, over the partition into $S_i$-classes), within $G[A_i]$ or within $G$ according to the localization mode.

\paragraph{Playing in batches.} For any of the above variants, it is sometimes useful to fix a function $g : \N \to \N$ and allow Flipper, at the $i$th round, to play in a batch of size $g(i)$ as follows:
\begin{itemize}
\item in the batched atomic-flips variants, Flipper applies an $\cal F$-flip with $|\cal F| \leq g(i)$;
\item in the batched separation variants, Flipper produces $\cal P_i$ by intersecting $\P_{i-1}$ with a partition of index $\leq g(i)$;
\item in the batched qf-definable separation variants, Flipper produces $S_i$ by adding at most $g(i)$ vertices to $S_{i-1}$.
\end{itemize}
Playing in batches has little influence over the outcome of the game: in any variant, it is easy to see that if Flipper wins the batched game in $k$ rounds then he wins the non-batched one in $\sum_{i=1}^k g(i)$ rounds.

\paragraph{Terminology.} The above variants can be combined, for instance one may consider the confining game with qf-definable separation.
By default, when no extra attribute is added, by ``Flipper game'' we mean the shrinking game with atomic flips.

\subsection{Relating the different variants}\label{sec:variants}

We now discuss easy implications between the game variants.
More precisely, from the point of view of Flipper, up to changing the bound on the number of rounds and/or the radius, it holds that
\begin{itemize}
\item shrinking variants are easier than corresponding confining variants;
\item separating variants are easier than corresponding qf-definable separating variants;
\item atomic-flip variants are easier than corresponding separating variants.
\end{itemize}

In particular, this proves that the shrinking game with atomic flips is easier than the confining game with qf-definable separation, which is the implication~\eqref{it:sg}$\rightarrow$\eqref{it:fg} in Theorem~\ref{thm:main}.

For the first item, it suffices to notice that fewer moves are available to the Localizer in shrinking variants, so Flipper strategies transfer directly.
Likewise for the second item: since fewer moves are available to Flipper in the qf-definable variants, Localizer strategies transfer directly.
Therefore, there remains to prove the third item, which is stated as follows.

\begin{lemma}\label{lem:atomic_flip_to_separation}
There is a function $f:\N \to \N$ such that for any graph $G$ and radius $r$, if Flipper wins a separating variant of the game with radius $r$ in $\leq k$ rounds, then he wins the corresponding atomic-flip variant with radius $2r$ in $\leq f(k)$ rounds.
\end{lemma}

\newcommand{\sep}{\text{sep}}
\newcommand{\af}{\text{af}}

\begin{proof}
Let $G$ be a graph and $r \geq 1$.
Fix a Flipper strategy $S^\sep$ in the radius $2r$ separating variant, which wins in $\leq k$ rounds for any choices of Localizer.
We define a Flipper strategy $S^\af$ in the radius $r$ atomic-flip variant, by sequentially applying, at round $i$, all possible $\cal P_i$ flips in the partition $\cal P_i$ prescribed by $S^\sep$ (a formal definition follows).
Since $\cal P_i$ has index $\leq 2^i$, we work in the batched variant of the game.
There are $b_i=O(2^{|\cal P_i|^2})$ such flips, so each round in $S^\sep$ will require using $b_i$ rounds in $S^\af$.

Formally, write $\cal P_i^{\sep}(c_1\dots c_i)$ for the partition produced at round $i$ by the strategy $S^\sep$ against Localizer moves $c_1 \dots c_i$.
By definition, $\cal P_0^{\sep}$ is the partition with one part, and $\cal P_i^{\sep}(c_1 \dots c_i)$ is the intersection of $\cal P_i^{\sep}(c_1 \dots c_{i-1})$ with a bipartition.
For $\tup c = c_1 \dots c_i$, we let $b_i(\tup c) = O(2^{2^{2i}})$ be the total number of $\cal P_i^{\sep}(\tup c)$-flips, which we enumerate as $\flip F_1(\tup c), \dots, \flip F_{b_i(\tup c)}(\tup c)$ (since $|\cal P_i^{\sep}(\tup c)| \leq 2^i$, these flips are valid moves for Flipper in the batched atomic-flips game).
Note that for $i=0$, Localizer has played the empty tuple $\eps$ of moves, $\cal P_0^\sep(\eps)$ is the partition of $V(G)$ with one part, so the number $b_0(\eps)$ of $\cal P_0^\sep(\eps)$-flips is $2$: the identity, and the one reversing all edges.

Fix $\ell$ and a sequence $\tup c^\af = c_1  \dots c_\ell$ of $\ell$ Localizer moves; we now formalize the definition of the strategy $S^\af$ against $\tup c^\af$.
Define $j_0=0$ and $j_{i+1}$ to be $j_{i} + b_{i}(c_{j_1} \dots c_{j_i})$ if this number is $\leq \ell$, or undefined otherwise, and let $i_0$ be maximal such that $j_{i_0}$ is defined.
Write $\tup c^\sep = c_{j_1} \dots c_{j_{i_0}}$.
Then let $p=\ell-j_{i_0}+1$; note that $1\leq p \leq b_{i_0}(\tup c^\sep)$.
We define the flip $\flip F_\ell^\af(\tup c^\af)$ played by the strategy $S^\af$ against $\tup c^\af$ to be $\flip F_{p}(\tup c^\sep)$, the $p$th flip in our fixed enumeration of all $\cal P^\sep_{{i_0}}(c^\sep)$-flips.

We now prove that $S^\af$ wins the batched atomic-flips game in $\leq f(k)=\sum_{i=1}^{k}2^{2^{2i}}$ rounds.
We say that a sequence $c_1 \dots c_i$ of Localizer moves is valid against a given Flipper strategy if for each $j \leq i$, $c_j$ belongs to the arena $A_{j-1}$ obtained from the strategy (in the variant under consideration).
We prove by induction on $\ell$ that for any sequence of Localizer moves $\tup c^\af=c_1 \dots c_\ell$ which is valid against $S^\af$ and such that, using the notation in the paragraph above, $\ell = j_{i_0}$, it holds that the arena $A_{\ell}^\af(\tup c^\af)$ obtained from the strategy $S^\af$ against $\tup c^\af$ is a subset of the arena $A_{i_0}^\sep(\tup c^\sep)$ obtained from $S^\sep$ against $\tup c^\sep$.
In particular, this implies that $\tup c^{\sep}$ is valid in the separating game against the strategy $S^\sep$.
Therefore, $|A_{i_0}^\sep(\tup c^\sep)|$ reaches $1$ before $i_0$ reaches $k$, and thus $|A_\ell^\af(\tup c^\af)|$ reaches $1$ before $\ell$ reaches $f(i_0)$, as required.

Consider a sequence of Localizer moves $\tup c^\af = c_1 \dots c_\ell$ which is valid against $S^\af$ and such that $\ell = j_{i_0} \geq 0$, and assume the result known for proper prefixes of $\tup c^\af$ (if any).
For concreteness, we assume that we are in confining mode (the proof is very easily adapted to the shrinking mode).
Let $v \in A_{\ell}^\af(\tup c^\af)$; we aim to prove that $v \in A_{i_0}^{\sep}(c^\sep)$, which rewrites as
\[
	v \in A_{i_0-1}^\sep(\tup c^\sep_{<i_0}) \qquad \text{ and} \qquad \text{for any } \cal P_{i_0-1}^\sep(\tup c^\sep_{<i_0})\text{-flip } \flip G' \text{ of G}, \quad \dist_{G'}(v,c_{\ell}) \leq 2r,
\]
where $\tup c^\sep_{<i_0} = c_{j_1} \dots c_{j_{i_0-1}}$.
By induction, we immediately get that the condition on the left is satisfied.
Let $G'$ be a $\cal P_{i_0-1}^\sep(\tup c^\sep_{<i_0})$-flip of $G$.
By definition of $S^\af$, there is $j<\ell$ such that $G'$ is the graph obtained by applying the $\cal P^\sep(\tup c^{\sep}_{<i_0})$-flip $F_{j}^\af(c_1 \dots c_{j})$ to $G$.
Since $v \in A_\ell^\af(\tup c^\af)$, we also have $v \in A_{j+1}^\af(c_1 \dots c_{j+1})$ thus $\dist_{G'}(v,c_{j+1}) \leq r$.
But we also have $c_\ell \in A_{j+1}^\af(c_1 \dots c_{j+1})$ and therefore $\dist_{G'}(c_\ell,c_{j+1}) \leq r$, which gives $\dist_{G'}(v,c_\ell) \leq 2r$, as required.
\end{proof}

It follows from closing the cycle of implications in Theorem~\ref{thm:main-main}, that the confining game with qf-definable separation is itself also easier than the shrinking game with atomic flips, but this is not at all obvious.
Therefore, all variants of the Flipper game are equivalent.

%% file: equivalences.tex
\section{Relations to flip-flatness and existential monadic stability}\label{sec:flip-to-mstab}

In this section we prove that the existence of short winning strategies in the Flipper game implies two properties that are known to be equivalent to monadic stability: flip-flatness and existential monadic stability.

\subsection{From Flipper game to flip-flatness}
\label{sec:pfg_to_fw}

This subsection is devoted to proving the following lemma.
\begin{lemma}
\label{lem:flipper_to_stable}
Let $\CC$ be a class of graphs such that for every $r$ there exists $k$ such that Flipper wins the confining separating game of radius $r$ on any $G \in \CC$, in at most $k$ rounds.
Then $\CC$ is flip-flat.
\end{lemma}
Together with Lemma~\ref{lem:atomic_flip_to_separation}, 
this yields the following corollary, proving the implication \eqref{it:sg}$\rightarrow$\eqref{it:fw} in Theorem~\ref{thm:main}.
\begin{corollary}\label{cor:game-to-flatness}
    Let $\CC$ be a class of graphs such that for every $r$ there exists $k$ such that Flipper wins the confining game with qf-definable separation of radius $r$ on any $G \in \CC$, in at most $k$ rounds.
    Then $\CC$ is flip-flat.        
\end{corollary}

\begin{proof}[Proof of Lemma~\ref{lem:flipper_to_stable}]


Let $r \in \N$ be arbitrary. Let $k$ be a number of rounds such that Flipper wins the confining separation game of radius $r$ on any $G \in \CC$ in at most $k$ rounds; such $k$ exists by our assumptions. 
Let $c$ be the maximal possible number of different $\F$-flips on any graph $G$ with respect to any $\F$ with $|\F| \le k$; it is easily seen that such number exists (the upper bound does not depend on any particular $G$ and $\F$, only on $k$).
For $m \in \N$ denote by $R(m)$ the least number $N$ such that any coloring of the edges of a clique with $N$ vertices using $c$ colors yields a monochromatic clique of size $m$. Such a number exists by Ramsey's theorem.
Set $s_r:=k^2$ and for $m \in \N$ set $N_r(m):=R(m)^k$.
Let $m \in \N$ and let $G \in \CC$ and $A \subseteq V(G)$ with $|A| > N_r(m)$.
From the definition of the game it follows that whenever Flipper can win such game in $k$ rounds, then he can also win in $k$ rounds if we decide that the initial arena $A_0$ is a subset of $V(G)$.
Indeed, in every round we intersect the previous arena with something that depends only on the partition played by Flipper.
Therefore, set $A_0 :=A$.
We will extract a partition $\F$ and an $\F$-flip $H$ of $G$ with the desired properties from a play of the Flipper game played on $G$. We define a strategy for Localizer in the Flipper game played on $G$ as follows. Let $\F_i$ denote the partition of $V(G)$ resulting from Flipper's choices after $i$ rounds. For $i>0$, in the $i$th round, 
if there is a vertex $v \in A_{i-1}$ such that $|B^r_{{\F}_{i-1}}(v) \cap A_{i-1}|  > R(m)^{k-i}$ then Localizer picks $v$ as $c_i$, otherwise Localizer picks any vertex in $A_{i-1}$. In other words, Localizer tries to make sure that $|A_i|> R(m)^{k-i}$ in $i$th round whenever possible.

Consider now a play of the Flipper game played on $G$ in which Localizer plays according to the strategy described above and Flipper plays according to an optimal strategy which leads to a win in at most $k$ rounds. We then get the following.
\begin{claim}
\label{claim:MF-bounds}
There exists $i < k$ such that$A_{i-1} \ge R(m)^{k-i+1}$ and $|B^r_{\F_{i-1}}(v) \cap A_{i-1}| \le R(m)^{k-i}$ for each $v \in A_{i-1}$.
\end{claim}
\begin{claimproof}
Since Flipper wins in $k$ rounds, there has to exist $i < k$ such that $|B^r_{\F_{i-1}}(v) \cap A_{i-1}| \le R(m)^{k-i}$ for each $v \in A_{i-1}$ (otherwise after $k-1$ rounds we would have that $|A_{k-1}|> R(m) \geq 1$, which would contradict Flipper winning after $k$ rounds).
Consider the smallest $i$ with this property. This means that no matter which vertex $v \in A_{i-1}$ Localizer plays, the arena $A_i$ has size less than $R(m)^{k-i}$. Since $i$ was the smallest such number, we have that  $A_{i-1} \ge R(m)^{k-i+1}$.
\end{claimproof}
Let $i$ be the number from Claim~\ref{claim:MF-bounds}. We greedily construct a subset $A'$ of $A_{i-1}$ by repeatedly picking a vertex $v$ in $A_{i-1}$ and removing $B^r_{\F_{i-1}}(v) \cap A_{i-1}$ from $A_{i-1}$ for as long as possible. Because of the bounds given by Claim~\ref{claim:MF-bounds}, set $A'$ will have size at least $R(m)$, and by construction all vertices in  $A'$ will be pairwise $r$-disconnected over $\F$ in $G$.

Assign to each unordered pair $u,v$ of distinct vertices in $A'$ a color which represents an $\F$-flip $H$ of $G$ such that  $\dist_H(u,v) > r$ (if there is more than one such $\F$-flip, pick one arbitrarily). This way we assign at most $c$ colors to pairs of vertices from a set of size at least $R(m)$, and therefore by the definition of $R(m)$ there exists a subset $A''$ of $A'$ with $|A''|\ge m$ such that each pair  $u,v \in A''$ is assigned the same $\F$-flip $H$. Thus we have $\dist_H(u,v) > r$ for each $u,v \in A''$, and so $\F$ and $H$ have the properties required by the definition of flip-flatness. This means that $\CC$ is flip-flat, which finishes the proof.
\end{proof}

%% file: flip_to_mstab.tex
\subsection{From Flipper game to existential monadic stability}\label{sec:flip-to-mstab-new}

In this subsection we will show that a winning strategy for Flipper in the shrinking game with atomic flips implies existential monadic stability. Since existential monadic stability implies 
stability of the edge relation, and existential monadic dependence, this will prove the implication \eqref{it:fg}$\rightarrow$\eqref{it:ems} in \cref{thm:main}.
For simplicity, we will concentrate on the case when we can define an infinite ladder in a single graph.

We will say that a formula $\phi(x, y)$ defines an infinite ladder in a structure $\str M$ if there is a sequence $(a_i, b_i)_{i \in \N}$ of pairs of elements of $\str M$ such that for every $i, j \in \N$
\[
    \str M \models \phi(a_i, b_j) \iff i < j.
\]

\begin{lemma}\label{lem:mstab_to_flip}
    Let $\str M$ be a graph and let $\widehat{\str M}$ be a monadic lift of $\str M$.
    Let $\phi(x, y)$ be an existential formula that defines an infinite ladder in $\widehat{\str M}$.
    Then there exists $r \in \N$ such that Localizer can play infinitely many rounds in the Flipper game of radius $r$ without losing.
\end{lemma}

The statement of Lemma \ref{lem:mstab_to_flip} can be restated for the case when we can define arbitrarily long ladders in graphs from a given class.

\begin{lemma}[Finitary version of Lemma \ref{lem:mstab_to_flip}]\label{lem:fin_mstab_to_flip}
    Let $\Cc$ be a class of graphs and let $\phi(x, y)$ be an existential formula such that for every $k \in \N$ there exists a graph $\str M_k \in \Cc$ and a monadic lift $\widehat{\str M_k}$ in which $\phi(x, y)$ defines a ladder of length at least~$k$.
    Then there exists $r \in \N$ such that for every $\ell \in \N$ there is a graph $\str N_\ell \in \Cc$ with the following property: in the Flipper game of radius $r$ with the initial current graph $\str N_\ell$ Localizer can play at least $\ell$ rounds without losing.
\end{lemma}

Of course Lemma \ref{lem:fin_mstab_to_flip} gives us implication \eqref{it:fg}$\rightarrow$\eqref{it:ems} from \cref{thm:main}.
For the rest of this section we will concentrate on proving Lemma \ref{lem:mstab_to_flip}.
It can be easily observed that the proof can be restated for Lemma \ref{lem:fin_mstab_to_flip}.


Let $\str M$ be a monadic lift of a graph.
For an integer $r$, we say that a set $A$ of vertices of $\str M$  is \emph{$r$-close} if the vertices of $A$ are pairwise at distance at most $r$.
In particular, if $(a_i, b_i)_{i \in \N}$ are the vertices of a ladder defined by some formula, we say that this ladder is $r$-close if the set $\setof{a_i}{i \in \N} \cup \setof{b_i}{i \in \N}$~is.

A useful tool for the proof of \cref{lem:mstab_to_flip} will be an easy corollary from Gaifman's locality theorem.
The theorem was originally proven in \cite{gaifman}, but we will use a corollary of it, similar to the one from \cite[Lemma 2.1]{DBLP:conf/lics/BonnetDGKMST22}.
\begin{theorem}[Corollary of Gaifman's locality theorem]\label{thm:gaifman}
    Let $\phi(x, y)$ be an FO formula in the vocabulary of graphs with a number of additional unary predicates.
    Then there are numbers $t, s \in \N$ with $t$ depending only on the quantifier rank of $\phi$ such that for every graph $G$ with a number of additional unary predicates, $G$ can be vertex-colored using $s$ colors in such a way that for any two vertices $u, v \in V(G)$ at distance more than $t$, whether or not $\phi(u, v)$ holds depends only on the color of $u$ and the color of $v$.
\end{theorem}


Using \cref{thm:gaifman} we can prove that whenever $\phi$ defines an infinite ladder, then it also defines an infinite $d$-close ladder.

\begin{lemma}
    \label{lem:d_close_ladder}
    Let $\str M$ be a monadic lift of a graph and let $\phi(x, y)$ be a formula which defines an infinite ladder in $\str M$.
    Then, there is an integer $d$ depending only on the quantifier rank of $\phi(x, y)$ such that $\phi(x, y)$ defines in $\str M$ an infinite $d$-close ladder.
\end{lemma}
\begin{proof}
    Observe, that by \cref{thm:gaifman}, there exists a constant $t$ depending only on the quantifier rank of $\phi$ and a vertex-coloring of $\str M$ with $s$ colors with the following property: if $a, b, a', b'$ are four vertices of $\str M$ such that $\mathrm{dist}(a, b), \mathrm{dist}(a', b') > t$, and colors of $a$ and $a'$ are equal, and colors of $b$ and $b'$ are equal, then
    \[
        \str M \models \phi(a, b) \iff \str M \models \phi(a', b').
    \]

    Take an infinite ladder $(a_i, b_i)_{i \in \N}$ defined by $\phi$.
    Let us assign to every pair of natural numbers $(i, j)$ with $i < j$ with one of three colors which correspond to three possible scenarios (if more than one scenario holds, we pick an arbitrary one):
    \begin{itemize}
        \item $\mathrm{dist}(a_i, b_j) \le t$,
        \item $\mathrm{dist}(b_i, a_j) \le t$,
        \item $\mathrm{dist}(a_i, b_j) > t$ and $\mathrm{dist}(b_i, a_j) > t$.
    \end{itemize}
    By Ramsey's theorem, there exists an infinite subladder $(a_{i_j}, b_{i_j})_{j \in \N}$ defined by $\phi$ such that every pair $(i_j, i_k)$ for $j < k$ has the same color.
    It is straightforward to see that in the first two cases the vertices $\setof{a_{i_j}}{j \ge 2} \cup \setof{b_{i_j}}{j \ge 2}$ are pairwise at distance at most $3t$.
    It remains to deal with the last case -- we will show that, in fact, it cannot hold.
    
    Color the vertices of the graph into a~finite number of colors according to \cref{thm:gaifman}.
    By the pigeonhole principle, we can assume that all $a_{i_j}$ have been assigned the same color and, separately, all $b_{i_j}$ have been assigned the same color.
    However, that means
    \[
        \str M \models \phi(a_{i_j}, b_{i_k}) \iff \str M \models \phi(a_{i_k}, b_{i_j})
    \]
    for any $j \neq k$.
    This is clearly a contradiction, so we conclude it is enough to take $d \coloneqq 3t$.
\end{proof}

To prove \cref{lem:mstab_to_flip}, we will present a strategy for Localizer. We start with an infinite ladder defined in $\str M$ by an existential formula $\phi(x,y)$ in \emph{prenex normal form} of quantifier rank at most $q$ (i.e., $\phi(x, y) \equiv \exists \tup z. \alpha(x, y, \tup z)$ where $|\tup z| \le q$ and $\alpha$ is a quantifier-free formula).
The strategy will maintain the following invariant: after each round there is an existential formula $\phi'(x, y)$ which defines an infinite ladder in some monadic lift of the current graph.
The proof will follow from two lemmas, corresponding to the moves of Flipper and Localizer.

\begin{lemma}
    \label{lem:flipper_moves}
    Let $\str M$ be a graph and let $\widehat{\str M}$ be a monadic lift of $\str M$.
    Let $\phi(x, y)$ be an existential formula of quantifier rank $q$ in prenex normal form which defines an infinite ladder in $\widehat{\str M}$.
    Let $\str N$ be a flip of $\str M$.
    Then there exists a monadic lift $\widehat{\str N}$ of $\str N$ and a formula $\psi(x, y)$ of quantifier rank $q$ in prenex normal which defines an infinite ladder in $\widehat{\str N}$.
\end{lemma}
\begin{proof}
    Let $(a_i, b_i)_{i \in \N}$ be an infinite ladder defined by $\phi(x, y)$ in $\widehat{\str M}$.
    We define $\widehat{\str N}$ by adding the same unary predicates as in $\widehat{\str M}$ and two additional unary predicates that mark the sets that were flipped in $\str M$.
    We also define $\psi(x, y)$ by changing the atomic check $E(u, v)$ in $\phi(x, y)$ to a more complicated quantifier-free formula verifying if there exists an edge between $u$ and $v$ in $\str M$ and whether $u$ and $v$ were included in the flipped sets.
    Of course, $\psi(x, y)$ has the same quantifier rank as $\phi(x, y)$ and is in prenex normal form.
    Moreover, for every $i, j \in \N$,
    \[
        \widehat{\str M} \models \phi(a_i, b_j) \iff \widehat{\str N} \models \psi(a_i, b_j).\qedhere
    \]
\end{proof}

\begin{lemma}
    \label{lem:connector_moves}
    Let $\str M$ be a graph and let $\widehat{\str M}$ be a monadic lift of $\str M$.
    Let $\phi(x, y)$ be an existential formula of quantifier rank $q$ in prenex normal form which defines an infinite ladder in $\widehat{\str M}$.
    There exists an integer $r$ depending only on $q$, a formula $\psi(x, y)$ of quantifier rank at most $q$ in prenex normal form, a monadic lift $\widehat{\str M}'$ of $\str M$, and an element $m \in \str M$ such that $\psi(x, y)$ defines an infinite ladder in $\widehat{\str M}'[B^{r}(m)]$.
\end{lemma}
\begin{proof}
    By \cref{lem:d_close_ladder} we can assume that $\phi(x, y)$ defines a $d$-close ladder $(a_i, b_i)_{i \in \N}$.
    As $\phi(x, y) \equiv \exists \tup z. \alpha(x, y, \tup z)$, for every $i < j$ we can find a tuple $\tup c^{ij}$ such that $\alpha(a_i, b_j, \tup c^{ij})$ holds in $\widehat{\str M}$.
    
    For every tuple $\tup c^{ij}$ we define its {\em{profile}} as a function
    \[ \pi_{ij}\,\colon\, \tup z \to [q] \cup \set{\infty}, \]
    where $\pi_{ij}(z)$ for a~variable $z \in \tup z$ is defined as follows.
    Let $c^{ij}_z$ be the element of the tuple $\tup c^{ij}$ corresponding to the variable $z$.
    For $\ell \in [q]$, we set $\pi_{ij}(z) = \ell$ if the distance between $a_i$ and $c^{ij}_z$ is at least $10(\ell - 1)d$ and at most $10\ell d - 1$.
    However, if the distance between $a_i$ and $c^{ij}_z$ is at least $10qd$, we set $\pi_{ij}(z) = \infty$.
    

    By a standard Ramsey argument, we can assume that for every $i < j$ all tuples $(a_i, b_j, \tup c^{ij})$ have the same atomic type $\tau$ and all $\tup c^{ij}$ have the same profile function $\pi$ (possibly by going to an infinite subladder of $(a_i, b_i)_{i \in \N}$).
    If $\pi$ does not assign $\infty$ to any coordinate, then all $\tup c^{ij}$ are in the ball of radius $r := 10qd + d$ around $a_1$.
    Therefore, after restricting $\str M$ to the ball of radius $r$ around $a_1$ we still have for all $i, j \in \N$ that
    \[
        \widehat{\str M}[B^{r}(a_1)] \models \phi(a_i, b_j) \iff i < j.
    \]
    (Note that it is important here that $\varphi$ is an~existential formula; if not for that assumption, $\phi(a_i, b_j)$ could have become true in $\widehat{\str M}[B^{r}(a_1)]$ for some $i \ge j$.)
    
    Now assume that $\pi$ does assign $\infty$ to some coordinate.
    Therefore, by the pigeonhole principle, there exists $s \in [q]$ such that no variable is assigned $s$ by $\pi$.
    We split the variables $z \in \tup z$ into two parts -- these where $\pi(z) < s$ (call these variables \emph{close}), and those where $\pi(z) > s$ (call those \emph{far}).
    Observe now that if $i < j$ and $z$ is close, then $\dist(a_i, c^{ij}_z) < 10(s - 1)d$ by definition, and thus $\dist(a_1, c^{ij}_z) < (10s - 9)d$ by triangle inequality.
    On the other hand, if $z$ is far, then $\dist(a_i, c^{ij}_z) \ge 10sd$, and therefore $\dist(a_1, c^{ij}_z) \ge (10s - 1)d$, again by triangle inequality.
    In particular, the atomic type $\tau$ specifies that there are no edges between the vertices assigned to close variables and the vertices assigned to far variables: if $z$ is close and $z'$ is far, then $\dist(c^{ij}_z, c^{ij}_{z'}) > 1$.

    Let us create $\widehat{\str M}'$ by adding to $\widehat{\str M}$ one more unary predicate $U$ satisfied for the vertices that are at distance at most $(10s-9)d$ from $a_1$.
    By construction, every vertex of the ladder is in $U$, and for every pair $i < j$ and every variable $z \in \tup z$, the vertex $c^{ij}_z$ is in $U$ if and only if $z$ is close.

    Let $\tau'$ be the atomic type obtained from $\tau$ by additionally specifying which variables should satisfy $U$.    
    Consider also $\phi'(x, y) \equiv \exists \tup z. \tau'(x, y, \tup z)$.
    It is immediate that for any $i, j \in \N$,
    \[
        \widehat{\str M}' \models \phi'(a_i, b_j) \iff i < j.
    \]
    Now, simplify $\phi'$ to $\phi''$ by removing the quantifiers that correspond to far variables (of course, we also need to simplify $\tau'$ to $\tau''$ by removing the same variables).
    It is again obvious that $\widehat{\str M}' \models \phi''(a_i, b_j)$ for $i < j$.
    We will show that $\widehat{\str M}' \not \models \phi''(a_i, b_j)$ for $i \ge j$.

    Assume by contradiction that for some $i \ge j$ we have a tuple $\tup c$ such that $\widehat{\str M}' \models \tau''(a_i, b_j, \tup c)$.
    By the construction of $\tau''$, the vertices $a_i$ and $b_j$ and each element of $\tup c$ must satisfy $U$ and thus each of them is at distance at most $(10s - 9)d$ from $a_1$.
    Consider now extending this tuple to $\tup c'$ by adding the vertices of $\tup c^{12}$ corresponding to the far variables of $\tup z$; recall that these vertices are at distance at least $(10s - 1)d$ from $a_1$.
    Clearly, $\widehat{\str M}' \models \tau'(a_i, b_j, \tup c')$, as the vertices which we used for extending $\tup c$ do not neighbor any vertex from $\tup c$, $a_i$ or $b_j$.
    This is a contradiction to $\widehat{\str M}' \not \models \phi'(a_i, b_j)$.
    Therefore, again we have
    \[
        \widehat{\str M}' \models \phi''(a_i, b_j) \iff i < j,
    \]
    and by using the same argument as previously we have
    \[
        \widehat{\str M}'[B^{r}(a_1)] \models \phi''(a_i, b_j) \iff i < j.\qedhere
    \]
\end{proof}

Using \cref{lem:flipper_moves} and \cref{lem:connector_moves} we can easily prove \cref{lem:mstab_to_flip}.
\begin{proof}[Proof of \cref{lem:mstab_to_flip}]
    Rewrite formula $\phi(x, y)$ to a formula $\phi'(x, y)$ in prenex normal form.
    Denote its quantifier rank by $q$ and take $r$ as in \cref{lem:connector_moves}.
    Assume that we consider the Flipper game of radius $r$ on $\str M$.

    We call a graph $\str A$ \emph{winning} if there exists a monadic lift $\widehat{\str A}$ of $\str A$ and a formula $\psi(x, y)$ of quantifier rank at most $q$ in prenex normal form which defines an infinite ladder in $\widehat{\str A}$.
    Obviously, if a graph is winning, then it cannot be a single vertex.

    Observe that by our assumption, the initial current graph $\str M$ is winning.
    Then, by \cref{lem:flipper_moves}, if Flipper does a flip on a winning current graph, the resulting current graph is also winning.
    Finally, by \cref{lem:connector_moves}, for every winning current graph there exists a move of Localizer such that the resulting current graph is also winning.
    Therefore, if Localizer always picks such moves, she plays infinitely many rounds without losing.
\end{proof}

%% file: mt-prelims.tex
\section{Additional model-theoretic preliminaries}
\label{sec:mt-prelims}

A \emph{theory} $T$ (over $\Sigma$) is a set of $\Sigma$-sentences. A \emph{model of} a theory $T$ is a model $\str M$ such that $\str M\models \phi$ for all $\phi\in T$. When a theory has a model, it is said to be \emph{consistent}. The \emph{theory of} a class of $\Sigma$-structures $\CC$ is the set of all $\Sigma$-sentences $\phi$ such that $\str M\models \phi$ for all $\str M\in \CC$. The \emph{elementary closure} $\overline {\CC}$ of $\CC$ is the set of all models $\str M$ of the theory of $\CC$. Thus $\CC\subset \overline {\CC}$, and $\CC$ and $\overline \CC$ have equal theories. 


We will use compactness for first-order logic and the Tarski-Vaught test, recalled~below.

\begin{theorem}[Compactness]\label{thm:compactness}
  A theory $T$ is consistent if and only if every finite subset $T'$ of $T$ is consistent.
\end{theorem}

Let $\str M$ and $\str N$ be two structures with $\str M\subset \str N$, that is, the domain of $\str M$ is contained in the domain of $\str N$.
Then $\str N$ is an \emph{elementary extension} of $\str M$, written $\str M\prec \str N$,
if for every formula $\phi(\tup x)$ (without parameters) and tuple $\tup m\in \str M^{\tup x}$, the following equivalence holds:
\[\str M\models \phi(\tup m)\iff \str N\models \phi(\tup m).\]
\noindent We also say that $\str M$ is an \emph{elementary substructure} of $\str N$.
In other words, $\str M$ is an elementary substructure of $\str N$ if $\str M$ is an induced substructure of $\str N$, where we imagine that $\str M$ and 
$\str N$ are each equipped with every relation 
$R_\phi$ of arity $k$ (for $k\in\N$) that is defined by any fixed first-order formula $\phi(x_1,\ldots,x_k)$. 
In this intuition, formulas of arity $0$ correspond to Boolean flags, with the same valuation for both $\str M$ and $\str N$.

\begin{theorem}[Tarski-Vaught Test]\label{thm:tarsk-vaught}
    The following conditions are equivalent for any structures $\str M$ and $\str N$ with $\str M\subset \str N$.
    \begin{itemize}
        \item The structure $\str N$ is an elementary extension of $\str M$.
        \item For every formula $\phi(y;\tup x)$ and tuple $\tup m\in \str M^{\tup x}$,
    if $\str N \models \phi(n;\tup m)$ holds for some $n\in \str N$, then $\str N \models \phi(n';\tup m)$ holds for some $n'\in \str M$.
    \end{itemize}
\end{theorem}

Fix a model $\str M$ over a signature $\Sigma$.
A $\Sigma$-formula $\phi(\tup x)$ \emph{with parameters} from a set $A\subset \str M$ is a formula $\phi(\tup x)$ over the signature $\Sigma\uplus A$, where the elements of $A$ are treated as constant symbols (which are interpreted by themselves). If $\phi(\tup x)$ is a formula (with or without parameters) and $U\subset \str M$, then by $\phi(U)$ we denote the set of all $\tup x$-tuples $\tup u\in U^{\tup x}$ such that $\str M\models \phi(\tup u)$. Now let $A,B\subset \str M$ be sets, and let $\phi(x;y)$ a formula
(here $x$ and $y$ are single variables).
 A \emph{$\phi$-type} of $A$ over $B$ is an equivalence class of the relation $\sim$ on $A$ such that
for  $a,a'\in A$ we have 
$a\sim a'$ if and only if 
$\phi(a;B)=\phi(a';B)$, that is, 
\[\str M\models \phi(a;b)\iff \phi(a';b)\text{\quad for all $b\in B$}.\] We denote the set of $\phi$-types of $A$ over $B$ by $\Types[\phi](A/B)$. 

The following result is a fundamental fact about stable formulas (see e.g. \cite[Lemma~2.2 (i)]{pillay1996geometric}).

\begin{theorem}[Definability of types]\label{thm:definability-of-types}
  Let $\str M\prec \str N$ be two models and $\phi(x;y)$ be a formula that is stable in $\str M$.
  For every $n\in \str N$ there is some formula 
  $\psi(x)$ with parameters from $\str M$,
  which is a positive boolean combination of formulas of the form $\phi(x;m)$ for $m\in\str M$,
  such that the following conditions are equivalent for every $a\in \str M$:
  \begin{itemize}
    \item $\str N\models \phi(n;a)$ holds,
    \item $\str M\models \psi(a)$ holds.
  \end{itemize}
\end{theorem}

The following lemma is reminiscent of the classic notion of Morley sequences from model theory (see e.g.
 \cite[Def.~2.27]{pillay1996geometric}).

\begin{lemma}\label{lem:kind_of_morley}
  Let $\str M\prec \str N$ be graphs, let $\tup n \in \str N^{\tup y}$, and let $A \subseteq \str M$ be a finite set.
  There is an infinite sequence $\tup b_0,\tup b_1,\ldots \in \str M^{\tup y}$ such that
  \begin{itemize} 
    \item for each $i \in \N$, $\tup b_i$ has the same atomic type as $\tup n$ over $A \cup \tup b_0 \cup \cdots \cup \tup b_{i-1}$; and 
    \item the atomic types of the tuples $\tup b_i \tup b_j$ are the same for all $i<j$.
  \end{itemize}
  If moreover $\str N$ has a stable edge relation, one may choose the sequence $(\tup b_i)_{i \in \N}$ so that the atomic types of $\tup b_i \tup b_j$ are the same for all $i \neq j$.
\end{lemma}
  
\begin{proof}
We construct the sequence $\tup b_0,\tup b_1,\dots \in \str M^{\tup y}$ satisfying the first condition by induction.
Let $i \in \N$ and assume that the elements $\tup b_0, \dots, \tup b_{i-1}$ have already been constructed and follow the first condition (this assumption is vacuous for $i=0$).
Let $\alpha(\tup y)$ be the conjunction of all (finitely many) formulas in the atomic type of $\tup n$ over $A\cup\tup b_0 \cup \dots \cup \tup b_{i-1}$.
Since $\str N \models \alpha(\tup n)$, and
$A\cup\tup b_0 \cup \dots \cup \tup b_{i-1}\subset \str M$, 
and $\str M \prec \str N$, it follows that there exists a tuple in $\str M^{\tup y}$ satisfying $\alpha(\tup y)$.
(Formally, let $\tup u$ be a~tuple enumerating the elements of $A \cup \tup b_0 \cup \dots \cup \tup b_{i-1}$, and $\varphi(\tup y, \tup z)$ be the formula without parameters for which $\varphi(\tup y, \tup u) = \alpha(\tup y)$.
We apply the definition of an~elementary extension to the formula $\exists \tup y. \varphi(\tup y, \tup u)$ and infer that the formula holds in $\str M$.
Denote by $\tup b_i\in \str M^{\tup y}$ any tuple satisfying $\varphi(\tup b_i, \tup u)$, concluding the induction step.)

By construction, the sequence we have constructed satisfies the first item in the statement of the lemma.
For the second item, it suffices to apply the infinite Ramsey theorem,
as there are finitely many possible atomic types for $\tup b_i\tup b_j$'s.

Assume now that $\str N$ has a~stable edge relation.
By the second item, we may assume the atomic types of tuples $\tup b_i\tup b_j$ for $i<j$ are all equal to each other, and hence, the atomic types of tuples $\tup b_j\tup b_i$ for $i < j$ are also equal to each other; assume for contradiction that these are two different types.
Since the atomic types of all $\tup b_i$'s are identical, this implies that there are two different coordinates $y_1,y_2$ of $\tup y$ such that for all $i<j$, the corresponding coordinates $b_{i,1},b_{i,2},b_{j,1},b_{j,2} \in \str M$ of $\tup b_i$ and $\tup b_j$ are such that
\begin{center}
$b_{i,1}$ and $b_{j,2}$ are adjacent in $\str M$, whereas $b_{i,2}$ and $b_{j,1}$ are not.
\end{center}
Thus the semi-induced bipartite graph in $\str M$ between the $b_{i,1}$'s and the $b_{i,2}$'s is an infinite ladder, a contradiction to the stability of $\str N$'s edge relation.
\end{proof}

For two sets $A$, $B$ of vertices in a given graph, we say that $A$ \emph{dominates} $B$ if for any $b \in B$ there is $a \in A$ such that $ab$ is an edge. Likewise, we say that $A$ \emph{antidominates} $B$ if it dominates $B$ in the complement graph: for each $b \in B$ there is $a \in A$ such that $ab$ is not an edge.

\begin{lemma}\label{lem:dominators_or_antidominators}
  Let $\str M$ be a graph with a stable edge relation, and $A,B \subseteq \str M$ be two sets of vertices. Then one of the following holds:
  \begin{itemize}
    \item there is a finite subset $S \subseteq A$ that dominates $B$,
    \item there is a finite subset $S \subseteq B$ that antidominates $A$.
  \end{itemize}
\end{lemma}
We note that a 
variant of the lemma holds even in dependent models~\cite[Corollary 6.13]{simon_book}, \cite[Theorem 2.4]{DBLP:conf/lics/BonnetDGKMST22},
under the additional assumption that $A$ and $B$ are finite.
We only require it for (monadically) stable models, for which it admits a simple proof given below.

\begin{proof}
We describe an iterative process that either gives one of the two desired outcomes, or constructs an infinite ladder.
Let $i \geq 1$, assume that $a_1,\dots, a_{\ell-1}$ as well as $b_{1}, \dots, b_{\ell-1}$ are already constructed so that for $i,j<\ell$, $a_ib_j$ is an edge if and only if $i \geq j$.
Then
\begin{itemize}
\item if $\{a_1, \dots, a_{\ell-1}\}$ dominates $B$, then we are done; otherwise there is $b \in B - \{b_1, \dots, b_{\ell-1}\}$ with no adjacencies to $\{a_1,\dots,a_{\ell-1}\}$.
Pick such a $b$ and call it $b_\ell$; then
\item if $\{b_1,\dots, b_{\ell}\}$ antidominates $A$, then we are done; otherwise there is $a \in A - \{a_1, \dots, a_{\ell-1}\}$ adjacent to each $b_1, \dots, b_{\ell}$.
Pick such an $a$ and call it $a_\ell$.
\end{itemize}
This extends our ladder, as required.
\end{proof}

An {\em{infinite matching}} is the bipartite graph on vertices $\{a_i,b_i\colon i\in \N\}$ such that $a_ib_j$ is an edge if and only if $i=j$. An {\em{infinite co-matching}} is defined in the same way, except there is an edge $a_ib_j$ if and only if $i\neq j$. (Recall that an infinite ladder is again defined similarly, but with the condition $i<j$.)
The following result is folklore.
An equivalent, finitary formulation can be found e.g. in \cite[Corollary~2.4]{DBLP:journals/jct/DingOOV96}.

\begin{theorem}\label{thm:folklore}
  Let $E\subset A\times B$ be an infinite bipartite graph.
  Then one of the following cases holds:
  \begin{enumerate}
    \item $\Types[E](A/B)$ is finite,
    \item $E$ contains an infinite induced matching,
    \item $E$ contains an infinite induced co-matching,
    \item $E$ contains an infinite induced ladder.
  \end{enumerate}
\end{theorem}

\begin{lemma}
  Let $\CC$ be a graph class with a stable edge relation. Then every graph $\str M$ in the elementary closure $\overline{\CC}$ of $\CC$ has a stable edge relation.
  \label{lem:stab}  
\end{lemma}
\begin{proof}
  For $k\in\N$, let $\phi_k$ 
be the sentence that holds in a graph $G$ if and only if there 
are vertices $a_1,\ldots,a_k$ and $b_1,\ldots,b_k$ in $G$ such that for all $1\le i,j\le k$,
$G\models E(a_i,b_j)$ if and only if $i\le j$.
Since $\CC$ has a stable edge relation, there is a number $k$ such that $G\models \neg \phi_k$ for all $G\in\CC$.
Hence, $\str M\models \neg \phi_k$,
proving that $\str M$ has a  stable edge relation.
\end{proof}

    The following lemmas will be useful in 
simplifying the inductive proof of Theorem~\ref{thm:model-separator}.

\begin{lemma}
\label{lem:flip-substructure}
    Let $\str M$ and $\str N$ be graphs such that $\str N$ is an elementary extension of $\str M$. Further, let $S \subseteq \str M$ be any finite set and $\str N'$ be any $S$-flip of $\str N$. Then $\str N'$ is an elementary extension of the subgraph of $\str N' $ induced by the domain of $\str M$. 
\end{lemma}

\begin{proof}
    Let $\str{M}'$ denote the subgraph of $\str N' $ induced by the domain of $\str M$. Observe first that there is a quantifier-free formula $\eta(x,y)$ with parameters from $S$ such that for all $u,v\in \str N$,
    \[\str N\models \eta(u,v)\iff \str N'\models E(u,v).\]

    Let $\phi(\tup x)$ be a formula and $\tup m\in \str M^{\tup x}$ a tuple. We need to show that 
    \begin{align}\label{eq:flip-substr}
        \str M'\models\phi(\tup m)\iff \str N'\models \phi(\tup m).  
    \end{align}

    Rewrite $\phi(\tup x)$ to a formula $\phi'(\tup{x})$ with parameters from $S$, by replacing each occurrence of an atom $E(x,y)$ with $\eta(x,y)$. Then we have that
    \[\str N'\models \phi(\tup m)\iff \str N\models \phi'(\tup m),\]
    and similarly,
    \[\str M'\models \phi(\tup m)\iff \str M\models \phi'(\tup m).\]
    Since $\str N$ is an elementary extension of $\str M$ and $S\subset \str M$, we have that
    \[\str M\models \phi'(\tup m)\iff \str N\models \phi'(\tup m).\]
    Putting together the equivalences yields~\eqref{eq:flip-substr}.
\end{proof}

%% file: patterns.tex
\section{Pattern-free classes}\label{sec:pattern-free}

If $\phi(x,y)$ is a first-order formula and $G$ is a structure, by $\phi(G)$ we denote the graph with vertices $V(G)$ and edges $uv$, for distinct $u,v\in V(G)$ such that $G\models \phi(u,v)\lor \phi(v,u)$. If $\CC$ is a class of structures, then denote $\phi(\CC):=\setof{\phi(G)}{G\in\CC}$.
Say that a class $\CC$ of graphs \emph{transduces} a class $\DD$ of graphs 
if there is a unary expansion $\wh\CC$ of $\CC$ and  a formula $\phi(x,y)$ in the signature of $\wh\CC$,
such that for every $H\in \DD$ there is some $\wh G\in\wh\CC$ such that $H$ is an induced subgraph of $\phi(\wh G)$.
If above, the formula $\phi$ is existential, resp. quantifier-free, then we say that $\CC$ \emph{existentially transduces} $\DD$, resp. \emph{quantifier-free transduces} $\DD$.

The \emph{$r$-subdivision} of a graph $G$, denoted $G^{(r)}$ below,
is obtained by replacing each edge of $G$ by a path of length $r+1$.
A \emph{$(1,r)$-subdivision} of a graph $G$ 
is a graph obtained from $G$ by replacing 
each edge of $G$ by a path of length at least $2$, and at most $r+1$.

\begin{definition}\label{def:pattern-free}
    Say that a graph class $\CC$ is \emph{pattern-free}
    if for every $r\ge 1$, unary expansion $\wh\CC$ of $\CC$, and quantifier-free formula $\phi(x,y)$ in the signature of $\wh\CC$,
    there is some $n\ge 1$ 
    such that 
    $\phi(\wh\CC)$ avoids the $r$-subdivision of $K_n$ as an induced subgraph.
    Say that a graph $\str M$ is \emph{pattern-free}
    if the class $\set{\str M}$ is pattern-free.
\end{definition}
Note that $\CC$ is not pattern-free if and only if 
$\CC$ quantifier-free transduces the class of $r$-subdivisions of all cliques, for some fixed $r\ge 1$.

In this section we study pattern-free classes.
In \cref{prop:pattern-free} 
we prove that every simply existentially monadically dependent class is pattern-free, proving implication \eqref{it:ems}$\rightarrow$\eqref{it:pat} in Theorem~\ref{thm:main}.
In Lemma~\ref{lem:patt-lim} we prove that if $\CC$ is pattern-free then every $\str M\in\overline{\CC}$ is pattern-free.
In Lemma~\ref{lem:flip-MS} we prove that if $\str M$ is a pattern-free graph and $\str M'$ is obtained from $\str M$ by applying a finite set of flips, then $\str M'$ is also pattern-free.
In Lemma~\ref{lem:bad_construction} we describe certain obstructions that are forbidden in pattern-free graphs $\str M$.

\begin{proposition}\label{prop:pattern-free}
    If a graph class $\CC$ is simply existentially monadically dependent, then $\CC$ is pattern-free.
\end{proposition}

Proposition~\ref{prop:pattern-free} immediately follows from the next three lemmas.
\begin{lemma}\label{lem:lll}
    If $\CC$ is not pattern-free, then for some $r\ge 1$, the class $\CC$ quantifier-free transduces the class $\setof{G^{(r)}}{G\textit{ is a graph}}$ of $r$-subdivisions of all graphs.
\end{lemma}
\begin{lemma}\label{lem:subdiv-emd}
    Fix $r\ge 0$.
    Let $\CC$ be a hereditary graph class
    that quantifier-free transduces the class of $r$-subdivisions of all graphs. Then $\CC$ existentially transduces the class of all graphs.
\end{lemma}
\begin{lemma}\label{lem:mmm}
    Let $\CC$ be a graph class that existentially transduces the class of all graphs. 
    Then $\CC$ is not simply existentially monadically dependent.
\end{lemma}

\begin{proof}[Proof of Lemma~\ref{lem:lll}]
    Suppose $\CC$ is not pattern-free.
    Then there is a unary expansion $\wh\CC$
    and a quantifier-free formula $\phi(x,y)$,
    such that for every $n\ge 1$ there is some $\wh G_n\in\wh\CC$ of $G$, such that 
    $K_n^{(r)}$ is an induced subgraph of $\phi(\wh G)$. 

    We argue that $\CC$ quantifier-free transduces the class of $r$-subdivisions of all graphs.

    Let $G$ be a graph with vertices $\set{1,\ldots,n}$. Let $G_n$ and $\wh G_n$ be as above, so that 
    $K_n^{(r)}$ is an induced subgraph of $\phi(\wh G)$. Since $G^{(r)}$ is an induced subgraph of 
    $K_n^{(r)}$, it is also an induced subgraph of $\phi(\wh G)$. Therefore, $\CC$ quantifier-free transduces the class of $r$-subdivisions of all graphs.
\end{proof}

\begin{proof}[Proof of Lemma~\ref{lem:subdiv-emd}]
    Assume that $\CC$ quantifier-free transduces the class of $r$-subdivisions of all graphs.
    Then there is a unary expansion $\wh\CC$ 
    of $\CC$ and a quantifier-free formula 
    $\phi(x,y)$ 
    such that for every graph $G$
    for some $\wh H_G\in \CC$ such that $G^{(r)}$ 
    is an induced subgraph of $\phi(\wh H_G)$.
    Without loss of generality, the class $\wh\CC$ is hereditary, since $\CC$ is.

We may assume that $V(\wh H_G)=V(G^{(r)})$. This is because $V(G^{(r)})\subset V(\wh H_G)$, and if  $\wh H_G'$ denotes the substructure of $\wh H_G$ induced by $V(G^{(r)})$, then $\phi(\wh H_G')=\phi(\wh H_G)[V(G^{(r)})]=G^{(r)}$,
where the first equality holds since $\phi$ is quantifier-free. Moreover, $\wh H_G'\in\wh\CC$  since $\wh\CC$ is hereditary.

We therefore assume that $V(\wh H_G)=V(G^{(r)})$,
for all graphs $G$.
We may also assume that the formula $\phi(x,y)$ is symmetric and irreflexive, that is, 
$\forall x,y.\phi(x,y)\rightarrow (x\neq y)\land \phi(y,x)$ holds in every $k$-colored graph.
(Otherwise replace $\phi(x,y)$ 
by the formula $(\phi(x,y)\lor \phi(y,x))\land x\neq y$.)

We show that $\CC$ existentially transduces the class of all graphs.

 Fix any finite graph $G$.
Then for $u,v\in G^{(r)}$
we have that  $\wh H_G\models \phi(u,v)$ 
if and only if $G^{(r)}\models E(u,v)$.

Let $W\subset V(G^{(r)})$ denote the set of vertices of $G^{(r)}$ that correspond to 
the original vertices of $G$.
In particular,
if $\eta(x,y)$ denotes 
the formula 
\[\eta(x,y):= \exists x_1\ldots x_{r}.
E(x,x_1)\land E(x_1,x_2)\land\cdots\land E(x_{r-1},x_r)\land E(x_r,y),\]
then $\eta(G^{(r)})[W]$ is isomorphic to $G$.
Now consider the formula 
\[\psi(x,y):= \exists x_1\ldots x_{r}.
\phi(x,x_1)\land \phi(x_1,x_2)\land\cdots\land \phi(x_{r-1},x_r)\land \phi(x_r,y).\]
It follows that for $u,v\in W$ we have that 
$\wh H_G\models \psi(u,v)$ if and only if 
$G^{(r)}\models \eta(u,v)$.
In particular, 
$\psi(\wh H_G)[W]$ is isomorphic to $G$.

Therefore, the existential formula $\psi(x,y)$ (which does not depend on $G$), witnesses that $\CC$ existentially transduces the class of all graphs.
\end{proof}

\begin{proof}[Proof of Lemma~\ref{lem:mmm}]
Suppose that $\CC$ existentially transduces the class of all graphs.     
Then 
there is an existential formula $\psi(x,y)$ such that every finite graph $G$ 
is an induced subgraph of $\psi(\wh H)$, for some $\wh H\in\wh\CC$. For $n\ge 1$, let $G_n$ be the bipartite graph with 
parts $\set{1,\ldots,n}$ and $2^{\set{1,\ldots,n}}$, and edges $iJ$ such that $i\in J$.
Let $\wh H_n\in\wh\CC$ be such 
that $G_n$ is an induced subgraph of $\psi(\wh H_n)$. This means that there are vertices $v_1,\ldots,v_n$ and $w_J$, for $J\subset \set{1,\ldots,n}$,
such that $\wh H_n\models \psi(v_i,w_J)$ if and only if $i\in J$, for $i\in\set{1,\ldots,n}$ and $J\subset \set{1,\ldots,n}$.
Since $n$ is arbitrary, this shows that $\CC$ is not simply existentially monadically dependent. 
\end{proof}

\begin{proof}[Proof of Proposition~\ref{prop:pattern-free}]
    We may assume without loss of generality that $\CC$ is hereditary, since if a class $\CC$ is simply existentially monadically dependent, then so 
    is the class of all induced subgraphs of $\CC$.
This is because that taking an induced subgraph may be simulated by using a unary predicate.
Proposition~\ref{prop:pattern-free} now follows from Lemmas~\ref{lem:lll}, \ref{lem:subdiv-emd}, and \ref{lem:mmm}.
\end{proof}

We now prove that every model in the elementary closure of a pattern-free class, is pattern-free as well. 

\begin{lemma}\label{lem:patt-lim}
    Let $\CC$ be a hereditary pattern-free graph class and let $\str M\in \overline\CC$. Then $\str M$ is pattern-free.
  \end{lemma}
  \begin{proof}
    Let $\str M\in\overline\CC$,
and let $\wh\Sigma$ be a unary expansion of the signature of graphs.
    Fix an integer $r\geq 1$ and a quantifier-free formula $\phi(x,y)$ in the signature $\wh\Sigma$. 
    Let $\wh\CC$ be the class of all monadic lifts of graphs in $\CC$ in the signature $\wh\Sigma$.
    By assumption, there is a number $n$ such that 
    $K_n^{(r)}$ is not an induced subgraph of $\phi(\wh{\str A})$, for all $\wh{\str A}\in\wh\CC$.
    
    Let $\wh{\str M}$ be a monadic lift of $\str M$ in the signature of $\phi$.
    We show that $\phi(\wh{\str M})$ does not contain $K_n^{(r)}$ as an induced subgraph.
    Suppose otherwise.

    Consider the substructure $\wh{\str A}$
    of $\wh{\str M}$ induced by the elements of $V(K_n^{(r)})\subset V(\wh{\str M})$.
Then $\wh{\str A}$ is a finite $\wh\Sigma$-structure.
Let $\str A$ be the graph such that $\wh{\str A}$ is a monadic lift of $\str A$.

We claim that $\str A$ is an induced subgraph of some graph $\str B\in\CC$.
Write a sentence $\phi_{\str A}$
that holds in a graph $\str B$ if and only if 
$\str A$ is isomorphic to an induced subgraph of $\str B$. Since $\str M\models \phi_{\str A}$ and $\str M\in\overline{\CC}$, there is some $\str B\in \CC$ such that $\str B\models \phi_{\str A}$, proving the claim.

Since $\CC$ is hereditary, we conclude that $\str A\in\CC$ as well, and hence $\wh{\str A}\in\wh\CC$.
In particular, $\phi(\wh{\str A})$ is isomorphic to $K_n^{(r)}$, a contradiction.
  \end{proof}

It is convenient to use a relaxed form of patterns,
where instead of $r$-subdivisions, we have $(1,r)$-subdivisions.

\begin{lemma}\label{lem:pat1}
    Fix $r\ge 1$, and let $\CC$ be a graph class that quantifier-free transduces 
    a class $\DD$ that contains 
    some $(1,r)$-subdivision of every clique. Then $\CC$ is not pattern-free.
\end{lemma}
 Lemma~\ref{lem:pat1} follows easily by Ramsey's theorem which we recall below.
\begin{theorem}[Ramsey's theorem]
    Fix $k\ge 1$.
    There is a monotone, unbounded function $f\from\N\to\N$, such that 
    for every $n\ge 1$,
    if the edges of $K_{n}$ are colored using $k$ colors, then there is a set $W\subset V(K_{n})$ with $|W|\ge f(n)$ such that $W$ is \emph{monochromatic}, that is, all edges $uv$ of $K_{n}$ with $u,v\in W$, have the same color.
\end{theorem}

Lemma~\ref{lem:pat1} follows immediately from the next lemma.
\begin{lemma}\label{lem:ramsey}
    Fix $r\ge 1$.
    Let $\DD$ be a hereditary class that contains some $(1,r)$-subdivision of every clique.
    Then there is some $s\in [1,r]$ such that 
    $\DD$ contains the $s$-subdivision of every clique.
\end{lemma}
\begin{proof}
We want to assign to each $n \ge 1$ a subdivision number $c_n\in [1,r]$ as follows.
Let $G_n$ be some $(1,r)$-subdivision of $K_n$ that belongs to $\DD$.
For each edge $e$ of $K_n$,
let $\ell(e)\in[1,r]$ 
be the number of times it was subdivided in $G_n$.
Then $\ell$ is a coloring of the edges of $K_n$ using $r$ colors.
By Ramsey's theorem, 
there is a subset $W_n\subset V(K_n)$ 
which is monochromatic, 
that is,  there is a color $c_n\in[1,r]$ such that for every pair $u,v$ of distinct vertices of $W_n$, we have  $\ell(uv)=c_n$. 
Moreover, the size of $W_n$ is at least $f(n)$,
where $f\from \N\to\N$ is some fixed unbounded, monotone function (depending only on $r$).

By the pigeonhole principle, in the infinite sequence $(c_1,c_2,\ldots)$, some element $s\in[1,r]$ 
occurs infinitely many times.
We claim that $\DD$ contains the $s$-subdivision of every clique $K_m$.
Indeed, pick any $m\ge 1$,
and let $n\in\N$ be such that $f(n)\ge m$
and $c_n=s$. Then $W_n\subset V(K_n)$ 
is such that $|W_n|\ge m$ and 
$\ell(u,v)=s$ for all distinct $u,v\in W_n$. 
Recall that $G_n\in \DD$ is a $(1,r)$-subdivision of $K_n$, and for all distinct $u,v\in W_n$ 
we have that edges $uv$ of $K_n$ are subdivided $s$ many times in $G_n$.
As $|W_n|\ge m$, it follows that $G_n$ contains, as an induced subgraph, the $s$-subdivision of $K_m$.
\end{proof}

\begin{figure}[h]
        \centering
        \includegraphics[width=0.75 \linewidth]{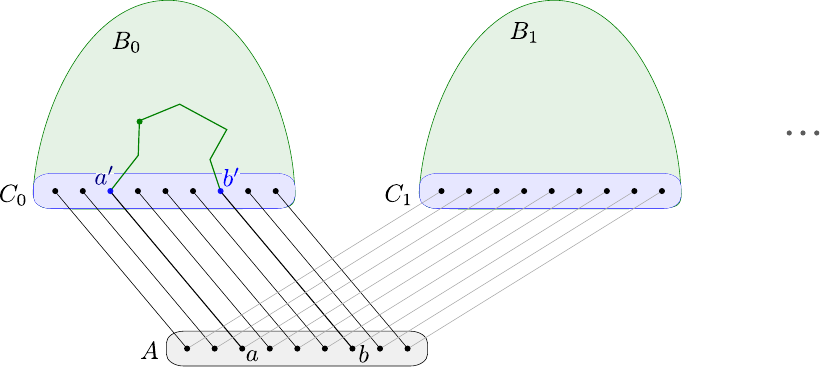}
        \caption{The obstruction pattern, and the setting of the proof of Lemma~\ref{lem:bad_construction}. We assume that $B_{ab}=B_0$.}\label{fig:obstruction}
    \end{figure}

The following lemma describes certain obstructions 
(depicted in Fig.~\ref{fig:obstruction}) that are forbidden in pattern-free graphs. 

\begin{lemma}\label{lem:bad_construction}
    Fix $r \geq 2$.
    Let $\str M$ be a graph such that for any $k \in \N$, there is an infinite collection of pairwise disjoint finite sets $A,B_0,B_1,B_2,\dots$ of vertices of $\str M$ with the following properties:
    \begin{enumerate}[(i)]
    \item \label{item:card} the set $A$ has cardinality $k$;
    \item \label{item:matching} there is a semi-induced matching between $A$ and a subset $C_i$ of $B_i$ for all $i$;
    \item \label{item:no_edges_to_base} there are no edges between $A$ and $B_i - C_i$ for all $i$; 
    \item \label{item:no_edges_between_inner_gadgets} there are no edges between $B_i - C_i$ and $B_j$ for $i \neq j$; and
    \item \label{item:stupid} for all $i$ and every pair of disjoint vertices $u,v\in C_i$, there is a path of length $\geq 2$ and $\leq r$ that connects $u$ and $v$ and whose all internal vertices belong to $B_i - C_i$
    \end{enumerate}
    Then $\str M$ is not pattern-free.
    \end{lemma}

    \begin{proof}
    We construct a quantifier-free formula $\phi(x,y)$ using three additional unary predicates, denoted $A,B,C$,
    and for each $k$ a lift $\lift M_k$ of $\str M$,
    such that $\phi(\lift M_k)$ 
    contains an $(1,r+1)$-subdivision of $K_k$,
    as an induced subgraph.
    Then Lemma~\ref{lem:pat1} implies that $\set{\str M}$ is not pattern-free.

    Fix $k \geq 1$ and consider sets $A,B_0,B_1,\dots$ given by the assumption of the lemma,
    so that in particular $|A|=k$.
    Pick a subfamily of the sets $B_0,B_1,\ldots$  containing exactly $|A|\choose 2$ sets,
    and reindex those sets as $B_{ab}$, for $ab\in \binom{A}{2}$. Likewise, denote the distinguished subsets $C_{ab}\subset B_{ab}$,
    so that each $C_{ab}$ semi-induces a matching with $A$. 
    
    We now describe the construction of $\lift M_k$.
    Mark the vertices in $A$ with the unary predicate $A$, the vertices in $B_{ab}$, for some $ab\in \binom{A}2$, with the unary predicate $B$,
    and the vertices in $C_{ab}\subset B_{ab}$,
    for some $ab\in \binom{A}2$, with the unary predicate $C$.
    Define a formula $\phi_0(x,y)$ 
    as follows:
    \[\phi_0(x,y)\quad\equiv \quad 
    (A(x)\land C(y))\quad \lor \quad 
    (A(y)\land C(x))\quad \lor \quad
    (B(x)\land B(y)\land \neg (C(x)\land C(y))).\]
    Let $\phi(x,y)\equiv \phi_0(x,y)\land E(x,y)$.
    We argue that 
    $\phi(\lift M_k)$ 
    contains an $(1,r+1)$-subdivision of $K_k$,
    as an induced subgraph.

    For each $ab\in {A\choose 2}$, let $a',b'$ be the unique elements in $C_{ab}$ connected to $a$ and $b$, respectively.
    Consider a path $\pi_{ab}$ between $a'$ and $b'$, of length $\geq 2$ and $\leq r$, that is internally contained in $B_{ab} - C_{ab}$.
    Let $W$ denote the vertices in $A\cup\setof{V(\pi_{ab})}{ab\in {A\choose2}}$.
    Now, it is easy to check that   $\phi(\lift M_k)[W]$ 
    is a $(1,r+1)$-subdivision of $K_k$
    (see Figure~\ref{fig:obstruction}).
    \end{proof}

Finally, we state two lemmas that will be useful later.

\begin{lemma}
\label{lem:flip-MS}
    For any pattern-free graph $\str M$, any graph $\str{M}'$ obtained from $\str M$ by applying a finite set of flips, is also pattern-free.
\end{lemma}
\begin{proof}
  It suffices to consider the case when $\str M'$ is obtained from $\str M$ by applying an atomic flip, specified by a pair $(A,B)$ 
of subsets of $\str M$. Add the sets $A$ and $B$ as unary predicates $U_A$ and $U_B$ to the graph $\str M$, obtaining a unary expansion $\wh{\str M}$ of $\str M$.
Then the quantifier-free formula $\eta(x,y):=
E(x,y)\triangle (U_A(x)\land U_B(y))$ is 
such that for any so that for every $a,b\in \str M$,
    \[\str M'\models E(a,b)\iff \wh{\str M}\models \eta(a,b). \]
    
    Suppose $\str M'$ is not pattern-free.
    Then there is $r\ge 1$, a class 
    $\wh\CC$ of unary expansions of $\set{\str M'}$,
    and a quantifier-free formula $\phi(x,y)$ such that $\phi(\wh\CC)$ contains the $r$-subdivision of every clique, as an induced subgraph.
Replace each atom $E(z,t)$ by the formula $\eta(z,t)$ in $\phi(x,y)$, yielding a quantifier-free formula $\phi'(x,y)$.
Then for every $\wh{\str M}'\in \wh\CC$ 
there is a unary expansion $\wh{\str M}''$ of $\wh{\str M}$, 
such that $\wh{\str M}'\models \phi(a,b)\iff \wh{\str M}''\models\phi'(a,b)$ for all $a,b\in \str M$.
It follows that there is some class~$\wh\CC'$ of unary expansions of $\set{\wh{\str M}}$ 
such that $\phi'(\wh\CC')$ contains the $r$-subdivision of every clique as an induced subgraph. Hence, by transitivity of unary expansions, $\str M$ is not pattern-free, a contradiction.
\end{proof}

A similar statement holds for graphs with a stable edge relation.

\begin{lemma}\label{lem:flip-stab}
  For any graph $\str M$ with a stable edge relation, any graph $\str{M}'$ obtained from $\str M$ by applying a finite set of flips, also has a stable edge relation.
\end{lemma}
\begin{proof}Again it is enough to assume that $\str M'$ is obtained from $\str M$ by performing a single atomic flip $\flip F=(A,B)$.
We prove the lemma by contraposition -- we argue that if $\str M'$ contains ladders of arbitrary length, then so does $\str M$.  

Let $a_1,a_2, \ldots, a_k, b_1,b_2,\ldots, a_k$ be the vertices of a ladder of length $k$ in $\str M'$ (we have $a_ib_j \in E(\str M)$ iff $i \le j$). 
Since each $a_i$ has four possibilities of being included or non-included in the sets $A,B$ and the same holds for $b_i$, there exists a subset $Z$ of $\set{1,\ldots, k}$ of size at least $k':=\frac{k}{16}$ such that for all $i,i',j,j'\in Z$ we have $(a_i, b_j) \in (A \times B) \cup (B \times A)$ if and only if $(a_{i'}, b_{j'}) \in (A \times B) \cup (B \times A)$. Without loss of generality assume that $Z = \set{1,\ldots, k'}$ (this can be achieved by forgetting all $a_i,b_i$ such that $i \not\in Z$ and renaming the indices which are left). Then there are two options -- either $(a_1,b_1) \not\in (A \times B) \cup (B \times A)$ and then $a_1,a_2, \ldots, a_{k'}, b_1,b_2,\ldots, a_{k'}$ is a ladder in $\str M' \oplus \flip F$ of length $\frac{k}{16}$, or $(a_1,b_1) \in (A \times B) \cup (B \times A)$ and then the edges between the sides of the ladder semi-induced by $a_1,a_2, \ldots, a_{k'}, b_1,b_2,\ldots, a_{k'}$ in $\str M'$ get complemented, which results in a ladder of length $\frac{k}{16}-1$ in $\str M' \oplus \flip F$ given by vertices $c_1,\ldots, c_{k'-1}, d_1,\ldots, d_{k'-1}$, where $c_i = a_{k'-i+1}$ and $d_i = b_{k'-i}$. 
Since by our assumption on $\str M'$ we can choose  $k$ to be arbitrarily large, this means that there are arbitrarily large ladders in $\str M' \oplus \flip F$, which finishes the proof.
\end{proof}

%% file: separators.tex
\section{Finite separators in pattern-free stable models}
\label{sec:separators}

Say that a graph $\str M$ 
is \emph{$r$-separable}
if for every elementary extension $\str N$ of $\str M$,
and every $v\in \str N\setminus \str M$,
there is a finite set $S\subset \str M$ such that $v$ and $\str M$ are $r$-separated over $S$ in $\str N$.
This section is dedicated to proving the following theorem.
\begin{theorem}
\label{thm:model-separator}Let 
      $\str M$ be a pattern-free graph with a stable edge relation.
    Then $\str M$ is $r$-separable, for every $r\in\N$.
\end{theorem}
By Lemmas~\ref{lem:stab} and~\ref{lem:patt-lim}, we immediately get the following corollary, proving the implication \eqref{it:pat}$\rightarrow$\eqref{it:sep} in Theorem~\ref{thm:main}.
\begin{corollary}\label{cor:separability}
    If $\CC$ is a pattern-free class of graphs with stable edge relation and $r\in\N$,
    then every $\str M\in \overline{\CC}$ is $r$-separable.
\end{corollary}

\medskip
We will prove Theorem~\ref{thm:model-separator} by induction on $r$.

\subsection{Case of finitely many types}
The following lemma 
is a generalization of the case $r=1$ 
of Theorem~\ref{thm:model-separator},
where instead of one vertex $v$ we have 
a set of vertices $U$ with a bounded number of types over $\str M$.

\begin{lemma}\label{lem:type_magic}
    For any graphs $\str M$ and $\str N$  
    with $\str M\prec \str N$ and such that the edge relation is stable in $\str N$, and for any set $U\subseteq \str N \setminus \str M$ such that $\Types[E](U/\str M)$ is finite, there exists a finite set $S \subseteq \str M$ and an $S$-flip which:
    \begin{itemize}
      \item $1$-separates $U$ from $\str M$; and
      \item does not flip the $S$-class $T \coloneqq \set{v \in \str N: \forall s \in S. \,\neg E(v, s)}$ with any other $S$-class (including itself), as long as $T \cap U$ is nonempty.
    \end{itemize}
\end{lemma}

\begin{proof}
    We will construct the sought set $S$ in a~series of steps.
    \begin{claim}
    There is a finite set $S_{U} \subseteq \str M$ such that any two vertices in $\str M$ with the same $S_U$-class also have the same $U$-class.
    \end{claim}    
        \begin{claimproof}
        For each set in $\Types[E](U/\str M)$, select one vertex $v \in U$ of that type. Then apply Theorem~\ref{thm:definability-of-types} to the formula $E(x,y)$ and the element $v \in \str{N}$: there is a positive boolean combination $\psi(x)$ of formulas of the form $E(x,m)$ for $m \in\str{M}$, and for every $u \in \str{M}$,
    \[\str N\models E(v, u) \iff \str M\models \psi(u).\]
    It follows that there is a finite set $S_v \subseteq \str M$, namely the set of parameters of $\psi(x)$, so that whether or not a vertex $u \in \str M$ is adjacent to $v$ depends only on its $S_v$-class. The finite set $S_U \coloneqq \bigcup_v S_v$ has the desired properties.
        \end{claimproof}
    
    
    \begin{claim}
    There is a finite set $S_{\str M} \subseteq \str M$ such that any two vertices in $U$ with the same $S_{\str M}$-class also have the same $\str M$-class.
    \end{claim}    
    \begin{claimproof}
      Notice that since $\Types[E](U/\str M)$ is finite, $\Types[E](\str M/ U)$ is also finite. Thus we can obtain $S_{\str M}$ by including one vertex from each set in $\Types[E](\str M/ U)$.
    \end{claimproof}


    Now, for each of the finitely many ordered pairs $(A,B)$ of distinct $(S_U \cup S_{\str M})$-classes, applying Lemma~\ref{lem:dominators_or_antidominators} on $A \cap \str M$ and $B \cap \str M$ yields a finite set $S_{A,B}$ which is either contained in $A \cap \str M$ and dominates $B \cap \str M$, or contained in $B \cap \str M$ and antidominates $A \cap \str M$.
    Note that it may be that $S_{A,B}$ and $S_{B,A}$ are different.
    We now set $S$ to be
    \[
      S \coloneqq S_U \cup S_{\str M} \cup \bigcup_{(A,B)} S_{A,B}.
    \]
    Note that $S$ is finite and contained in $\str M$, as required.

    Now we define an $S$-flip $\str{N}'$ of $\str N$.  Let $\str{N}'$ be the graph obtained from $\str N$ by flipping between $S$-classes $A$ and $B$ (possibly with $A = B$) if there exists an edge in $\str N$ which has one end in $\str M$ and the other end in $U$, and at the same time has one end in $A$ and the other end in $B$.
    
    We will now show that $\str{N}'$ satisfies the required properties.
    We begin by considering the $S$-class $T \coloneqq \set{v \in \str N: \forall s \in S.\, \neg E(v, s)}$ of vertices non-adjacent to all of $S$.
    
    \begin{claim}
    If $T \cap U$ is nonempty, then $T$ is not flipped with any other $S$-class.
    \end{claim}
    \begin{claimproof}
        Assume by contradiction that there is some $T'$ (possibly with $T' = T$) such that $T$ and $T'$ are flipped. Moreover, let $w$ be some vertex of $T \cap U$.

        First, observe that there is no edge in $\str N$ between $w$ and $\str M$.
        Indeed, as $S_{\str M}$ contains a representative of every equivalence class in $\Types[E](\str M/ U)$ and yet $w$ is not connected in $\str N$ to any vertex in $S_{\str M}$, then it is not connected to any vertex in $\str M$.
        Therefore, there is no edge in $\str N$ between $T \cap U$ and $\str M$, so in particular there is no edge in $\str N$ between $T \cap U$ and $T' \cap \str M$.
        However, a flip was made between $T$ and $T'$, so there must be an edge in $\str N$ between $T \cap \str M$ and $T' \cap U$.

        Denote by $v$ and $u$ two vertices connected by an edge in $\str N$ such that $v \in T \cap \str M$ and $u \in T' \cap U$.
        We know that the neighborhood of $u$ in $\str M$ can be defined by a positive boolean combination of the formulas of the form $E(x, s)$ for $s \in S$.
        Since $v$ satisfies none of these formulas (and yet is connected to $u$), we infer that the neighborhood of $u$ must be described by a formula that is always true.
        Therefore, the neighborhood of $u$ contains all of $\str M$.
        It follows that $U$ contains two vertices $u$ and $w$ such that in $\str N$ one of them is connected to every vertex in $\str M$ and the other is disconnected from every vertex in $\str M$.

        However, this cannot happen.
        Indeed, by \cref{lem:dominators_or_antidominators} applied to $A = B = \str M$ there is a finite $R \subseteq \str M$ that either dominates every vertex in $\str M$ or antidominates every vertex in $\str M$.
        As $\str M$ is an elementary substructure of $\str N$, then $R$ also dominates every vertex in $\str N$ or antidominates every vertex in $\str N$.
        This is a contradiction, as $R$ neither dominates $w$ nor antidominates $u$.
        Therefore, $T$ is not flipped with any other $S$-class (including itself).
    \end{claimproof}

    
    To complete the proof, we now show that $\str{N}'$ has no edge connecting a vertex in $\str M$ with a vertex in $U$.
    Suppose towards a contradiction that $\str{N}'$ has an edge $ab$ so that $a \in \str M$ and $b \in U$. Let $A$ and $B$ denote the $S$-classes of $\str N$ which contain $a$ and $b$, respectively (so maybe $A=B$). If $ab \in E(\str {N})$, then we would have flipped between $A$ and $B$, removing the edge. Thus $ab \notin E(\str{N})$, and we flipped between $A$ and $B$. So there must have been some other edge $a'b' \in E(\str{N})$ such that $a' \in A$, $b' \in B$, one of $a',b'$ is in $\str{M}$, and the other is in $U$.
    
    \begin{claim}
    \label{claim:same-class-inference}
    Let $x \in \str M$, $y \in U$.
    Assume that $a$ and $x$ have the same $S_U \cup S_{\str M}$-class, and that $b$ and $y$ have the same $S_U \cup S_{\str M}$-class.
    Then $xy \notin E(\str N)$.
    \end{claim}
    \begin{claimproof}
      Since, in $\str N$, the vertices $a$ and $x$ have the same $S_U \cup S_{\str M}$-class, they also have the same $S_U$-class and therefore the same $U$-class.
      As $ab \notin E(\str N)$, $a \in \str M$ and $b \in U$, we infer that $xb \notin E(\str N)$.
      Likewise, $b$ and $y$ have the same $\str M$-class.
      Therefore, $xy \notin E(\str N)$.
    \end{claimproof}

    First suppose that $a' \in \str M$ and $b' \in U$.
    Then Claim~\ref{claim:same-class-inference} applies with $(x, y) = (a', b')$, contradicting the fact that $a'b' \in E(\str N)$.
    
    Thus we may assume that $a' \in U$ and $b' \in \str{M}$.
    Let $A_0$ and $B_0$ denote the $S_U \cup S_{\str M}$-classes of $a'$ and $b'$, respectively; note that $A \subseteq A_0$ and $B \subseteq B_0$.
    If $A_0 = B_0$, then Claim~\ref{claim:same-class-inference} applies with $(x, y) = (b', a')$, again contradicting that $a'b' \in E(\str N)$.
    Thus $A_0 \neq B_0$, and $a, b, a', b'$ are four different vertices.
    


    \begin{figure}[h]
        \centering
        \includegraphics[width=0.4\linewidth]{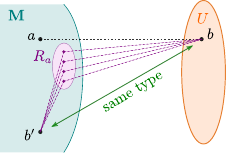}
        \caption{Illustration for the end of the proof of Lemma~\ref{lem:type_magic}.}\label{fig:type_magic}
    \end{figure}
    \begin{claim}
    \label{claim:s-not-dominating}
    $S$ does not contain a~dominating set of $B_0 \cap \str M$ entirely contained in $A_0 \cap \str M$.
    \end{claim}
        \begin{claimproof}
        Consider the intersection $R_a$ of $S$ and $A_0$; see Figure~\ref{fig:type_magic} for an illustration.
        If $R_a = \varnothing$, then the claim is trivial.
        Otherwise, let us pick an~arbitrary element $r$ of $R_a$, aiming to show that $rb'$ is a~nonedge in $\str M$.
        As $b' \in B_0 \cap \str M$, this immediately finishes the proof.
        
        Since $r$ and $a$ belong to $\str M$ and have the same $S_U$-class, they have the same $U$-class and therefore $rb$ is a non-edge in $\str N$, just as $ab$.
        Now since $b$ and $b'$ have the same $S$-class (as they both belong to $B$), they have the same connection to $r \in S$, and therefore $rb'$ is a nonedge as well.
        \end{claimproof}

    However, we also have that:
    \begin{claim}
    \label{claim:s-not-antidominating}
    $S$ does not contain an~antidominating set of $A_0 \cap \str M$ entirely contained in $B_0 \cap \str M$.
    \end{claim}
        \begin{claimproof}
        Similarly, we let $R_{b'}$ be the intersection of $S$ and $B_0$.
        A symmetric argument shows that there are all possible edges between $a$ and $R_{b'}$, and thus $S$ cannot contain an antidominating set for $A_0 \cap \str M$ which is contained in $B_0 \cap \str M$.
        \end{claimproof}
    Claims~\ref{claim:s-not-dominating} and~\ref{claim:s-not-antidominating} directly contradict the construction of $S_{A_0,B_0}$.
    This finishes the proof.
\end{proof}

\subsection{Balls have finitely many types over a model}
Observe that the base case of $r=1$ of Theorem~\ref{thm:model-separator}, which follows by Lemma~\ref{lem:type_magic},
only assumes that the edge relation in $\str M$ is stable. In the inductive argument 
we will use the assumption that $\str M$ is pattern-free.
By Lemma~\ref{lem:flip-MS}, every graph obtained from $\str M$ by performing some flips remains pattern-free.
Using Lemma~\ref{lem:bad_construction}, we prove the following lemma.
\begin{lemma}\label{lem:fin-types}
    Fix $r\in\N$.
    Let $\str M$ be pattern-free graph with a stable edge relation,
let $\str N$ be its elementary extension,
    and let $v\in \str N$ be such that the $r$-ball $B^r(v)$ around $v$ in $\str N$ is disjoint from $\str M$.
    Then $\Types[E](B^{r}(v) / \str M)$ is finite.
\end{lemma}
\begin{proof}
First note that $\str N$ is pattern-free and a stable edge relation.
This is because $\str N\in\overline{\set{\str M}}$,
and we can apply 
Lemma~\ref{lem:patt-lim} and Lemma~\ref{lem:stab}, respectively.

    Suppose, going for a contradiction, that $\Types[E](B^{r}(v) / \str M)$ is infinite. By Theorem~\ref{thm:folklore}, the bipartite graph semi-induced in $\str N$ between $B^{r}(v)$ and $\str M$ either contains an infinite induced matching, an infinite induced co-matching, or an infinite induced ladder.
  Since $\str M$ has a stable edge relation, the last case is excluded.
  Moreover, up to performing a flip (over $\varnothing$) which exchanges edges and non-edges, we may assume without loss of generality (again thanks to Lemmas~\ref{lem:flip-substructure} and~\ref{lem:flip-MS} and \ref{lem:flip-stab}) that we are in the first case: there is an infinite induced matching between $B^{r}(v)$ and $\str M$.
  We now show that the assumptions of Lemma~\ref{lem:bad_construction} are satisfied with radius $2r$: for any $k \in \N$, we will construct $A$ and the $B_i$'s as in the statement of the lemma. By Lemma~\ref{lem:bad_construction}, this implies that $\str N$ is not pattern-free, a contradiction.

  Let $A \subseteq \str M$ and $C \subseteq B^{r}(v)$ be sets of cardinality $k$ that semi-induce a matching in $\str N$.
  Now for each $c \in C$, consider a path of length $\le r$ from $v$ to $c$, let $B \subseteq B^r(v)$ be the union of these paths (as sets of vertices); note that $|B| \le kr + 1$.
  Let $\tup y$ be a tuple of variables with $|\tup y| = |B|$, and let $\tup n \in \str N^{\tup y}$ be a $\tup y$-tuple comprised of all elements of $B$ as its components.
  Note that since $B^r(v)$ avoids $\str M$, there can be no edges in $\str N$ between $\str M$ and vertices at distance $<r$ from $v$, therefore $C$ is exactly the set of vertices which are at distance $r$ from $v$ in $\str N[B]$ (and they are also at distance $r$ from $v$ in $\str N$).
  Observe that for every pair of distinct vertices $c_1$, $c_2$ in $C$, there exists a~simple path of length at least $2$ and at most $2r$ internally contained in $B - C$, constructed as the~concatenation of a nonempty suffix of the path from $v$ to $c_1$ and a nonempty suffix of the path from $v$ to $c_2$.

  We now apply Lemma~\ref{lem:kind_of_morley}, which yields a sequence of tuples $\tup b_0, \tup b_1, \dots$ in $\str M$ such that for each $i$, $\tup b_i$ has the same atomic type as $\tup n$ over $A \cup \{\tup b_0, \dots, \tup b_{i-1}\}$, and such that the atomic types of $(\tup b_i, \tup b_j)$ are the same for $i \neq j$.
  For each $i$, we let $B_i \subseteq \str N$ be the set of vertices appearing as components of $\tup b_i$.
  Note that in particular, the $\tup b_i$'s have same atomic type as $\tup n$ (in other words, the $\str N[B_i]$'s are all isomorphic copies of $\str N[B]$).
  Thus for each $i$, we let $v_i \in B_i$ be the vertex corresponding to $v \in B$, and we let $C_i \subseteq B_i$ be the set of vertices at distance $r$ from $v_i$ in $\str N[B_i]$, which are also the vertices corresponding to $C \subseteq B$.
  Note that conditions~\eqref{item:card} and~\eqref{item:stupid} from the statement of Lemma~\ref{lem:bad_construction} hold.

  By definition, there is a semi-induced matching between $A \subseteq \str M$ and $C$ in $\str N$, and moreover there are no edges between $B - C$ and $\str M$.
  Now since each $\tup b_i$ has the same atomic type as $\tup n$ over $A$, there is also a semi-induced matching between $A$ and $C_i$ for all $i$ and there are no edges between $B_i - C_i$ and $A$: conditions \eqref{item:matching} and \eqref{item:no_edges_to_base} are satisfied.

  Recall from Lemma~\ref{lem:kind_of_morley} that for $i \neq j$, the atomic type of $(\tup b_i,\tup b_j)$ is the same as the atomic type of $(\tup b_1, \tup b_0)$.
  Moreover, $\tup b_1$ and $\tup n$ have the same atomic type over $A \cup \tup b_0$, so the atomic type of $(\tup b_1, \tup b_0)$ is the same as the atomic type of $(\tup n, \tup b_0)$.
  Now as noticed above, there are no edges in $\str N$ between $\str M$ and $B - C$.
  As also $\tup b_0$ is contained in $\str M$, there are no edges between the vertices of $\tup n$ (outside of $C$) and the vertices of $\tup b_0$.
  By the equality of atomic types of $(\tup b_i, \tup b_j)$ and $(\tup n, \tup b_0)$, it follows that there are no edges between the vertices of $\tup b_i$ (outside of $C_i$) and the vertices of $\tup b_j$.
  Hence, there are no edges in $\str N$ between $B_i - C_i$ and $B_j$. Therefore, \eqref{item:no_edges_between_inner_gadgets} holds. 
  
  Therefore Lemma~\ref{lem:bad_construction} tells us that $\str N$ is not pattern-free, a contradiction, which  proves that $\Types[E](B^{r}(v) / \str M)$ is finite.
\end{proof}


\subsection{Inductive proof}
An inductive proof of Theorem~\ref{thm:model-separator} now follows by putting together Lemma~\ref{lem:type_magic} and Lemma~\ref{lem:fin-types}.

\begin{proof}[Proof of Theorem~\ref{thm:model-separator}]
    We proceed by induction on $r$.
  The base case $r=0$ is immediate as we may take $S$ to be $\varnothing$ since $v \notin \str M$.
In the inductive step,  assume that the result is proved for the distance $r\in \N$; that is, there is a finite $S \subseteq \str M$ such that $v \ind[S] r \str M$.
  Stated differently, there is an $S$-flip $\str N'$ of $\str N$ in which the $r$-ball around $v$ is disjoint from~$\str M$.
  By working in $\str N'$ instead of $\str N$, we may assume, thanks to Lemmas~\ref{lem:flip-substructure} and~\ref{lem:flip-MS}
  and~\ref{lem:flip-stab}, that the $r$-ball $B^r(v)$ around $v$ in $\str N$ is disjoint from~$\str M$.
By Lemma~\ref{lem:fin-types}, $\Types[E](B^{r}(v) / \str M)$ is finite.
  Now Lemma~\ref{lem:type_magic} applied to $B^r(v)$
   finishes the inductive step and the proof (we are using the fact that we obtain a set $S$ and an $S$-flip which doesn't flip the $S$-class which contains $B^{r - 1}(v)$).
\end{proof}


%% file: separation_game.tex

%
\newcommand{\FlipperStrategyClassName}{Localizer-complete\xspace}

\section{Flipper wins in monadically stable classes}\label{sec:sep-game}

\medskip

The goal of this section is to prove the following theorem.

\begin{theorem}
\label{thm:SepWins}Fix $r\in \N$, and  
let $\Cc$ be a class of graphs 
such that every $G\in\overline{\CC}$ is $r$-separable.
Then there exists $k\in\N$ such that Flipper wins the confining game with qf-definable separation with radius $r$ in $k$ rounds on every $G\in\CC$.
\end{theorem}
This proves the implication \eqref{it:sep}$\rightarrow$\eqref{it:sg} in Theorem~\ref{thm:main}.
In \cref{sec:variants} we have already observed that winning strategies for the confining game with qf-defineable separation can be translated to winning strategies for the shrinking game with atomic flips (i.e., ``the'' Flipper game).
Together with \cref{cor:separability}, this immediately yields the following.
\begin{corollary}
    \label{cor:stable_to_flipper}
    Let $\CC$ be a monadically stable class of graphs. Then for any $r \in \N$, there exists $k \in \N$ such that Flipper wins the Flipper game with radius $r$ in $k$ rounds on every $G \in \CC$.
\end{corollary}
    
The rest of Section~\ref{sec:sep-game} is devoted to the proof of Theorem~\ref{thm:SepWins}.
Fix an~enumeration $\phi_1,\phi_2,\ldots$ of all formulas (in the signature of graphs) of the form $\phi(y,x_1,\ldots,x_\ell)$, with $\ell\ge 0$.
We define a~strategy of Flipper in any graph $G$.
In the $k$th round, after Localizer picks $c_k\in A_{k-1}$, Flipper first sets $S:=S_{k-1}\cup \set{c_k}$ and marks $c_{k}$. Then, for every $i=1,\ldots,k$, for the formula $\phi_i(y,\tup x)$, Flipper does the following.
\begin{quote}
   For each valuation $\tup a\in S^{\tup x}$ such that $G\models\exists y.\phi_i(y,\tup a)$, Flipper marks any vertex $b\in V(G)$ such that $G\models \phi_i(b,\tup a)$.
\end{quote} We say that any strategy of Flipper which has this property is \emph{\FlipperStrategyClassName}. The marked vertices form Flipper's response in the $k$th round, and we set $S_{k}$ to be the union of $S_{k-1}$ and all the marked vertices. Note that there is a function $f\from\N\to\N$ such that $|S_k|\le f(k)$ for all $k\in\N$, regardless of which vertices Localizer picks or which of the formulas $\exists y.\phi_i(y,\tup a)$ hold.

\medskip
We prove that there is a number $k\in\N$ such that 
when Flipper plays according to any \FlipperStrategyClassName strategy on a graph $G\in\CC$,
then he wins in at most $k$ rounds.
Assume that the conclusion of the theorem does not hold.
Then, there exists a~sequence of graphs $G_1, G_2, \ldots \in \CC$, where in $G_n$ Localizer has a~strategy ensuring that Flipper does not win for at least $n$ rounds.
We shall now prove that there is some graph $G$ in the elementary closure of $\Cc$ and a vertex in the graph that survives in the arena indefinitely,
when Flipper plays according to a \FlipperStrategyClassName strategy.
We will then show that this contradicts $r$-separability of $G$.

\begin{claim}
  \label{claim:exists-indefinite-vertex}
  There exists a~graph $G \in \overline{\CC}$, a~strategy of Localizer, and a~\FlipperStrategyClassName strategy of Flipper for which the confining game with qf-definable separation on $G$ lasts indefinitely and $\bigcap_{n < \omega} A_n$ is nonempty.
\end{claim}
\begin{claimproof}
For every graph $G_n \in \Cc$, choose any \FlipperStrategyClassName strategy of Flipper, and any strategy of Localizer ensuring the game continues for more than $n$ rounds.
We define a new class $\CC'$ of structures by adding constants to graphs $G_1, G_2, \ldots$ as follows.
\begin{itemize}
    \item Add constant symbols $c_1,c_2,\ldots,c_\omega$. For each graph $G_n$, interpret $c_k$ as:
      \begin{itemize}
        \item Localizer's move in the $k$th round if $k \leq n$;
        \item any vertex remaining in the arena after $n$ rounds if $k = \omega$;
        \item an~arbitrary vertex of $G_n$ otherwise.
      \end{itemize}
    \item Add constant symbols $s_{n,i}$ for $1\le n<\omega$ and $1\le i\le f(n)$. For each graph $G_n \in \CC$, interpret $s_{k,1}, \ldots, s_{k, f(k)}$ as:
    \begin{itemize}
        \item the vertices marked by Flipper in the $k$th round if $k \leq n$. We allow these symbols to be interpreted as the same vertex if Flipper plays fewer than $f(k)$ vertices, and we ensure that $s_{k,1}$ is interpreted in the same way as $c_k$;
        \item arbitrary vertices of $G_n$ otherwise.
    \end{itemize}
\end{itemize}
Now, for convenience, for each $1 \le n < \omega$, we write $S_{n-1}\coloneqq \setof{s_{k,i}}{k<n, 1\le i\le f(k)}$.

Finally, let $T$ be the theory which is obtained by including:
\begin{itemize}
    \item the sentences $\phi$ which hold in all structures $G\in\CC'$ (that is, the theory of $\CC'$);
    \item sentences which express that $c_n$ is a valid move in the $n$th round, namely, that $c_n\nind[S_{n-1}]r c_{k}$ for $1\le k<n$ (we remark that these sentences are existential);
    \item analogous sentences which express that $c_\omega$ is a valid move in each round $n$;
    \item sentences which express that Flipper plays according to a~\FlipperStrategyClassName strategy; that is, $s_{n,1}=c_n$, and for each $1 \le i \le n$, for the formula $\phi_i(y, \tup{x})$ and for each valuation $\tup{c} \in (S_{n-1} \cup \{s_{n,1}\})^{\bar{x}}$, the sentence $\exists y. \phi_i(y, \tup{c}) \implies \phi_i(s_{n,2}, \tup{c}) \lor \ldots \lor \phi_i(s_{n,f(n)}, \tup{c})$.
\end{itemize}

We now show that $T$ is consistent.
To this end, pick a finite subset $T'$ of $T$. Then there is a~number $n_0\in\N$ such that no constants $c_{n}$ or $s_{n,i}$, with $n_0<n<\omega$, occur in a sentence in $T'$. Since Localizer avoids losing in $G_{n_0 + 1}$ for at least $n_0+1$ rounds in our fixed strategies, the structure corresponding to $G_{n_0 + 1}$ in $\CC'$ models $T'$. It follows by compactness (Theorem~\ref{thm:compactness}) that $T$ is consistent.

Since $T$ is consistent, there exists a model $G'$ of $T$,
which is a graph equipped with constant symbols $c_1,c_2,\ldots,c_\omega$ and $s_{k,i}$.
Let $G$ be the graph obtained from $G'$ by forgetting these constants.
As $T$ contains the theory of $\CC'$, which in turn contains the theory of $\CC$, we infer that $G \in \overline{\CC}$. Now consider the instance of the game where in the $n$th round, Localizer picks the vertex $c_n$ and Flipper picks the vertices $s_{n,i}$ with $1\le i\le f(n)$. This is a valid strategy for Localizer by the fact that $G' \models T$. For the same reason, the vertex $c_{\omega}$ remains in the arena after each round. Finally, in $G$, Flipper's strategy as defined above is \FlipperStrategyClassName.
\end{claimproof}

Let $G \in \overline{\CC}$ be the graph produced by Claim~\ref{claim:exists-indefinite-vertex}, along with the strategies of Localizer and Flipper.
By assumption, $G$ is $r$-separable.
Recall that $A_0\supseteq A_1\supseteq \ldots$ is the sequence of arenas in the play, $c_1,c_2,\ldots$ is the sequence of moves of Localizer, and $S_0\subseteq S_1\subseteq \ldots$ is the sequence of sets of vertices marked by Flipper. Denote $A_\omega:=\bigcap_{n<\omega} A_n$, and  $S_\omega:=\bigcup_{n<\omega} S_n$.
We will get a contradiction with the previous claim by proving the following claim:

\begin{claim}
  \label{claim:not-exists-indefinite-vertex}
  $A_\omega$ is empty.
\end{claim}
\begin{claimproof}
    Observe that for each $k \in \N$, we have $c_k \notin S_{k-1}$: as soon as Localizer plays $c_k$ in $S_{k-1}$, the arena $A_k$ shrinks to a~single vertex and Flipper wins in the following round.
    Then, $A_k$ is disjoint from $S_{k-1}$: since Localizer plays $c_k$ outside of $S_{k-1}$, each vertex of $S_{k-1}$ becomes separated from $c_k$ and thus is removed from the arena.
    It follows that $A_\omega \cap S_\omega = \emptyset$.
    
    Since Flipper follows a~\FlipperStrategyClassName strategy, $S_\omega$ induces an elementary substructure of $G$ by the Tarski-Vaught test (Theorem~\ref{thm:tarsk-vaught}). We also have that $c_1,c_2,\ldots\in S_\omega$ by construction. Now suppose for a contradiction that there exists some $c_\omega\in A_\omega$.
We remark that $c_\omega \notin S_\omega$.

    By Theorem~\ref{thm:model-separator}, there exists a finite set $S\subset S_\omega$ such that $c_\omega\ind[S]rS_\omega$. As $S$ is finite, there is some $n<\omega$ such that $S\subset S_n$, so in particular, $c_\omega\ind[S_n]rS_\omega$. On the other hand, $c_\omega\nind[S_n]rc_{n+1}$, as $c_\omega\in A_{n+1}$. This is a contradiction since $c_{n+1}\in S_\omega$.
\end{claimproof}

However, this means that there exists a~graph $G \in \overline{\CC}$ and strategies of Localizer and Flipper, for which $A_\omega$ is simultaneously nonempty (Claim~\ref{claim:exists-indefinite-vertex}) and empty (Claim~\ref{claim:not-exists-indefinite-vertex}).
This contradicts the existence of the graphs $G_1, G_2, \ldots \in \CC$ and completes the proof of Theorem~\ref{thm:SepWins}.

%% file: afg.tex
\input{afg/afg_macros.tex}
\input{afg/afg_outline.tex}
\input{afg/afg_canonic.tex}

\input{afg/afg_strategy.tex}

%% file: afg/afg_macros.tex
\newcommand{\qr}[0]{\mathrm{qr}}
\renewcommand{\AA}[0]{\mathcal{A}}
\newcommand{\BB}[0]{\mathcal{B}}
\newcommand{\VV}[0]{\mathcal{V}}
\newcommand{\NN}[0]{\mathrm{\mathbb{N}}}
\newcommand{\LL}[0]{\mathrm{\mathcal{L}}}
\newcommand{\struc}[1]{\mathfrak{#1}}

\newcommand{\FF}{\mathcal{F}}
\newcommand{\GG}{\mathcal{G}}
\newcommand{\KK}{\mathcal{K}}
\newcommand{\RR}{\mathcal{R}}
\newcommand{\II}{\mathcal{I}}
\newcommand{\JJ}{\mathcal{J}}
\newcommand{\col}{\mathrm{col}}
\renewcommand{\phi}{\varphi}

%% file: afg/afg_outline.tex
\section{Outline}

In this part we prove \cref{thm:afg_main}, recalled below for convenience.
Recall that by default ``Flipper game'' means ``shrinking Flipper game with atomic flips''; this will be the only variant considered in the part.

\afgmain*

Recall here that notation $\Oh_{\Cc,r}(\cdot)$ hides multiplicative factors that depend only on the class $\Cc$ and the radius $r$.

Let us first sketch a natural approach to use the flip-flatness 
characterization of monadic stability (see \Cref{def:wideness}) to derive a winning strategy for Flipper.
Consider the radius-$r$ Flipper game on a graph $G$ from a monadically stable class $\Cc$.
For convenience we may assume for now that we work with an extended version of the game where at each round Flipper can apply a bounded (in term of the round's index) number of flips, instead of just one (see the discussion in \cref{sec:flippers}). As making a vertex isolated requires one flip --- between the vertex in question and its neighborhood --- we can always assume that the flips applied by Flipper in round $i$ make all the $i$ vertices previously played by Localizer isolated. Hence, Localizer needs to play a new vertex in each round, thus building a growing set $X$ of her moves.


Fix some constant $m\in \N$.
According to flip-flatness, there exists some number $N\coloneqq N_{2r}(m)$ with the property that once $X$ has grown to the size $N$, we find a set of flips~$F$ --- whose size is bounded independently of $m$ --- and a set $Y$ of $m$ vertices in $X$  that are pairwise at distance greater than $2r$ in $G\oplus F$. 
It now looks reasonable that Flipper applies the flips from $F$ within his next move. Indeed, since after applying $F$ the vertices of $Y$ are at distance more than $2r$ from each other, the intuition is that $F$ robustly ``disconnects'' the graph so that the subsequent move of the Localizer will necessarily localize the game to a simpler setting. This intuition is, however, difficult to capture: flip-flatness a priori does not provide any guarantees on the disconnectedness of $G\oplus F$ other than that the vertices of $Y$ are far from each other.

The main idea for circumventing this issue is to revisit the notion of flip-flatness and strengthen it with an additional {\em{predictability}} property. Intuitively, predictability says that being given any set of $5$ vertices in $Y$ as above is sufficient to uniquely reconstruct the set of flips $F$. Formally, in  \Cref{sec:canonical_wideness} we prove the following strengthening of the results of \cite{dreier2022indiscernibles}. Here and later on, $O(G)$ denotes the set of linear orders on the vertices of $G$.

\renewcommand{\FF}{\mathsf{FF}}

\begin{theorem}[Predictable flip-flatness]\label{lem:canonic_fuqw}
	Fix radius $r\in\NN$ and a monadically stable class of graphs~$\CC$.
	Then there exist the following:
	\begin{itemize}
        \item An unbounded non-decreasing function $\wsize_r\colon \N \rightarrow \N$ and a bound $\wflips_r\in \N$.
		\item A function $\FF_r$ that maps each triple $(G\in \Cc,\preceq~\in O(G),X\subseteq V(G))$ to a pair $(Y,F)$ such that:
		\begin{itemize}
		 \item $F$ is a set of at most $\wflips_r$ flips in $G$, and
		 \item $Y$ is a set of $\wsize_r(|X|)$ vertices of $X$ that is distance-$r$ independent in $G\oplus F$.
		\end{itemize}
		\item A function $\guessflips_r$ that maps each triple $(G\in \Cc,\preceq~\in O(G),Z\subseteq V(G))$ with $|Z|=5$ to a set $F$ of flips in $G$ such that the following holds:
		\begin{itemize}
        \item For every $X\subseteq V(G)$, if $(Y,F)=\FF_r(G,\preceq,X)$ and $Z\subseteq Y$, then $F=\guessflips_r(G,\preceq,Z)$.
		\end{itemize}
	\end{itemize}
	Moreover, given $G$, $\preceq$, and $Z$, $\guessflips_r(G,\preceq, Z)$ can be computed in time $\Oh_{\Cc,r}(|V(G)|^2)$.
\end{theorem}


Let us explain the intuition behind the mappings $\FF_r$ and $\guessflips_r$ provided by \cref{lem:canonic_fuqw}. The existence of bounds $\wsize_r$ and $\wflips_r$ and of the function $\FF_r$ with the properties as above is guaranteed by the standard flip-flatness, see \cref{def:wideness} and \cref{thm:fuqw}. However, in the proof we pick the function $\FF_r$ in a very specific way, so that the flip set $F$ is defined in a somewhat minimal way with respect to a given vertex ordering $\preceq$. This enables us to predict what the flip set $F$ should be given any set of $5$ vertices from $Y$. This condition is captured by the function $\guessflips_r$.

We remark that the predictability property implies the following condition, which we call {\em{canonicity}}, and which may be easier to think about. (We assume the notation from \cref{lem:canonic_fuqw}.)
\begin{itemize}
 \item For every $G\in \Cc$, $\preceq\in O(G)$, and $X,X'\subseteq V(G)$, if we denote $(Y,F)=\FF_r(G,\preceq,X)$ and $(Y',F')=\FF_r(G,\preceq,X')$, then $|Y\cap Y'|\geq 5$ entails $F=F'$.
\end{itemize}
Indeed, to derive canonicity from predictability note that $F=\guessflips_r(G,\preceq,Z)=F'$, where $Z$ is any $5$-element subset of $Y\cap Y'$. Predictability strengthens canonicity by requiring that the mapping from $5$-element subsets to flip sets is governed by a single function $\guessflips_r$, which is moreover efficiently computable.

%

\medskip

We now outline how Flipper can use predictable flip-flatness for radius $2r$ to win the radius-$r$ Flipper game in a bounded number of rounds. Suppose the game is played on a graph~$G$; we also fix an arbitrary ordering $\preceq$ of vertices of $G$. Flipper will keep track of a growing set $X$ of vertices played by the Localizer. The game proceeds in a number of {\em{eras}}, where at the end of each era $X$ will be augmented by one vertex. In an era, Flipper will spend $2\cdot \binom{|X|}{5}$ rounds trying to robustly disconnect the current set $X$. To this end, for every $5$-element subset $Z$ of $X$ Flipper performs a pair of rounds:
\begin{itemize}
 \item In the first round, Flipper computes $F\coloneqq \guessflips_{2r}(G,\preceq,Z)$ and applies the flips from $F$. Subsequently, Localizer needs to localize the game to a ball of radius $r$ in the $F$-flip of the current graph.
 \item In the second round, Flipper reverses the flips by applying $F$ again, and Localizer again localizes.
\end{itemize}
Thus, after performing a pair of rounds as above, we end with an induced subgraph of the original graph, which moreover is contained in a ball of radius $r$ in the $F$-flip.
Having performed all the $\binom{|X|}{5}$ pairs of rounds as above, Flipper makes the last round of this era: he applies flips that isolates all vertices of $X$, thus forcing Localizer to play any vertex outside of $X$ that is still available. This adds a new vertex to $X$ and a new era begins.

Let us sketch why this strategy leads to a victory of Flipper within a bounded number of rounds. Suppose the game proceeds for $N$ eras, where $N$ is such that $\alpha_{2r}(N)\geq 7$. Then we can apply predictable flip-flatness to the set $X$ built within those eras, thus obtaining a pair $(Y,F)\coloneqq \FF_{2r}(G,\preceq,X)$ such that $|Y|=7$ and $F$ is a set of flips such that $Y$ is distance-$2r$ independent in $G\oplus F$. 
Enumerate $Y$ as $\{v_1,\ldots,v_7\}$, according to the order in which they were added to $X$ during the game. Let $Z\coloneqq \{v_1,\ldots,v_5\}$ and note that $F=\guessflips_{2r}(G,\preceq,Z)$. Observe that in the era following the addition of $v_5$ to $X$, Flipper considered $Z$ as one of the $5$-element subsets of the (current) set $X$. Consequently, within one of the pairs of rounds in this era, he applied flips from $F$ and forced Localizer to localize the game subsequently. Since $v_6$ and $v_7$ are at distance larger than $2r$ in $G\oplus F$, this necessarily resulted in removing $v_6$ or $v_7$ from the graph. This is a contradiction with the assumption that both $v_6$ and $v_7$ were played later in the game.


In \cref{sec:canonical_wideness} we prove \cref{lem:canonic_fuqw}. In \cref{sec:afgstrat} we formalize the strategy outline presented above and analyze the time complexity needed to compute the moves. This will amount to proving \cref{thm:afg_main}.

%% file: afg/afg_canonic.tex
\section{Predictable flip-flatness}\label{sec:canonical_wideness}

This section is devoted to proving \Cref{lem:canonic_fuqw}: monadic stability implies predictable flip-flatness.
We will first collect some facts from the work of Dreier et al.~\cite{dreier2022indiscernibles} and shape them to our convenience.

\input{afg/afg_ball_classifiers.tex}

\input{afg/afg_wideness_new.tex}

%% file: afg/afg_ball_classifiers.tex
\subsection{Classifiers}

The main idea behind the proof of~\cite{dreier2022indiscernibles} is to perform gradual classification of vertices while showing that this classification needs to satisfy very rigid conditions, imposed by monadic stability. To formalize this we will rely on the notion of a {\em{classifier}}, presented below.

For a vertex $u$, $N(u)$ denotes the {\em{(open) neighborhood}} of $u$, that is, the set comprising all the neighbors of $u$. The neighborhood of $u$ in a set of vertices $B$ is $N(u)\cap B$. Further, $s$ is {\em{adjacent}} to $B$ if $N(u)\cap B\neq \emptyset$. Usually, the graph, in which neighborhoods and adjacencies are evaluated, will be clear from the context. Otherwise, we specify it in the subscript.

\newcommand{\Bf}{\mathfrak{B}}
\newcommand{\exc}{\mathsf{exc}}
\newcommand{\rep}{\mathsf{rep}}

\begin{definition}\label{def:bc}
	A \emph{classifier} in a graph $G$ is a quadruple $\Bf=(\BB,S,\exc,\rep)$, where $\BB$ is a family of pairwise disjoint vertex subsets of $G$, called further {\em{blobs}}, $S$ is a non-empty subset of vertices of $G$, and $\exc\colon V(G)\to \BB\cup \{\bot\}$ and $\rep\colon V(G)\to S$ are mappings satisfying the following properties:
\begin{enumerate}[(a)]
	\item \label{obs:normal} $S\cap \bigcup \BB=\emptyset$; that is, no vertex of $S$ belongs to any blob.
	\item \label{obs:k_props_empty_full}  Every $s\in S$ is adjacent either to all the blobs in $\BB$ or to no blob in $\BB$.
	\item \label{obs:k_props_different_colors} For all distinct $s,s'\in S$ and each blob $B\in\BB$, $N(s)\cap B\neq N(s')\cap B$.
	\item \label{obs:exception} For each $v\in\bigcup\BB$, we have $\exc(v)\neq \bot$ and $v\in \exc(v)$.
	\item \label{obs:reps}  For all $v\in V(G)$ and $B\in \BB\setminus \{\exc(v)\}$, we have $N(v) \cap B = N(\rep(v))\cap B.$
\end{enumerate}
The \emph{size} of a classifier $(\BB,S,\exc,\rep)$ is $|\BB|$, and its \emph{order} is $|S|$.
\end{definition}

Let us give some intuition. In a classifier we have a family of disjoint blobs $\BB$ and a set of representative vertices $S$. Further, with every vertex $v$ we can associate its {\em{exceptional blob}} $\exc(v)\in \BB$ and its {\em{representative}} $\rep(v)\in S$. The key condition \eqref{obs:reps} says the following: every vertex $v$ behaves in the same way as its representative $\rep(v)$ with respect to all the blobs in $\BB$, except for its (single) exceptional blob $\exc(v)$. We allow $\exc(v)$ to be equal to $\bot$, which indicates that $v$ has no exceptional blob (this will be convenient in notation). Condition \eqref{obs:exception} says that if $v$ is contained in some blob $B\in \BB$, then in fact $B$ must be the exceptional blob of $v$. Conditions \eqref{obs:normal}, \eqref{obs:k_props_empty_full}, and~\eqref{obs:k_props_different_colors} are technical assertions that expresses that the representative set $S$ is reasonably chosen.

A classifier naturally partitions the vertex set of the graph, as formalized below.

\begin{definition}
For a classifier $\Bf=(\BB,S,\exc,\rep)$, the {\em{partition raised}} by $\Bf$ is the partition $\Pi_{\Bf}$ of the vertex set of $G$ defined as follows: 
$$\Pi_{\Bf}\coloneqq \{\rep^{-1}(s)\colon s\in S\}.$$
For $s\in S$, we write $\Pi_{\Bf}(s)\coloneqq \rep^{-1}(s)$ to indicate the part of $\Pi_{\Bf}$ associated with $s$.
\end{definition}

%

The following observation is easy, but will be the key to our use of classifiers.

\begin{observation}\label{obs:partition}
	Let $\Bf=(\BB,S,\exc,\rep)$ be classifier of size at least five in a graph $G$.
	Then for every pair of vertices $u,v$ of $G$, the following conditions are equivalent.
	\begin{enumerate}[(i)]
		\item\label{i:same-part} $u$ and $v$ are in the same part of $\Pi_\mathfrak{B}$.
		\item\label{i:geq-3} $u$ and $v$ have the same neighborhood in at least three blobs from $\BB$.
		\item\label{i:leq-2} $u$ and $v$ have different neighborhoods in at most two blobs from $\BB$.
	\end{enumerate}
\end{observation}
\begin{proof}
 Implication \eqref{i:same-part}$\Rightarrow$\eqref{i:leq-2} follows by observing that since $\rep(u)=\rep(v)$, $u$ and $v$ must have exactly the same neighborhood in every blob, possibly except for $\exc(u)$ and $\exc(v)$. Implication \eqref{i:leq-2}$\Rightarrow$\eqref{i:geq-3} is immediate due to $|\BB|\geq 5$. 
 
 Finally, for implication \eqref{i:geq-3}$\Rightarrow$\eqref{i:same-part}, observe that $u$ and $\rep(u)$ have the same neighborhood in all but at most one blob from $\BB$, and similarly for $v$ and $\rep(v)$. Since $u$ and $v$ have the same neighborhood in at least three blobs from $\BB$, it follows that $\rep(u)$ and $\rep(v)$ have the same neighborhood in at least one blob from $\BB$. By condition~\eqref{obs:k_props_different_colors} of \cref{def:bc}, this means that $\rep(u)=\rep(v)$, so $u$ and $v$ belong to the same part of $\Pi_{\Bf}$.
\end{proof}

From \cref{obs:partition} we can derive a canonicity property for classifiers: whenever two classifiers share at least five blobs in common, the associated partitions are the same. In the next lemma we show an even stronger property: (efficient) predictability for classifiers.

\newcommand{\Prt}{\mathsf{Partition}}

\begin{lemma}\label{lem:five}
	There exists an algorithm that given a graph $G$ and family $\BB^\circ$ consisting of five pairwise disjoint subsets of $V(G)$, computes a partition $\Pi^\circ$ of $V(G)$ with the following property: for every classifier $\Bf = (\BB,S,\exc,\rep)$ in $G$ with $\BB^\circ \subseteq \BB$, we have $\Pi^\circ = \Pi_{\Bf}$. The running time of the algorithm is $\Oh(|\Pi^\circ|\cdot n^2)$, where $n=|V(G)|$.
\end{lemma}
\begin{proof}
    We first present the construction of $\Pi^\circ$. Along the way we also construct a set of representatives $S$, and at each point vertices $s\in S$ are in one-to-one correspondence with parts $\Pi^\circ(s)$ of $\Pi^\circ$.
    We start with $\Pi^\circ=\emptyset$ and $S=\emptyset$. Then we iterate through the vertices of $G$ in any order, and when considering the next vertex $v$ we include it in the partition as follows:
    \begin{itemize}
     \item If there exists $s\in S$ such that $v$ and $s$ have the same neighborhood in at least three of the blobs of $\BB^\circ$, select such $s$ that was added the earliest to $S$ and add $v$ to $\Pi^\circ(s)$.
     \item Otherwise, if no $s$ as above exists, add $v$ to $S$ and associate with $v$ a new part $\Pi^\circ(v)=\{v\}$.
    \end{itemize}
    It is straightforward to implement the algorithm to work in time $\Oh(|\Pi^\circ|\cdot n^2)$, where $|\Pi^\circ|=|S|$ is the size of the output partition.
    
%

    We now verify that $\Pi^\circ$ constructed in this manner satisfies the requested properties.
	Let then $\Bf = (\BB,S,\exc,\rep)$ be any classifier with $\BB^\circ\subseteq \BB$; we need to argue that $\Pi^\circ = \Pi_{\Bf}$. Consider any pair $u,v$ of vertices of $G$. We need to prove that $u,v$ are in the same part of $\Pi^\circ$ if and only if they are in the same part of $\Pi_\Bf$.
	
	For the forward implication, suppose $u$ and $v$ belong to the same part of $\Pi^\circ$, say $\Pi^\circ(s)$ for some $s\in S$. By construction, $u$ and $s$ have the same neighborhood in at least three of the blobs of $\BB^\circ$. By \cref{obs:partition}, this implies that $u$ and $s$ are in the same part of $\Pi_\Bf$. Similarly, $v$ and $s$ are in the same part of $\Pi_\Bf$. By transitivity, $u$ and $v$ are in the same part of $\Pi_\Bf$.
	
	For the other direction, suppose $u\in \Pi^\circ(s)$ and $v\in \Pi^\circ(s')$ for some $s\neq s'$. By symmetry, we may assume that $s'$ was added to $S$ later than $s$. Since $v$ was included in $\Pi^\circ(s')$ instead of~$\Pi^\circ(s)$, by construction it follows that $v$ and $s$ must have different neighborhoods in at least three different blobs of $\BB^\circ$. So by \cref{obs:partition}, $v$ and $s$ belong to different parts of $\Pi_\Bf$. Since $u$ and $s$ belong to the same part of $\Pi^\circ$, by the forward implication they also belong to the same part of~$\Pi_\Bf$. Hence $u$ and $v$ belong to different parts of $\Pi_\Bf$.
\end{proof}	
%
%
%
%
%

We next observe that, at the cost of a moderate loss on the size of a classifier, we may choose the representatives quite freely.

\begin{lemma}
	\label{lem:choose_s}
	Let $\Bf=(\BB,S,\exc,\rep)$ be a classifier in a graph $G$ and let $S'$ be any set such that $\rep$ is a bijection from $S'$ to $S$.
	Then there is a classifier $\Bf'=(\BB',S',\exc',\rep')$ in $G$ such that $\BB'\subseteq \BB$ and $|\BB'|\geq |\BB|-|S|$. 

\end{lemma}	

\begin{proof}
Let $\BB'$ be obtained from $\BB$ by removing $\exc(s')$ for each $s'\in S'$. Note that by condition \eqref{obs:exception} of \cref{def:bc}, $S'$ is disjoint from $\bigcup \BB'$. Next, for each vertex $u$ set $\exc'(u)\coloneqq \exc(u)$, except for the case when $\exc(u)\in \BB\setminus \BB'$; then set $\exc'(u)\coloneqq \bot$. 
Since $\rep$ is a bijection from $S'$ to $S$, for every vertex $u$ there exists exactly one vertex $s'\in S'$ satisfying $\rep(u) = \rep(s')$, and we set $\rep'(u) \coloneqq s'$.


We claim that $\Bf'\coloneqq (\BB',S',\exc',\rep')$ is a classifier. For this, observe that for every $s'\in S'$, since $\exc(s')$ has been removed when constructing $\BB'$, we in fact have $N(\rep(s'))\cap B=N(s')\cap B$ for every $B\in \BB'$. With this observation in mind, all conditions of \cref{def:bc} for $\Bf'$ follow directly from those for $\Bf$. 
\end{proof}
%

Let $G$ be a graph and $\preceq$ be any linear order on $V(G)$. We shall say that a classifier $\Bf=(\BB,S,\exc,\rep)$ is \emph{canonical} with respect to $\preceq$ if the following condition holds: each $s\in S$ is the $\preceq$-minimum element of $\Pi_{\Bf}(s)$. We note the following.

\begin{corollary}\label{lem:canonize}
Let $G$ be a graph, $\preceq$ be a linear order on $G$, and $\Bf=(\BB,S,\exc,\rep)$ be a classifier in $G$. Then there is also a classifier $\Bf'=(\BB',S',\exc',\rep')$ such that $|S'|=|S|$, $\BB'\subseteq \BB$, $|\BB'|\geq |\BB|-|S|$, and $\Bf'$ is canonical with respect to $\preceq$. 
\end{corollary}
\begin{proof}
 It suffices to apply \cref{lem:choose_s} to $S'\coloneqq \{\min_{\preceq} \Pi_{\Bf}(s)\colon s\in S\}$. 
\end{proof}

%
%
%

The following lemma is the main outcome of this section. Here, an {\em{$r$-ball}} in a graph $G$ is a set of the form $\{w\in V(G)~|~\dist_G(v,w)\leq r\}$ for some vertex $v$ (the center of the ball).

\begin{lemma} \label{thm:disjoint_families} 
	Fix a monadically stable class of graphs $\CC$ and $r\in \N$. Then there exist a constant $\kappa\in \N$ and an unbounded non-decreasing function $\beta\colon \N\to \N$ such that the following holds. For every $G\in \CC$, linear order $\preceq$ on $V(G)$, and a non-empty family $\AA$ of pairwise disjoint $r$-balls in $G$, there exists a classifier $\Bf=(\BB,S,\exc,\rep)$ in $G$ such that:
	\begin{itemize}
	 \item $\Bf$ has size at least $\beta(|\AA|)$ and order at most $\kappa$;
	 \item $\BB\subseteq \AA$; and
	 \item $\Bf$ is canonical with respect to $\preceq$.
	\end{itemize}
\end{lemma}

To prove \cref{thm:disjoint_families} we need the following lemma, which follows from Theorem 4.2 of \cite{dreier2022indiscernibles} by setting $\phi(x,y)=\mathsf{adj}(x,y)$ and $\alpha(x,y)=\mathsf{dist}_{\leq r}(x,y)$, where $\mathsf{adj}(x,y)$ is the adjacency predicate and $\mathsf{dist}_{\leq r}(x,y)$ is the first-order formula expressing that the distance between $x$ and $y$ is at most $r$. 

\begin{lemma}[follows from Theorem 4.2 of \cite{dreier2022indiscernibles}]
	\label{lem:orig_indisc}
	Fix a monadically stable class of graphs $\Cc$ and $r\in \N$. Then there exist $k\in \N$ and a function $M\colon \N\to \N$ such that for every $m\geq 3$ and family $\Aa$ of size at least~$M(m)$ of pairwise disjoint $r$-balls in a graph $G\in \Cc$, the following holds. There exists a subfamily $\Bb\subseteq \Aa$ of size at least $m$ and a set $S \subseteq V(G)$ of at most $k$ vertices such that for every $v \in V(G)$ there exists a single exceptional ball $B_v\in \Bb$ and an element $s_v\in S$ such that for every ball $B\in \Bb\setminus\{B_v\}$ we have 
	\[
		N(v) \cap B = N(s_v)\cap B.
	\]
	Furthermore, if $v \in B$ for some $B\in \Bb$, then $B_v=B$.
\end{lemma}

In essence, \cref{lem:orig_indisc} already gives us the needed classifier, except that we need to massage it using \cref{lem:canonize} and a pigeonhole argument.

\begin{proof}[Proof of \cref{thm:disjoint_families}]
	Let $k$ and $M$ be the constant and the function provided by \cref{lem:orig_indisc} for the class $\CC$ and the radius $r$. We may assume that $M$ is non-decreasing. We set $\kappa\coloneqq k$. Further, for every $n\in \N$, set $\beta(n)$ to be the largest positive integer $m$ such that 
	$$n\geq M(k^k\cdot(k+1)\cdot (m+k)+k);$$
	if there is no such $m$, set $\beta(n)\coloneqq 0$.
	Clearly, $\beta$ defined in this way is non-decreasing and unbounded.
	
Consider any family $\AA$ of pairwise disjoint $r$-balls $\AA$ in $G$. Let $m\coloneqq \beta(|\AA|)$. If $m=0$, then 
we set $\Bf$ to be the unique canonical classifier of size $0$ and order $1$, that is
$\Bf \coloneqq (\varnothing, \{v_0\}, v \mapsto \bot, v \mapsto v_0)$ where $v_0 = \min_\preceq V(G)$.
So from now on we may assume that $m\geq 1$; thus we have $|\AA|\geq M(k^k\cdot(k+1)\cdot (m+k)+k)$. 

Apply \cref{lem:orig_indisc} to $\AA$, yielding suitable $\BB_1$, $S_1$, and mappings $v\mapsto B_v$ and $v\mapsto s_v$. Note that we have $|S_1|\leq k$ and $|\BB_1|\geq k^k\cdot(k+1)\cdot (m+k)+k$. We would like to claim that these objects constitute a classifier, but for this we have to make sure that conditions~\eqref{obs:normal},~\eqref{obs:k_props_empty_full}, and~\eqref{obs:k_props_different_colors} of \cref{def:bc} hold. This will be done using a pigeonhole argument as follows.
	
Construct $\BB_1'$ from $\BB_1$ by removing the exceptional ball $B_s$ for each $s\in S_1$; thus $|\BB_1'|\geq k^k\cdot(k+1)\cdot (m+k)$ and $S_1$ is disjoint from $\bigcup \BB_1'$.
For a ball $B\in \BB'_1$, let the {\em{profile}} of $B$ be the pair consisting of:
\begin{itemize}
 \item the following equivalence relation on $S_1$: $s,s'\in S_1$ are equivalent if they have the same neighborhood in $B$; and
 \item the unique equivalence class of the relation above whose members have empty neighborhood in $B$, or $\bot$ if there is no such equivalence class.
\end{itemize}
Note that the total number of profiles is at most $|S_1|^{|S_1|} \cdot (|S_1|+1) \leq k^k \cdot (k+1)$. Hence, there exists $\BB_2\subseteq \BB_1'$ with $|\BB_2|\geq m+k$ such that all balls from $\BB_2$ have the same profile. Say this profile is $(\equiv,C)$, where $\equiv$ is an equivalence relation on $S_1$ and $C$ is either an equivalence class of $\equiv$ or $\bot$. 

Let $S_2\subseteq S_1$ be any set consisting of one member of each equivalence class of $\equiv$. Note that for all distinct $s,s'\in S_2$ and $B\in \BB_2$, the neighborhoods of $s$ and $s'$ in $B$ are different. Furthermore, every member of $S_2$ is adjacent to all the balls from $\BB_2$, except possibly for the vertex chosen from $C$ (if existent), which is non-adjacent to all the balls in $\BB_2$.

We now define $\Bf_2\coloneqq (\BB_2,S_2,\exc_2,\rep_2)$ as follows. For each vertex $u$ of $G$, set $\exc_2(u)=B_u$, unless $B_u\notin \BB_2$, in which case set $\exc_2(u)\coloneqq \bot$. Finally, set $\rep_2(u)=\eta(s_u)$, where $\eta\colon S_1\to S_2$ maps every $s\in S_1$ to the unique $s'\in S_2$ such that $s\equiv s'$. It is now straightforward to verify that $\Bf_2$ is a classifier.

%
%
%
	It now remains to apply \cref{lem:canonize} to the classifier $\Bf_2$, yielding a classifier $\Bf=(\BB,S,\exc,\rep)$ that is canonical with respect to $\preceq$ and satisfies 
	\[|S|=|S_2|\leq k,\qquad \BB\subseteq \BB_2\subseteq \AA,\quad\textrm{and}\quad|\BB|\geq m=\beta(|\AA|).\qedhere\]
\end{proof}

%% file: afg/afg_wideness_new.tex
\subsection{Proof of the result}

We are ready to prove \Cref{lem:canonic_fuqw}. The proof follows closely the reasoning from~\cite{dreier2022indiscernibles}, except that we define the flip set somewhat more carefully in order to ensure the predictability property.
%
%
%
%
%

\begin{proof}[Proof of \cref{lem:canonic_fuqw}]
Throughout the proof we fix the monadically stable class $\CC$.
For $t\in \N$, by $\CC[t]$ we denote the class consisting of all $\leq t$-flips of graphs from $\CC$, that is,
$$\CC[t]\coloneqq \{G\oplus F~|~G\in \CC\textrm{ and }F\textrm{ is a flip set of size at most }t\textrm{ in }G\}.$$
Note that since flips can be simulated by unary predicates, monadic stability of $\CC$ entails monadic stability of $\CC[t]$, for every fixed $t\in \N$.

Since we will be using \cref{thm:disjoint_families} for different radii and different classes, in notation we follow the convention that $\kappa^\DD_s$ and $\beta^\DD_s$ denote the constant $\kappa$ and the function $\beta$ obtained from applying \cref{thm:disjoint_families} to a monadically stable class $\DD$ and radius $s\in \N$.

The proof proceeds by induction on $r$.
Recall that our goal is to define suitable functions $\FF_r$ and $\guessflips_r$, along with bounds $\wflips_r$ and $\wsize_r$, for all $r\in \N$. For this, we may use functions $\FF_{r'}$ and $\guessflips_{r'}$ and bounds $\wflips_{r'}$ and $\wsize_{r'}$ for $r'<r$, obtained from the induction assumption.

\paragraph{Case 1: Base Case.}
For $r=0$, we may simply set 
$$\FF_0(G,\preceq,X)\coloneqq (X,\emptyset)\qquad\textrm{and}\qquad \guessflips_0(G,\preceq,Z)\coloneqq \emptyset.$$ Thus, we can set the bounds 
$$\wflips_0\coloneqq 0\qquad\textrm{and}\qquad\wsize_0(n)\coloneqq n.$$

%

%
%
%
%
%

\paragraph{Case 2: Inductive Case.} We first define the function $\FF_r$. For this, let us consider any $G\in \CC$, $\preceq~\in O(G)$, and $X\subseteq V(G)$. We would like to find a pair $(Y,F)$ satisfying the requirements for the value $\FF_r(G,\preceq,X)$.

Let
$$(Y_{r-1},F_{r-1})\coloneqq \FF_{r-1}(G,\preceq,X)\qquad\textrm{and}\qquad H\coloneqq G\oplus F_{r-1}.$$
By induction, $F_{r-1}$ consists of at most $\wflips_{r-1}$ flips. Thus $H\in\DD$, where 
$$\DD\coloneqq \CC[\wflips_{r-1}].$$
Moreover, $Y_{r-1}$ has size at least $\wsize_{r-1}(|X|)$ and is $(r-1)$-independent in $H$.

For convenience, we denote
$$r'\coloneqq \lceil r/2\rceil-1.$$
Let $\AA$ be the family of $r'$-balls in $H$ whose centers are the vertices of $Y_{r-1}$. Note that the balls of $\AA$ are pairwise disjoint.

Apply \cref{thm:disjoint_families} to radius $r'$, graph $H\in \DD$, order $\preceq$, and family of $r'$-balls $\AA$, thus obtaining a classifier $\Bf=(\BB,S,\exc,\rep)$. Hence we have
$$|\BB|
\geq \beta_{r'}^\DD(|\AA|)\qquad\textrm{and}\qquad |S| \leq \kappa_{r'}^\DD.$$
By Ramsey's Theorem, we may select a subfamily $\BB^\star\subseteq \BB$ of size at least $\log |\BB|$ satisfying the following property: either the centers of the balls in $\BB^\star$ are pairwise at distance greater than $r$ in $H$, or they are pairwise at distance exactly $r$ in $H$. We set
$$\wsize_r(n)\coloneqq \left\lceil\log \beta^\DD_{r'}(\wsize_{r-1}(n))\right\rceil$$
and note that by the construction, we have
$$|\BB^\star|\geq \wsize_r(|X|).$$
We define $Y$ to be the set of centers of the balls in $\BB^\star$; thus $|Y|=|\BB^\star|\geq \wsize_r(|X|)$. 

It remains to construct a set of flips $F$ such that $Y$ is distance-$r$ independent in $G\oplus F$. We proceed by cases, each time exposing a flip set $F'$ such that we may set $F\coloneqq F_{r-1}\triangle F'$ and
$$|F'|\leq \max(4,k^2\cdot 4^k),\qquad \textrm{where } \qquad k\coloneqq \kappa_{r'}^\DD.$$ Note that thus we will have $|F|\leq \wflips_r$, assuming we set
$$\wflips_r\coloneqq \wflips_{r-1}+\max(4,k^2\cdot 4^k).$$
Also, we will have $G\oplus F=H\oplus F'$.

\paragraph{Case 2.1: The vertices of $\mathbf Y$ are pairwise at distance greater than $\mathbf r$ in $\mathbf H$.}
In this case $Y$ is already distance-$r$ independent, hence we may simply set $F'\coloneqq \emptyset$.

	\paragraph{Case 2.2: The vertices of $\mathbf Y$ are pairwise at distance exactly $\mathbf r$ in $\mathbf H$.} We may assume that $|Y|\geq 5$, for otherwise we can choose $F'$ to be a set of at most $4$ flips that isolate every vertex of $Y$ in $H$. In what follows, whenever speaking about adjacencies or distances, we mean adjacencies and distances in $H$.

	Recall that $\Bf$ is canonical with respect to $\preceq$, hence
	$$s=\min_{\preceq} \Pi_{\Bf}(s)\qquad\textrm{for every }s\in S.$$
	By definition, every vertex of $S$ is adjacent  either to all the balls in $\BB^\star$, or to none. Further, since vertices of $S$ have pairwise different neighborhoods in every ball $B\in\BB^\star$, there is at most one vertex of $S$ that is not adjacent to any ball of $\BB^\star$. Let $S'\subseteq S$ consist of those vertices of $S$ that are adjacent to every ball in $\BB^\star$; thus either $S'=S$ or $|S\setminus S'|=1$. Let
	$$W\coloneqq \bigcup_{s\in S'} \Pi_{\Bf}(s).$$
	We observe that the vertices of $W$ are the ones that keep the vertices of $Y$ at close distance, in the following sense.

	\begin{claim}
		\label{claim:dist}
		For every vertex $v \in V(G)$, the following conditions are equivalent:
		\begin{enumerate}
			\item\label{i:inW} $v$ belongs to $W$;
			\item\label{i:all} $v$ is at distance exactly $r'+1$ from all the vertices of $Y$, possibly except for one;
			\item\label{i:two} $v$ is at distance at most $r'+1$ from at least two vertices of $Y$.
		\end{enumerate}
	\end{claim}
	\begin{claimproof}
		Implication \eqref{i:all}$\rightarrow$\eqref{i:two} is trivial due to $|Y|\geq 5$.
		
		For implication \eqref{i:inW}$\rightarrow$\eqref{i:all}
		we use that $\Bf$ is a classifier.
		Let $s\in S'$ be such that $v\in \Pi_{\Bf}(s)$. By definition, $s$ is adjacent to all the balls in $\BB^\star$. Therefore, $v$ is adjacent to all the balls in $\BB^\star$, possibly except for $\exc(v)$. Recalling that $v\in \exc(v)$ in case $v\in \bigcup \BB^\star$, we conclude that $v$ is at distance exactly $r'+1$ from the centers of all the balls in $\BB^\star$, that is, vertices of $Y$, possibly except for one --- the center of $\exc(v)$.
		
		We are left with implication \eqref{i:two}$\rightarrow$\eqref{i:inW}. Let $s\in S$ be such that $v\in \Pi_{\Bf}(s)$. As $v$ is at distance at most $r'+1$ from the centers of two balls in $\BB^\star$, at least one of them, say $B$, is different from $\exc(v)$. In particular $v\notin B$, so $v$ being at distance $r'+1$ from the center of $B$ means that $v$ has to be adjacent to $B$. As $B\neq \exc(v)$, we infer that $s$ is also adjacent to $B$. It follows that $s\in S'$, implying that $v\in W$.
	\end{claimproof}

	Next, we make a case distinction depending on whether $r$ is odd or even.
	In both cases, we use the following notation. For $s\in S$ and $U\subseteq S$, we write
	$$Q_{s,U}\coloneqq \{v\in \Pi_{\Bf}(s)~|~N_H(v)\cap S=U\}.$$
	Further, we let
	$$\Qq\coloneqq \{Q_{s,U}~\colon s\in S, U\subseteq S\}.$$ 
	Note that $\Qq$ is a partition of the vertex set of $H$ into at most $k\cdot 2^k$ parts, and the definition of $\Qq$ only depends on the graph $H$, partition $\Pi_{\Bf}$, and set $S$.
	In order to later prove the predictability property, it will be crucial that, in both of the following two cases, the definition of the exposed set of flips $F'$ only depends on the partition $\Qq$ (and therefore on $H$, $\Pi_{\Bf}$, and $S$), the set $S'$, and the order $\preceq$.



	\medskip
	\paragraph{Case 2.2.1: $\mathbf r$ is odd.}
	We define $F'$ as the set of all pairs $(Q_{s_1,U_1},Q_{s_2,U_2})\in \Qq^2$ satisfying the following conditions:
	\begin{itemize}
	 \item $s_1,s_2\in S'$;
	 \item $Q_{s_1,U_1}\neq \emptyset$ and $Q_{s_2,U_2}\neq \emptyset$;
	 \item $s_1\in U_2$ or $s_2\in U_1$; and
	 \item $\min_\preceq Q_{s_1,U_1}\preceq\min_\preceq Q_{s_2,U_2}$.
	\end{itemize}
 Thus $|F'|\leq \binom{|\Qq|}{2}\leq k^2\cdot 4^k$. 
 As desired, $F'$ depends only on $\Qq$, $S'$, and $\preceq$. 
 The following claim explains the flip set $F'$ in more friendly terms.
 
 \begin{claim}\label{cl:char-flip}
     For any $u_1,u_2\in V(G)$, applying $F'$ flips the adjacency between $u_1$ and $u_2$ if and only if $u_1,u_2\in W$ and ($u_2\in N_H(\rep(u_1))$ or $u_1\in N_H(\rep(u_2))$).
 \end{claim}
 \begin{claimproof}
  Let $s_1,U_1,s_2,U_2$ be such that $u_1\in Q_{s_1,U_1}$ and $u_2\in Q_{s_2,U_2}$; in particular $Q_{s_1,U_1}\neq \emptyset$ and $Q_{s_2,U_2}\neq \emptyset$. By symmetry, we may assume that $\min_\preceq Q_{s_1,U_1}\preceq\min_\preceq Q_{s_2,U_2}$. By definition, the adjacency between $u_1$ and $u_2$ is flipped when applying $F'$ if and only if $(Q_{s_1,U_1},Q_{s_2,U_2})\in F'$, which in turn is equivalent to the conjunction of conditions $s_1,s_2\in S'$ and ($s_1\in U_2$ or $s_2\in U_1$). It now remains to note that condition $s_1,s_2\in S'$ is equivalent to $u_1,u_2\in W$, and condition ($s_1\in U_2$ or $s_2\in U_1$) is equivalent to ($u_2\in N_H(\rep(u_1))$ or $u_1\in N_H(\rep(u_2))$).
 \end{claimproof}
	
 Further, we note that the vertices of $W$ may only lie outside the balls of $\BB^\star$ or on their boundaries.
 
 \begin{claim}\label{cl:sphere}
	 If $v\in W$, then for every $y\in Y$ we have $\dist_H(v,y)\geq r'$.
 \end{claim}
 \begin{claimproof}
     Suppose $\dist_H(v,y)\leq r'-1$ for some $y\in Y$.
     As $v\in W$, by \Cref{claim:dist} there exists some other $y'\in Y$, $y'\neq y$, such that $\dist_H(v,y')=r'+1$. Hence $\dist_H(y,y')\leq 2r'=r-1$. This is a contradiction with the assumption that $Y$ is $(r-1)$-independent in $H$.
 \end{claimproof}

    We are now ready to argue the following: $Y$ is distance-$r$ independent in $H\oplus F'$. See \Cref{fig:case221} for an illustration. For contradiction, suppose in $H\oplus F'$ there exists a path $P$ of length at most~$r$ connecting some distinct $y_1,y_2\in Y$. Let $B_1,B_2\in \BB^\star$ be the $r'$-balls with centers $y_1,y_2$, respectively. Since the flips of $F'$ only affect the adjacency between the vertices of $W$, and these vertices have to be at distance at least $r'=\frac{r-1}{2}$ from $y_1,y_2$ due to \Cref{cl:sphere}, we infer the following: $P$ can be written as $$P=(y_1,\ldots,v_1,v_2,\ldots,y_2),$$ where $(y_1,\ldots,v_1)$ and $(v_2,\ldots,y_2)$ are paths of length $r'$ in $H$ that are entirely contained in $B_1$ in $B_2$, respectively. In particular, $P$ has length exactly $2r'+1=r$ and $v_1v_2$ is the only edge on $P$ that might have been flipped when applying $F'$.
    
    \begin{figure}[t]
    \centering
    \begin{tikzpicture}
     \node at (0,0) {\includegraphics[width=0.6\textwidth]{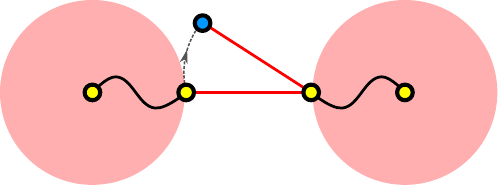}};
     \node at (-3.6,0.9) {$B_1$};
     \node at ( 3.6,0.9) {$B_2$};
     \node at (-2.75,-0.5) {$y_1$};
     \node at ( 2.75,-0.5) {$y_2$};
     \node at (-1.1, -0.5) {$v_1$};
     \node at ( 1.1, -0.5) {$v_2$};
     \node at (-0.7,  1.8) {$s_1=\rep(v_1)$};
    \end{tikzpicture}
    \caption{Case 2.2.1 in a nutshell: Up to symmetry, the (depicted in red) adjacency between $v_1$ and $v_2$ is the same as between $s_1$ and $v_2$, hence the edge $v_1v_2$ is flipped away when applying $F'$ if and only if it was present.}\label{fig:case221}
    \end{figure}
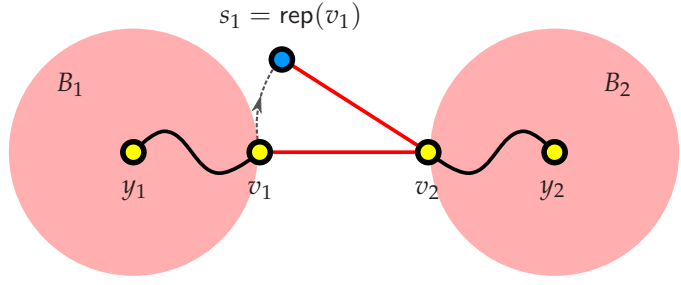

    Observe that if the edge $v_1v_2$ appeared when applying the flip $F'$, then we necessarily have $v_1,v_2\in W$. Otherwise, if $v_1v_2$ was present in $H$, then path $P$ witnesses that already in~$H$, both $v_1$ and $v_2$ are at distance at most $r'+1$ from both $y_1$ and $y_2$. By \Cref{claim:dist}, this implies that $v_1,v_2\in W$. So in any case, we have $v_1,v_2\in W$.
    
    Let $s_1\coloneqq \rep(v_1)$ and $s_2\coloneqq\rep(v_2)$. Since $v_1\in B_1$ and $v_2\in B_2$, we have $\exc(v_1)=B_1$ and $\exc(v_2)=B_2$, hence
    $$N_H(s_1)\cap B_2=N_H(v_1)\cap B_2\qquad\textrm{and}\qquad N_H(s_2)\cap B_1=N_H(v_2)\cap B_1.$$
    In particular,
    $$v_1,v_2\textrm{ are adjacent in }H\quad\Leftrightarrow\quad v_1,s_2\textrm{ are adjacent in }H\quad\Leftrightarrow\quad v_1\in N_H(s_2),$$
    and similarly 
    $$v_1,v_2\textrm{ are adjacent in }H\quad\Leftrightarrow\quad s_1,v_2\textrm{ are adjacent in }H\quad\Leftrightarrow\quad v_2\in N_H(s_1).$$
    Therefore,
    $$v_1,v_2\textrm{ are adjacent in }H\quad\Leftrightarrow\quad (v_1\in N_H(s_2)\textrm{ or }v_2\in N_H(s_1)).$$
    As $v_1,v_2\in W$, by \Cref{cl:char-flip} we conclude that $v_1$ and $v_2$ are adjacent in $H$ if and only if their adjacency gets flipped when applying $F'$. So $v_1$ and $v_2$ are non-adjacent in $H\oplus F'$, a contradiction with the existence of the edge $v_1v_2$ on $P$.

	\paragraph{Case 2.2.2: $\mathbf r$ is even.} 
	This time, $F'$ is defined as the set of all pairs $(Q_{s_1,U_1},Q_{s_2,U_2})\in \Qq^2$ satisfying the following conditions:
	\begin{itemize}
	 \item $Q_{s_1,U_1}\neq \emptyset$ and $Q_{s_2,U_2}\neq \emptyset$;
	 \item ($s_1\in S'$ and $s_1\in U_2$) or ($s_2\in S'$ and $s_2\in U_1$); and
	 \item $\min_\preceq Q_{s_1,U_1}\preceq\min_\preceq Q_{s_2,U_2}$.
	\end{itemize} 
	Again, $|F'|\leq |\Qq|^2\leq k^2\cdot 4^k$ and the definition of $F'$ depends only on $\Qq$, $S'$, and $\preceq$. Also, we may similarly explain flipping according to $F'$ as follows.

	\begin{claim}\label{cl:char-flip-2}
	For any $u_1,u_2\in V(G)$, applying $F'$ flips the adjacency between $u_1$ and $u_2$ if and only if ($u_1\in W$ and $u_2\in N_H(\rep(u_1))$) or ($u_2\in W$ and $u_1\in N_H(\rep(u_2))$).
	\end{claim}
	\begin{claimproof}
	 Analogous to the proof of \Cref{cl:char-flip}, we leave the details to the reader.
	\end{claimproof}

Note that \Cref{cl:char-flip} implies in particular that whenever the adjacency between two vertices is flipped when applying $F'$, at least one of them belongs to $W$. (However, contrary to the odd case, there might be flips in $F'$ that affect vertices outside of $W$.) In this vein, the following observation will be convenient.
	
	\begin{claim}\label{cl:W-only-outside}
	 $W\cap \bigcup \BB^\star=\emptyset$.
	\end{claim}
	\begin{claimproof}
	 For contradiction, suppose there exists $B\in \BB^\star$ and $v\in B$ such that $v\in W$. Letting $y$ be the center of $B$, we have $\dist_H(v,y)\leq r'$. By \Cref{claim:dist}, there exists another $y'\in Y$, $y'\neq y$, such that $\dist_H(v,y')\leq r'+1$. Hence $\dist_H(y,y')\leq 2r'+1=r-1$, contradicting the distance-$(r-1)$ independence of $Y$ in $H$.
	\end{claimproof}
	
	As in the odd case, we are left with arguing that $Y$ is distance-$r$ independent in $H\oplus F'$. For contradiction, suppose that there exist distinct $y_1,y_2\in Y$ and a path $P$ of length at most $r$ that connects $y_1$ and $y_2$ in $H\oplus F'$. As before, let $B_1,B_2\in \BB^\star$ be the balls with centers $y_1,y_2$, respectively.
		
    \begin{figure}[t]
    \centering
    \begin{tikzpicture}
     \node at (0,0) {\includegraphics[width=0.6\textwidth]{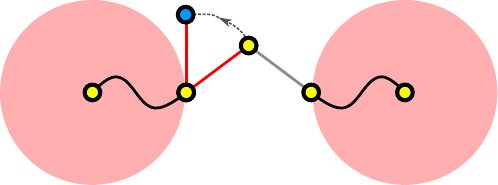}};
     \node at (-3.6,0.9) {$B_1$};
     \node at ( 3.6,0.9) {$B_2$};
     \node at (-2.75,-0.5) {$y_1$};
     \node at ( 2.75,-0.5) {$y_2$};
     \node at (-1.1, -0.5) {$v_1$};
     \node at ( 1.1, -0.5) {$v_2$};
     \node at (-1,    1.8) {$s=\rep(u)$};
     \node at ( 0.35,  1)   {$u$};
    \end{tikzpicture}
    \caption{Case 2.2.2 in a nutshell: Up to symmetry, the (depicted in red) adjacency between $v_1$ and $u$ is the same as between $v_1$ and~$s$, hence the edge $uv_1$ is flipped away when applying $F'$ if and only if it was present.}\label{fig:case221}
    \end{figure}
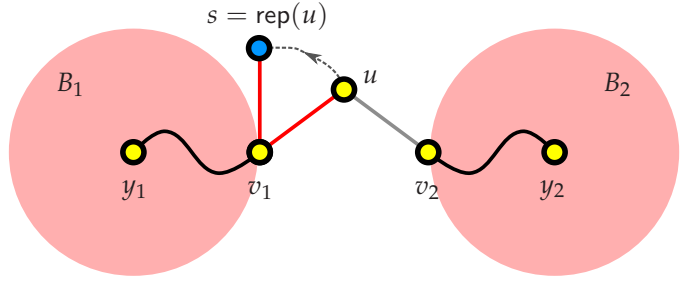

	By \Cref{cl:char-flip-2}, the flips of $F'$ affect only the vertices of $W\cup \bigcup_{s\in S'}N_H(s)$. By \Cref{cl:W-only-outside} and as $S'$ is disjoint with $\bigcup \BB^\star$, all vertices of $W\cup \bigcup_{s\in S'}N_H(s)$ are at distance (in $H$) at least $r'$ from all the vertices of $Y$. Since $r=2r'+2$, similarly as in Case~2.2.1 it follows that $P$ has length $2r'+1=r-1$ or $2r'+2=r$ and can be written as
	$$P=(y_1,\ldots,v_1,v_2,\ldots,y_2)\qquad \textrm{or}\qquad P=(y_1,\ldots,v_1,u,v_2,\ldots,y_2),$$
	where $(y_1,\ldots,v_1)$ and $(v_2,\ldots,y_2)$ are paths of length $r'$ in $H$ entirely contained in $B_1$ and $B_2$, respectively.
	
	Consider the first case: $P$ has length $r-1$ and is of the form $(y_1,\ldots,v_1,v_2,\ldots,y_2)$. Observe that edge $v_1v_2$ cannot be present in $H$, because then $P$ would be entirely contained in $H$, a contradiction with distance-$(r-1)$ independence of $Y$ in $H$. On the other hand, note that $v_1,v_2\notin W$ due to \Cref{cl:W-only-outside}, so by \Cref{cl:char-flip-2} the adjacency between $v_1$ and $v_2$ is not flipped when applying $F'$. We conclude that $v_1$ and $v_2$ remain non-adjacent in $H\oplus F'$, a contradiction with the presence of the edge $v_1v_2$ on $P$.
	
	We are left with the second case: $P$ has length $r$ and is of the form $(y_1,\ldots,v_1,u,v_2,\ldots,y_2)$.
	
	Let us first argue that $u\in W$. If $u$ is adjacent both to $v_1$ and to $v_2$ in $H$, then $u$ is at distance at most $r'+1$ from both $y_1$ and $y_2$ in $H$, hence that $u\in W$ follows directly from \Cref{claim:dist}. On the other hand, if $u$ is non-adjacent in $H$ to one of $v_1$ or $v_2$, say to $v_1$, then the adjacency between $u$ and $v_1$ must get flipped when applying $F'$. By \Cref{cl:char-flip-2} this means that at least one of $u$ and $v_1$ belongs to $W$, but it cannot be $v_1$ due to \Cref{cl:W-only-outside}. So $u\in W$ in this case as well.
	
	Let $s\coloneqq \rep(u)$. By symmetry, we may assume that $B_1\neq \exc(u)$. This means that
    $$v_1,u\textrm{ are adjacent in }H\quad\Leftrightarrow\quad v_1,s\textrm{ are adjacent in }H\quad\Leftrightarrow\quad v_1\in N_H(s).$$
    Since $u\in W$ and $v_1\notin W$ (due to \Cref{cl:W-only-outside}), by \Cref{cl:char-flip-2} we conclude that $u$ and $v_1$ are adjacent in $H$ if and only if their adjacency gets flipped when applying $F'$. So in any case, $u$ and $v_1$ are non-adjacent in $H\oplus F'$. This is a contradiction with the presence of the edge $uv_1$ on $P$.
    
    \bigskip
    
    This concludes the construction of the pair $(Y,F)$ that we set for $\FF_r(G,\preceq,X)$. We are left with defining a suitable inductive case for the function $\guessflips_r$ and showing that it can be computed efficiently, in time $\Oh_{\CC,r}(|V(G)|^2)$.
    For $G\in \Cc$ and $Z\subseteq V(G)$ with $|Z|=5$, $\guessflips_r(G,\preceq, Z)$ is defined as the flip set $F^\circ$ output by the following procedure.
    \begin{itemize}
     \item Let $F_{r-1}^\circ\coloneqq \guessflips_{r-1}(G,\preceq,Z)$, where the function $\guessflips_{r-1}$ is provided by the induction assumption. Let $H^\circ\coloneqq G\oplus F_{r-1}^\circ$.
     \item  If $Z$ is not distance-$(r-1)$ independent in $H^\circ$, output $F^\circ\coloneqq \emptyset$ and terminate. 
     \item Similarly, if $Z$ is distance-$r$ independent in $H^\circ$, output $F^\circ\coloneqq F^\circ_{r-1}$ and terminate. 
     \item Otherwise, let $\BB^\circ$ consist of the five $r'$-balls in $H^\circ$ with centers in vertices of $Z$; note that the balls of $\BB^\circ$ are pairwise disjoint. Apply the algorithm of \Cref{lem:five} to the family $\BB^\circ$, thus obtaining a partition $\Pi^\circ$.
     \item Let $S^\circ\coloneqq \{\min_{\preceq} A~\colon~A\in \Pi^\circ\}$ and let $S'^\circ$ be the subset of vertices of $S^\circ$ that are adjacent to every ball of $\BB^\circ$.
     \item Output the flip set $F^\circ\coloneqq F^\circ_{r-1}\triangle F'^\circ$, where $F'^\circ$ is defined from $H^\circ$, $\Pi^\circ$, $S^\circ$, and $S'^\circ$ exactly in the way described in Cases~2.2.1 and~2.2.2 above.
    \end{itemize}
    
    We now argue that provided $\FF_r(G,\preceq,X)=(Y,F)$ and $Z\subseteq Y$ is a set of size $5$, we have $\guessflips_r(G,\preceq,Z)=F$. Adopt the notation from the definition of $\FF_r$. We revisit the case study presented above and show that in each case, the set $F^\circ$ output by the procedure defining $\guessflips_r(G,\preceq,Z)$ coincides with the set of flips $F$ constructed in the definition of $\FF_r(G,\preceq,X)$. 
    
    By the induction assumption, we have $F_{r-1}^\circ=F_{r-1}$, which implies that $H^\circ=H$. In particular, as $Z\subseteq Y$ is distance-$(r-1)$ independent in $H$, the termination in the second point above cannot happen. 
    Also, if the vertices of $Y$ are pairwise at distance more than $r$ in $H$, then so is the case for $Z$, and we have $F^\circ=F^\circ_{r-1}$ (termination in the third point above). In the definition of $\FF_r(G,\preceq,X)$ Case 2.1 applies here, yielding $F=F_{r-1}$. So $F=F_{r-1}=F^\circ_{r-1}=F^\circ$, as required.
    
We are left with Case 2.2: the vertices of $Y$ are pairwise at distance exactly $r$ in $H$. By \Cref{lem:five} we have $\Pi^\circ=\Pi_{\Bf}$, where $\Bf=(\BB,S,\exc,\rep)$ is the classifier provided by \cref{thm:disjoint_families} that is used in the construction of $Y$. Since $\Bf$ is canonical with respect to $\preceq$, we have $$S=\{\min_{\preceq} A~\colon~A\in \Pi_{\Bf}\}=\{\min_{\preceq} A~\colon~A\in \Pi^\circ\}=S^\circ.$$
Similarly, as $\Bf$ is a classifier in $H = H^\circ$, we have that a vertex from $S = S^\circ$ is adjacent to every ball of $\BB^\star \subseteq \BB$ if and only if it is adjacent to every ball of $\BB^\circ$, and we conclude that $S' = S'^\circ$.
As now $H^\circ=H$, $\Pi^\circ=\Pi_{\Bf}$, $S^\circ=S$, and $S'^\circ = S'$, the construction presented in Cases~2.2.1 and~2.2.2 provides the same flip set for $H^\circ$, $\Pi^\circ$, $S^\circ$, and $S'^\circ$, as for $H$, $\Pi_{\Bf}$, $S$, and $S'$: we have $F^\circ=F$, as required.
    
    Finally, we need to argue that $\guessflips_r(G,\preceq,Z)$ can be computed in time $\Oh_{\CC,r}(|V(G)|^2)$. For this, observe that the procedure presented above executes $r$ inductive calls, each of which consists of internal computation that is easy to implement in time $\Oh_{\CC,r}(|V(G)|^2)$, and one call to the algorithm of \cref{thm:disjoint_families}. Since we a priori know that the partition $\Pi^\circ$ returned by this call should be of size $\Oh_{\Cc,r}(1)$, we may terminate this call once the elapsed running time exceeds $\Oh_{\Cc,r}(|V(G)|^2)$, and if so, return $\emptyset$ as $\guessflips_r(G,\preceq,Z)$. Therefore, each of the $r$ inductive calls runs in time $\Oh_{\Cc,r}(|V(G)|^2)$, giving a total time complexity of $\Oh_{\Cc,r}(|V(G)|^2)$ as well.
\end{proof}

%% file: afg/afg_strategy.tex
\section{Winning strategy}\label{sec:afgstrat}

We are almost ready to prove \Cref{thm:afg_main}. 
Before commencing to the proof, we will first clarify the notion of a {\em{strategy}} for Flipper, and what we mean by a {\em{running time}} of a strategy.

\begin{remark}
In this section, we will work with a variant of the shrinking Flipper game with atomic flips, in which we allow Localizer to localize the graph to an induced subgraph of an $r$-ball in the current arena, instead of requiring her to pick an entire $r$-ball.
Intuitively, ``losing'' additional vertices on purpose yields no benefit to Localizer.
However, the ability to work with induced subgraphs is useful for the design of algorithms, as exhibited in \cite{mcss}. 
For this reason we will explicitly prove the algorithmic winning strategy for Flipper for this variant of the game.
As the modification only expands the move pool of Localizer, the proven strategy then also works for the unmodified Flipper game and implies \Cref{thm:afg_main}.
\end{remark}

\input{afg/afg_strategy_prelims.tex}


\input{afg/afg_strategy_new.tex}

%% file: afg/afg_strategy_prelims.tex
\subsection{Strategies and runtimes}\label{sec:strategies-prelims}

\paragraph*{Strategies and runs.}
Strategies are commonly represented by functions mapping  the history of the game to a new (played) position. In our context, it will be convenient to use the following equivalent abstraction, which will fit better to our algorithmic perspective. 
Fix radius $r\in\N$. Graphs considered in consecutive rounds of the Induced-Subgraph-Flipper game will be often called {\em{arenas}}, for brevity.
A radius-$r$ Localizer strategy is a function 
$$ \stratcon\colon (G_i)\mapsto (G_i^{\text{loc}})$$
that maps the arena $G_i$ at round $i$
to a graph $G_i^{\text{loc}}$ that is an induced subgraph
of the ball of radius $r$ around some vertex $v$ in $G_{i}$.

A radius-$r$ Flipper strategy is a function
$$ \stratflip\colon (G_i^{\text{loc}},\mathcal I_i)\mapsto (\flip F,\mathcal I_{i+1})$$
that maps the graph $G^\mathrm{loc}_{i}$ obtained from Localizer's move to the atomic flip $\flip F$ chosen by Flipper; the next arena will be $G_{i+1}\coloneqq G_i^{\text{loc}}\oplus \flip F$.
Additionally, we allow Flipper to keep an auxiliary memory: the strategy takes, as the second argument, an {\em{internal state}} $\II_i$ from the previous round, and outputs an updated internal state $\II_{i+1}$. The initial state $\II_0 = \II_0(\stratflip,G)$ will be computed from the initial graph at the beginning of the game. 
The internal states will be used as memory and to precompute flips for future turns, which makes them convenient from an algorithmic point of view. 
Strategies operating with game histories instead of internal states can simulate the latter in the following sense: knowing the game history, Flipper can compute the current internal state by replaying the entire game up to the current round. Note that since we are interested in Flipper's strategies that work against {\em{any}} behavior of Localizer, it is not necessary to equip Localizer's strategies with memory as well.

Given radius-$r$ Localizer and Flipper strategies {\rm con} and {\rm flip}, and a graph $G$, we define the \emph{run} $\RR(\stratcon,\stratflip,G)$ to be the infinite sequence of \emph{positions}
$$\RR(\stratcon,\stratflip,G) \coloneqq (G_0,\II_0),(G_1,\II_1),(G_2,\II_2),(G_3,\II_3),\ldots$$
such that $G_0 = G$, $\II_0 = \II_0(\stratflip,G)$, and for all $i \geq 0$ we have
$$
G_{i+1} = \stratcon(G_{i}) \oplus \flip F,
    \qquad \text{where}\qquad
    (\II_{i+1},\flip F) = \stratflip(\stratcon(G_{i}),\II_{i}).
$$
A \emph{winning position} is a tuple $(G_{i},\II_i)$ where $G_i$ contains only a single vertex.
A radius-$r$ Flipper strategy $\stratflip$ is \emph{$\ell$-winning} on a class of graphs $\CC$, if for every $G\in \CC$ and for every radius-$r$ Localizer strategy $\stratcon$, the $\ell$th position of $\RR(\stratcon,\stratflip,G)$ is a winning position. 
Note that while $\RR(\stratcon,\stratflip,G)$ is an infinite sequence, once a winning position is reached, it is only followed by winning positions.

\paragraph*{Runtime.}
Let $r \in \N$ and let $\stratflip$ be a radius-$r$ Flipper strategy. For a function $f\colon \N\to \N$, we say that $\stratflip$ has {\em{runtime}} $f$ if the following holds:
\begin{itemize}
 \item given a graph $G$, the internal state $\II_0(\stratflip,G)$ can be computed in time $f(|G|)$; and
 \item given a graph $H$ and an internal state $\II$, the value $\stratflip(H,\II)$ can be computed in time~$f(|G|)$. 
\end{itemize}
Note that in the second item above, the time complexity is allowed to depend on the original graph $G$, which is possibly much larger than the current arena $H$.
On the other hand, we do not require a dependence on the size of the encoding of $\II$.
Namely, it will always be the case that in positions that may appear in runs of $\stratflip$ on graphs from the considered class of graphs, the encoding size of $\II$ will be linear in the encoding size of $G$. Hence, positions with larger encoding size of $\II$ can be just ignored. (Formally, the algorithm outputs anything on them while not reading the whole internal state.)

We will often say that a strategy has runtime $\cal F$ for a class of functions $\cal F$ to indicate that it has runtime $f\in \cal F$. (For instance, we may say that a strategy has runtime $\Oh_{\CC,r}(n^2)$.)

\paragraph*{Playing multiple flips.} As discussed in \cref{sec:flippers}, we may also consider the variant of Flipper game where in every round, Flipper can apply not a single atomic flip $\flip F$, but a set of flips $F$ of size at most $g(i)$, where $i$ is the index of the round. Here, $g\colon \N\to \N$ is a function and we call this variant of the game {\em{$g$-bounded}}. We may also speak about the {\em{$k$-bounded}} variant of the game where $k\in \N$, and by this we mean the $g$-bounded game for $g$ being the constant function $g(i)=k$. Thus, the standard game is $1$-bounded. The notions of strategies, runs, and runtimes translate to the setting of $g$-bounded Flipper game naturally.

The following simple lemma shows that when designing a strategy for Flipper on a graph class, it suffices to consider the setting where playing multiple flips in a single move is allowed.

\begin{lemma}\label{lem:multi-runtime}
 Let $\CC$ be a class of graphs and $r\in \N$ be a fixed radius. Suppose that for some $\ell\in \N$ and functions $f,g\colon \N\to \N$, Flipper has a strategy in the $g$-bounded radius-$r$ Induced-Subgraph-Flipper game that is $\ell$-winning on $\CC$, and moreover this strategy has runtime $f$. Then Flipper also has a strategy in the standard ($1$-bounded) radius-$r$ Induced-Subgraph-Flipper game that is $\ell'$-winning on $\CC$, where $\ell'=\sum_{i=1}^\ell g(i)$, and this strategy has runtime $\Oh_{g,\ell}(f)$. 
\end{lemma}
\begin{proof}
 Let $\stratflip$ be the assumed strategy in the $g$-bounded game. We define a strategy $\stratflip'$ in the $1$-bounded game as follows. When playing $\stratflip'$, Flipper simulates $\stratflip$ by replacing the $i$th move in the $g$-bounded game by $g(i)$ consecutive moves in the $1$-bounded game. More precisely, supposing that $\stratflip$ proposes to play a flip set $F$ of size at most $g(i)$, in $\stratflip'$ Flipper plays the atomic flips of $F$ one by one, in $|F|$ consecutive rounds. Within the formal framework of strategies, these atomic flips are saved in a queue within the internal state, and popped from the queue one by one until the queue is empty --- and the next move of $\stratflip$ in the simulated game needs to be computed. The moves of Localizer along the way are ignored, except for the last one, which is considered the next Localizer's move in the simulated $g$-bounded game for the purpose of computing the next Flipper's move proposed by $\stratflip$.
 
 A straightforward induction argument shows that for every $j\in \N$, the arena after $\sum_{i=1}^j g(i)$ rounds in the $1$-bounded game played according to $\stratflip'$ is an induced subgraph of the arena after $j$ rounds in the simulated $g$-bounded game played according to $\stratflip$. Consequently, $\stratflip'$ is $\ell'$-winning on $\CC$ for $\ell'=\sum_{i=1}^\ell g(i)$. As for the runtime, the algorithm computing the next move of $\stratflip'$ either pops the next atomic flip from the queue stored in the internal state, or, in case the queue is empty, invokes the algorithm to compute the next move of $\stratflip$. It is straightforward to see that this can be done in time $\Oh_{g,\ell}(f)$. 
\end{proof}


%% file: afg/afg_strategy_new.tex
\subsection{Finalizing the argument}

With the definitions above settled, we can now rephrase and prove \Cref{thm:afg_main} as follows.

\begin{theorem}\label{thm:flipper_game}
	For every monadically stable class of graphs $\CC$ and radius $r\in\N$, there exists $\ell \in \N$ and a 
    radius-$r$ Flipper strategy for the Induced-Subgraph-Flipper game that is $\ell$-winning on $\CC$ and has runtime $\Oof_{\CC,r}(n^2)$.
\end{theorem}
\begin{proof}
 In notation, we fix the objects provided by \cref{lem:canonic_fuqw} for the class $\CC$.
 Let then $$t\coloneqq\wsize_{2r}^{-1}(7), \qquad k\coloneqq \wflips_{2r},\qquad\textrm{and}\qquad \ell\coloneqq 2\cdot \left(\binom{t}{5}+1\right)^2,$$
 where by $\wsize_{2r}^{-1}(7)$ we mean the least integer $N$ such that $\wsize_{2r}(N)\geq 7$. We will describe a strategy $\flipstrat$ for Flipper in the $g$-bounded radius-$r$ game, where
 $$g(i)\coloneqq \max(i,k).$$
 Strategy $\flipstrat$ will be $\ell$-winning on $\CC$ and will have runtime $\Oh_{\CC,r}(n^2)$. By \cref{lem:multi-runtime}, this suffices to prove \cref{thm:flipper_game}.
 
 We explain now $\flipstrat$ in natural language; the easy translation to the formal layer of strategies with internal states, described in \cref{sec:strategies-prelims}, is left to the reader.
 We fix the graph $G\in \CC$ on which the game is played, together with an arbitrary linear order $\preceq$ on $V(G)$.
 
 First, Flipper will always play moves in {\em{move pairs}}: Having constructed some flip set $F$, Flipper first applies $F$ to the current arena $H$, then lets the Localizer localize the game to a radius-$r$ ball in $H\oplus F$, and finally he applies $F$ again. In this way, the following invariant will be satisfied:
  After applying every move pair, the arena is an induced subgraph of $G$ (cf.~\cref{obs:flips_cancel} and \cref*{obs:subgraph_flip}). We will say that a move pair as described above is {\em{defined}} by $F$.

 Second, Flipper proceeds in a sequence of {\em{eras}}, each consisting of a number of consecutive moves. Along the way, he keeps track of a growing chain of vertex subsets
 $$\emptyset=X_0\subsetneq X_1\subsetneq X_2\subsetneq X_3\subsetneq \ldots,$$
 where $X_i$ is obtained from $X_{i-1}$ at the end of era $i$ by adding one vertex that is still contained in the arena, but not contained in $X_{i-1}$. 
 Up until Flipper wins the game, we will ensure that such a vertex always exists, and therefore $|X_i|=i$ for every $i\in\NN$ until the game concludes.

 We now describe Flipper's moves in era $i$ ($i=1,2,3,\ldots$). For every $Z\subseteq X_{i-1}$ with $|Z|= 5$, we compute the flip set
 \[F_Z\coloneqq \guessflips_{2r}(G,\preceq, Z).\]
 Note that, instead of the current arena, the original graph $G$ is used to compute $F_Z$.
 First, Flipper performs $\binom{|X_{i-1}|}{5}$ move pairs, each defined by $F_Z$ for a different $Z$ as above. 
 Finally, let $F$ be a flip set of size $|X_{i-1}|=i-1$ such that in $G\oplus F$, every vertex of $X_{i-1}$ is isolated. 
 (Such $F$ can be obtained by iteratively isolating vertices of $X_{i-1}$ by performing a flip between a vertex and its neighborhood.) 

 At the end of the era, Flipper applies the move pair defined by $F$. 
 After the first application of $F$ within the move pair, the resulting arena is an induced subgraph of $G\oplus F$ where all the vertices of $X_{i-1}$ are isolated.
 Therefore, the induced subgraph chosen as Localizer's response must contain a vertex $x$ not belonging to $X_{i-1}$, otherwise Localizer loses immediately after making her move. 
 Followingly, we may set $X_i\coloneqq X_{i-1}\cup \{x\}$ and proceed to the next era.
 
 This concludes the description of $\flipstrat$. Clearly, $\flipstrat$ is a valid strategy in the $g$-bounded game. We now argue that following $\flipstrat$ leads to a quick victory.
 
 \begin{claim}\label{cl:termination}
  If Flipper follows $\flipstrat$, the game concludes within at most $t$ eras.
 \end{claim}
 \begin{claimproof}
  For contradiction, suppose the game enters era $t+1$ without termination. Denote $X\coloneqq X_t$; we have $|X|=t=\wsize_{2r}^{-1}(7)$. Let
  $$(Y,F)\coloneqq \FF_{2r}(G,\preceq,X).$$
  Thus, $|Y|\geq 7$ and $Y$ is distance-$2r$ independent in $G\oplus F$. Let $y_1,\ldots,y_7$ be any seven distinct vertices of $Y$, where $y_i$ was added earlier to $X$ than $y_j$ for all $i<j$.
  
  Let
  $$Z\coloneqq \{y_1,\ldots,y_5\}\qquad \textrm{and} \qquad F_Z\coloneqq \guessflips_{2r}(G,\preceq,Z).$$
  Further, let $s$ be the index of the era that concluded with adding $y_6$ to $X$. (That is, we have $X_s=X_{s-1}\cup \{y_6\}$ and in particular $X_{s-1} \cap \{y_6,y_7\} = \varnothing$.) Note that $Z\subseteq X_{s-1}$, hence within era $s$, Flipper applied the move pair defined by $F_Z$. 
  Let $H$ be the arena during that era right before the first application of $F_Z$.
  Note that $H$ is an induced subgraph of $G$.
  After the first application of $F_Z$,
  Localizer responded by restricting the arena to an induced subgraph $H'$ of some radius-$r$ ball in $H \oplus F_Z$. Clearly, the vertex set of $H'$ is entirely contained in some radius-$r$ ball in $G\oplus F_Z$. Since $Y$ is distance-$2r$ independent in $G\oplus F_Z$ and $y_6,y_7\in Y$, we conclude $\{y_6,y_7\}\nsubseteq V(H')$. In other words, at least one of the vertices $y_6,y_7$ got removed from the arena as a consequence of Localizer's move.
  This contradicts the assumption that both $y_6$ and $y_7$ were later added to $X$,
  which requires them to both be contained in the arena at the end of era $s$.
 \end{claimproof}

 Note that in era $i$, Flipper applies exactly $\binom{|X_{i-1}|}{5}+1=\binom{i-1}{5}+1$ move pairs. Hence, by \Cref{cl:termination}, the game terminates within at most
 $$\sum_{i=1}^{t} 2\cdot \left(\binom{i-1}{5}+1\right)\leq \ell\quad \textrm{rounds.}$$
 We conclude that $\flipstrat$ is $\ell$-winning on $\CC$, as promised. Finally, computing Flipper's moves for an era boils down to at most $\binom{t}{5}=\Oh_{\CC,r}(1)$ applications of the algorithm provided by \cref{lem:canonic_fuqw}, which runs in time $\Oh_{\CC,r}(n^2)$. It follows that $\flipstrat$ has runtime $\Oh_{\CC,r}(n^2)$ (in a suitable formalization of the strategy through internal states).
\end{proof}